\newif\ifcheckpagelimits
\checkpagelimitstrue
\checkpagelimitsfalse

\ifcheckpagelimits
 \documentclass[nofootinbib,pra,aps,twocolumn,showpacs,showkeys,%
 amsmath,amssymb,superscriptaddress,final,reprint,floatfix,longbibliography]{revtex4-1}
 \newcommand{\todo}[1]{}
\else
 \documentclass[pra,aps,twocolumn,showpacs,showkeys,%
 amsmath,amssymb,superscriptaddress,final,reprint,floatfix,longbibliography]{revtex4-1}
 \newcommand{\todo}[1]{{\pdfmargincomment[icon=Note,color=pink]{#1}}}
\fi

\usepackage{helvet} 
\usepackage{lineno}
  \usepackage{mathptmx}
\usepackage{subfigure}
\usepackage{hhline} 
\usepackage{dcolumn}
\usepackage{amsmath,amssymb}
\usepackage{bm}
\usepackage{color}
\usepackage{overpic}
\usepackage{latexsym}
\usepackage{epstopdf}
\usepackage{color}
\usepackage[english]{babel}
\usepackage{latexsym}
\usepackage{stmaryrd}

\usepackage{braket}

\definecolor{mygrey}{gray}{0.35}
\definecolor{myblue}{rgb}{0.2,0.2,0.8}
\definecolor{myzard}{cmyk}{0,0,0.05,0}
\definecolor{mywhite}{rgb}{1,1,1}
\definecolor{myred}{rgb}{1,0.,0.3}

\usepackage[colorlinks=true,citecolor=myblue,linkcolor=myred]{hyperref}
\DeclareMathAlphabet{\mathpzc}{OT1}{pzc}{m}{it}

 \def\ee{\mathord{\rm e}}
 
 \def\ii{\mathord{\rm i}}

\def\half{\textstyle\frac{1}{2}}

\renewcommand{\ii}{{\rm i}}
\renewcommand{\ee}{{\rm e}}
\def\beq{\begin{equation}}
\def\eeq{\end{equation}}

\def\barray{\begin{eqnarray}}
\def\earray{\end{eqnarray}}

\newcommand{\FontsizeTables}{\footnotesize} 

\begin{document}

\title{Assessing the progress of trapped-ion processors towards fault-tolerant quantum computation }

\author{A. Bermudez}
\affiliation{Department of Physics, College of Science, Swansea University, Singleton Park, Swansea SA2 8PP, United Kingdom}
\affiliation{Instituto de F\'{i}sica Fundamental, IFF-CSIC, Madrid E-28006, Spain}

\author{X. Xu}
\affiliation{Department of Materials, University of Oxford, Oxford OX1 3PH, United Kingdom}

\author{R. Nigmatullin}
\affiliation{Complex Systems Research Group, Faculty of Engineering and IT, The University of Sydney, Sydney,
Australia}
\affiliation{Department of Materials, University of Oxford, Oxford OX1 3PH, United Kingdom}

\author{J. O'Gorman}
\affiliation{Department of Materials, University of Oxford, Oxford OX1 3PH, United Kingdom}

\author{ V. Negnevitsky}
\affiliation{Institute for Quantum Electronics, ETH Z\"urich, Otto-Stern-Weg 1, 8093 Z\"urich, Switzerland}

\author{P. Schindler}
\affiliation{Institute for Experimental Physics, University of Innsbruck, 6020 Innsbruck, Austria}

\author{T. Monz}
\affiliation{Institute for Experimental Physics, University of Innsbruck, 6020 Innsbruck, Austria}

\author{U. G. Poschinger}
\affiliation{Institut f\"ur Physik, Universit\"at Mainz, Staudingerweg 7, 55128 Mainz, Germany}

\author{C. Hempel}
\affiliation{ARC Centre for Engineered Quantum Systems, School of Physics, The University of Sydney, NSW 2006 Australia}

\author{J. Home}
\affiliation{Institute for Quantum Electronics, ETH Z\"urich, Otto-Stern-Weg 1, 8093 Z\"urich, Switzerland}

\author{F. Schmidt-Kaler}
\affiliation{Institut f\"ur Physik, Universit\"at Mainz, Staudingerweg 7, 55128 Mainz, Germany}

\author{M. Biercuk}
\affiliation{ARC Centre for Engineered Quantum Systems, School of Physics, The University of Sydney, NSW 2006 Australia}

\author{R. Blatt}
\affiliation{Institute for Experimental Physics, University of Innsbruck, 6020 Innsbruck, Austria}
\affiliation{Institute for Quantum Optics and Quantum Information of the Austrian Academy
of Sciences, A-6020 Innsbruck, Austria}

\author{S. Benjamin}
\affiliation{Department of Materials, University of Oxford, Oxford OX1 3PH, United Kingdom}

\author{M.~M\"uller}
\affiliation{Department of Physics, College of Science, Swansea University, Singleton Park, Swansea SA2 8PP, United Kingdom}

\begin{abstract}
 A  quantitative assessment of the  progress of small prototype quantum processors towards fault-tolerant quantum computation  is a problem of current interest in experimental and theoretical quantum information science. We  introduce   a  necessary and fair criterion for quantum error correction (QEC), which must be achieved in the development of these  quantum processors before their sizes are  sufficiently big    to consider  the well-known QEC threshold. We apply this criterion to benchmark the ongoing effort in implementing  QEC with topological color codes using trapped-ion quantum processors and, more importantly, to guide the  future hardware developments that shall be required in order to demonstrate beneficial QEC with small topological quantum codes. In doing so, we present a thorough description of a realistic trapped-ion toolbox for QEC, and a physically-motivated error model that goes beyond standard simplifications in the QEC literature. We focus on laser-based quantum gates realised in  two-species trapped-ion crystals in high-optical aperture segmented traps. Our large-scale numerical analysis shows that, with the foreseen technological improvements hereby described, this platform is a very promising candidate for fault-tolerant quantum computation.
  
\end{abstract}


\maketitle

\makeatletter
\makeatother
\begingroup
\hypersetup{linkcolor=black}
\tableofcontents
\endgroup

\section{\bf Introduction}
\label{sec:intro}

Solving hard computational problems by exploiting  the quantum-mechanical laws of Nature is one of the goals of current scientific and technological research~\cite{nielsen-book}. To turn this idea into  experimental reality, intense research efforts are currently devoted  to scale  existing small prototypes, which have served for proof-of-principle demonstrations~\cite{qc_review}, into larger quantum devices capable of processing information quantum-mechanically even in the presence of  noise and processing errors (i.e. fault-tolerantly). This poses a significant challenge from both fundamental and technological perspectives. 

Fundamentally, the quantum-mechanical features responsible for the advantage of these  processors with respect to their classical counterparts,  also give rise to a different behavior with respect to noise and errors, which excludes the straightforward application of classical error correction schemes. Despite these difficulties, the theory of quantum error correction (QEC)~\cite{qec_shor,calderbank-pra-54-1098,steane-prl-77-793} has shown a well-defined route for the   development of large-scale quantum computers. The main ingredients of QEC to combat the detrimental impact   of  noise are:  {\it (i)} encoding  quantum information redundantly in ever-larger quantum registers, and   {\it (ii)} detecting and correcting errors during a computation without  altering the encoded quantum information. Exploiting these ingredients using particular  QEC codes, it has been shown theoretically that it is possible to perform quantum computing sequences of arbitrary complexity  fault-tolerantly if the noise/error of elementary operations is maintained below a certain threshold~\cite{FTQC}. The redundant encoding of the information in these QEC protocols, which is required to improve  the level of protection against noise for a fault-tolerant computation, can be achieved by either {\it (a)} concatenating elementary  codes in several layers~\cite{steane-prl-77-793}, or {\it (b)} storing the information in  topological features of  registers of increasing size~\cite{toric_code,bombin-prl-97-180501}.  The quest is therefore to implement these QEC ideas in quantum devices of increasing sizes. 

The first experiments on QEC have implemented the  3-qubit~\cite{3_qubit_code_nmr} and 5-qubit~\cite{5_qubit_code_nmr}  quantum codes in nuclear magnetic resonance. Starting from initial experiments on the 3-qubit code with trapped ions~\cite{3_qubit_code_nist_ions} and superconducting circuits~\cite{3_qubit_code_superconducting}, these two platforms have recently been employed to show  repetitive error correction~\cite{3_qubit_code_repetitive_qec_ions}, fault-tolerant error detection with a four-qubit code~\cite{ft_error_detection_monroe}, and small-scale versions of the topological color~\cite{nigg-science-345-302} and surface~\cite{surface_code_superconducting_line,surface_code_superconducting_square} codes. We note, however, that QEC is also being pursued in other platforms~\cite{repetitive_QEC_NV}. The   theory of QEC, described in the paragraph above, defines a clear roadmap towards the demonstration of fault tolerance in  large quantum processors. However, despite this remarkable     progress, the hardware platforms  are still far away from the  sizes that are required to render the errors on the encoded data negligibly small.  Hence, it would be desirable to define a set of intermediate QEC goals, which are necessary for the progress  towards the  fully-fledged fault-tolerant quantum computer, and  can serve as a guiding principle in the experimental   design by benchmarking the progress in building and scaling these smaller quantum codes.

\begin{figure}[t]
 \begin{centering}
  \includegraphics[width=1\columnwidth]{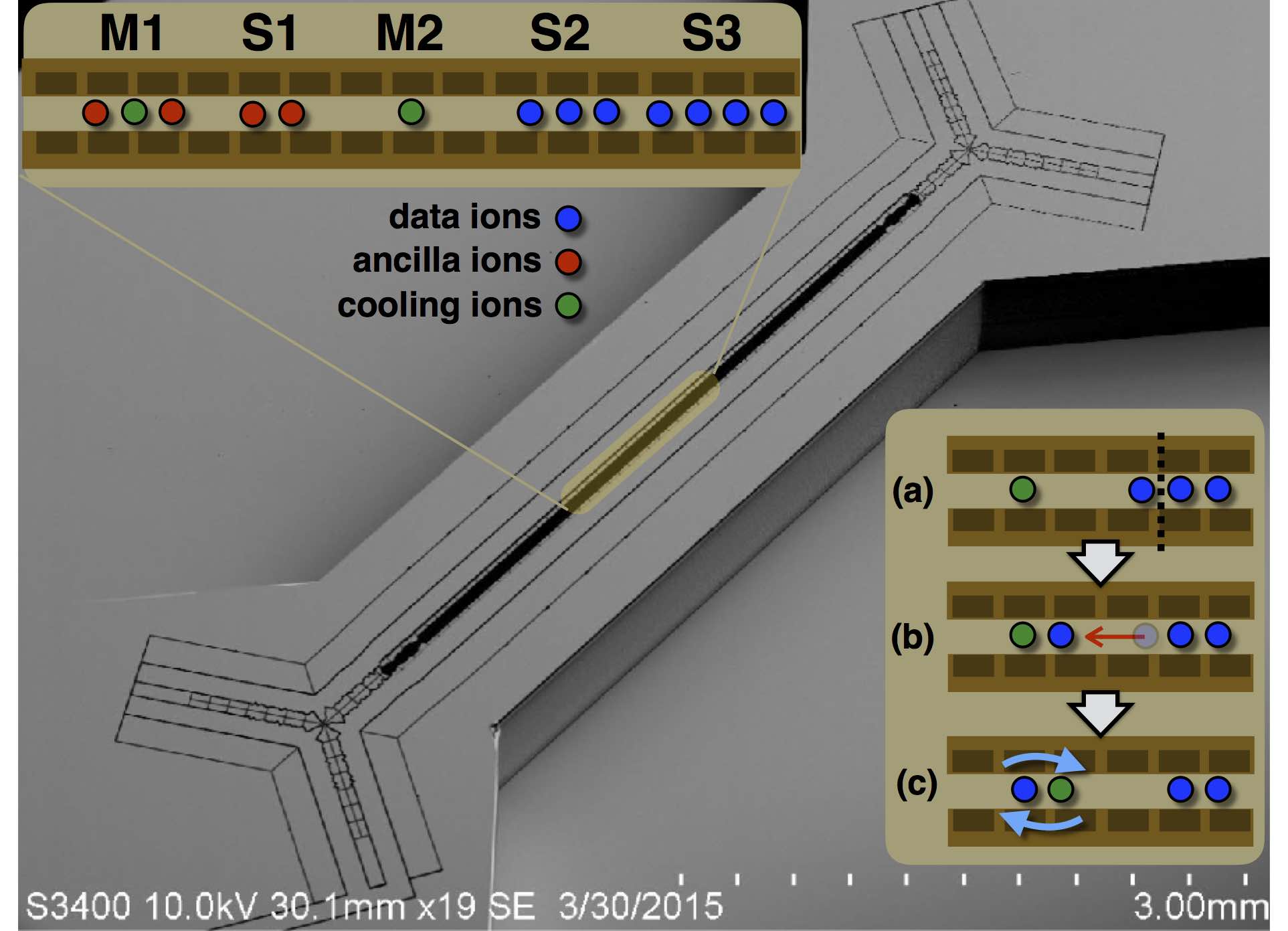}\\
  \caption{\label{Fig:ExpHoa} {\bf The Sandia HOA2 trap as a QEC platform:} In our envisioned scheme, 
 $^{40}$Ca$^+$ ions (blue and red dots) are co-trapped with $^{88}$Sr$^+$ ions (green dots) in a quantum zone divided in three storage regions $S_1,S_2,S_3$ and two manipulations zones $M_1,M_2$. Some of the $^{40}$Ca$^+$ ions can be used as data qubits to encode quantum information according to a QEC code (blue dots), while others (red dots) can be used as ancilla qubits for syndrome extraction. The $^{88}$Sr$^+$ ions (green dots)  are used as sympathetic coolants to reduce the number of phonons prior to the entangling gates.   Possible  crystal reconfiguration operations are shown in the panel of the lower right corner: {\bf  (a)} Splitting of an ion crystal, {\bf (b}) shuttling of an ion and subsequent merging with another ion to form a  crystal, and {\bf (c)} rotation (swapping) of a mixed species crystal. Schematics of the trap adapted from a micrograph in~\cite{hoa_source}.}
\end{centering}
\end{figure}

A necessary condition for  QEC  is that the effect of a complete round of error detection and correction must be {\em beneficial} for the encoded qubit. This is a non-trivial condition since the effect of an attempt at error correction, while  aiming to correct the existing errors, inevitably introduces risk of new ones. Accordingly, quantifying such a crossover into beneficial/useful QEC, and certifying that it is met in a particular QEC code,  will translate into specific requirements on the fidelities of the various gates, measurements, and other internal processes that conform the QEC cycle. This can establish a set of  goals that must be achieved by  future  hardware development. Once this is achieved, another necessary criterion is  to verify if the  encoding, followed by  a complete round of error detection and correction, is  { beneficial} in comparison to the degradation of an unprotected physical qubit subjected to the same sources of physical noise during the time required by the QEC cycle.

We note that the theoretical studies of the performance of different QEC strategies, quantified by the particular value of the threshold, depend crucially on the assumptions about the underlying platform capabilities and noise models. Using over-simplified noise models, or unrealistic platform capabilities, can lead to an overestimation or underestimation of the correcting power of a given QEC code. Therefore, if we are aiming at  assessing and guiding the progress of a particular experimental platform by the above intermediate QEC goals, a very careful microscopic modelling  of the noise and the operational errors is required. The objective of this work is to present a detailed study along these lines for  trapped-ion quantum processors with current and anticipated future capabilities in the near term.

\subsection*{Summary of the results of this work}

In this manuscript, we quantify the above intermediate  goals for beneficial QEC by introducing a  quantum-information protocol  with a  clear and intuitive operational meaning   in Sec.~\ref{sec:QEC_criterion}. This protocol can serve as a  benchmark scenario to assess the progress of experimental QEC codes.

We focus on trapped-ion implementations of small QEC codes~\cite{3_qubit_code_nist_ions,3_qubit_code_repetitive_qec_ions,ft_error_detection_monroe, nigg-science-345-302,phase_opt_color_code}, and use the above measure to assess theoretically the methodological and technological improvements that would be required to reach the break-even point for a logical qubit, i.e. to enter the regime of beneficial QEC. In order to reach this goal, it is of the utmost importance  to choose and adapt QEC schemes according to the particular technological advantages and disadvantages of the hardware platform at hand.  In particular, one must  exploit the particular technological strengths and simultaneously mitigate the dominant sources of noise.  This requires a detailed knowledge of the state of the art and foreseeable technological improvements, which we discuss in Sec.~\ref{sec:expt_system}. We present a thorough description of an experimental  toolbox for QEC using a high-optical access segmented ion trap to manipulate dual-species ion crystals in a cryogenic environment (see Fig.~\ref{Fig:ExpHoa}).  We consider  a universal set of single-qubit and multi-qubit entangling gates~\cite{schindler-njp-15-123012} that differs from  the more standard CNOT-based approaches~\cite{nielsen-book}. The current and anticipated performance of these elementary operations, as discussed in detail below, is summarized in Table~\ref{tab:summary_gates}. These tools shall be combined with spectroscopic decoupling of a subset of ions (i.e. {\it hiding-based approach}), and with crystal-reconfiguration techniques (i.e. {\it shuttling-based approach}) summarized in Table~\ref{tbl:shuttlingops}, which include splitting ion crystals, shuttling ions across trap segments, and merging two sets of ions into a larger crystal. Together with the possibility of using a dual-species crystal for sympathetic re-cooling of the ions and stabilizer readout, this toolbox contains all the ingredients required for trapped-ion QEC. In addition to this  knowledge of  experimental capabilities, assessing the prospects of QEC  also requires a detailed modelling of the main  sources of noise and errors for these operations, which  we address in Sec.~\ref{sec:error_models}.

 Equipped with this toolbox, we  develop different approaches for trapped-ion QEC in Sec.~\ref{sec:cnot_alternative}. We start by describing how multi-qubit M\o lmer-S\o rensen (MS)  gates~\cite{PhysRevA.62.022311,molmer-prl-82-1835}  can be exploited    for efficient stabilizer readout~\cite{mueller-njp-13-085007}, as   experimentally demonstrated in~\cite{nigg-science-345-302}. In the context of fault-tolerant QEC, however,  different  schemes based on one-qubit gates and two-qubit MS gates would be required. Since  fault-tolerant QEC schemes have been typically conceived using single-qubit and two-qubit CNOT gates~\cite{shor_ft_qec,aliferis_ft_qec}, it would be desirable  to devise trapped-ion circuits that exploit MS gates directly, and to study how errors propagate on those circuits  to demonstrate fault tolerance. We address these points by presenting a detailed description of a generic MS-based toolbox for QEC.  We  apply this toolbox to  the  7-qubit topological color code with trapped ions,  either using a  {\it non-fault-tolerant stabilizer readout} with 7 data qubits and 1 additional ancillary qubit (i.e. 7+1-qubit scheme) based on  multi-qubit/ sequential two-qubit MS gates, or by using {\it  fault-tolerant stabilizer readout} via  MS-based schemes that realize the equivalent of the CNOT DiVincenzo-Shor (DVS) protocol for 7+5 qubits~\cite{shor_ft_qec}, or the DiVincenzo-Aliferis (DVA) protocol for 7+4 qubits~\cite{aliferis_ft_qec}. Although we have focused on this particular code, we remark that this trapped-ion QEC toolbox for stabilizer readout can be generalized to any other stabilizer QEC code of interest, and scaled to larger-size codes in a modular fashion.

The MS-based  stabilizer readout is used, in combination with some of the elementary operations of Tables~\ref{tab:summary_gates} and~\ref{tbl:shuttlingops}, as a building block for the development of  trapped-ion QEC protocols in Sec.~\ref{sec_qec_protocols}. As already outlined above, we explore different  scenarios according to varying experimental capabilities: 
\begin{enumerate}
\item
\textbf{Shuttling-based protocol}: Here, we consider trapped-ion crystals with either a single or two ion species, i.e.~data  and  ancillary qubits being encoded in the same/different atomic  species. We develop sequences of crystal-reconfiguration operations and stabilizer mappings to perform a full QEC cycle on a single logical qubit. We explore how the ability of crystal re-cooling    by sympathetic cooling via the ancillary ion at intermediate stages  affects the performance of the  protocol. 
\item
\textbf{Hiding-based protocol}: Here, we consider the  protocols  realized in a  static ion crystal. Qubits are selectively addressed by shelving inactive ions via spectroscopic de- and re-coupling pulses,  and combined with stabilizer mappings to perform a full QEC cycle on a single logical qubit. We consider encoding of data and  ancillary qubits in  two different species, and the possibility to apply re-cooling after the readout. 
\end{enumerate}

 These QEC protocols are complemented with the with the error model introduced in Sec.~\ref{sec:error_models}, which improves upon  customary circuit-error models that consider a unique quantum channel for all  elementary operations in a QEC cycle. This allows us to perform a detailed study that goes beyond standard, albeit not very realistic, assumptions: {\it (i)} we consider that the different gates (including the identity), the state preparation, and the measurements, do not take the same amount of time. {\it (ii)} We use distinct error channels affecting the different stages of the QEC protocols. For instance, idle qubits are subjected to dephasing in a trapped-ion setup, whereas single- and multi-qubit gates are subjected to depolarising noise. More importantly, {\it (iii)} the different channels are not all characterized by a unique  error  probability. Certainly, single- and multi-qubit gates do not have the same error in any known experimental platform. We use a microscopic modelling of the ion crystals to derive the particular expressions/values of the corresponding error rates for each operation. Therefore, our treatment does not only go beyond  models that do not consider, or simplify, the occurrence of errors on the syndrome readout, but it also goes beyond the standard so-called circuit-level noise model, which typically makes these over-simplifications.

 These sections set the stage for a  large-scale numerical analysis that investigates the performance of such protocols in Sec.~\ref{sec:numerical_studies}. The introduced  criterion for beneficial/useful QEC is used to  quantify the three essential requirements that will need to be
met in forthcoming experiments for trapped-ion QEC: {\it (i)} sufficiently small  natural physical error rates from fundamental error sources; {\it (ii)}  to detect and dynamically correct errors at a fast enough rate; and 
{\it (iii)} sufficiently accurate  realizations of unavoidably imperfect
error-correction routines, so that there still remains an overall gain
of applying (imperfect) QEC procedures.

Finally, in Sec.~\ref{sec:conclusions}, we present our conclusions.

\section{\bf Assessing the progress on quantum error correction (QEC) by small  quantum codes}
\label{sec:QEC_criterion}

While small quantum codes have already been demonstrated on different platforms \cite{3_qubit_code_nmr,5_qubit_code_nmr,3_qubit_code_nist_ions,3_qubit_code_superconducting,3_qubit_code_repetitive_qec_ions,ft_error_detection_monroe, nigg-science-345-302,surface_code_superconducting_line,surface_code_superconducting_square}, it would be of interest to ultimately demonstrate fault tolerance on existing or near-future hardware. However,  this would require showing the supremacy (i.e. reducing the error rate)   of the small codes with respect to the  {\it best-possible un-encoded qubits on all representative quantum circuits} belonging to a large set of protocols~\cite{gottesman_small_codes}. Depending on the platform, this comparison can be very stringent. For instance, ion-trap processors can use  decoherence-free qubits~\cite{dfs} or the $\mu$-metal shielded Zeeman qubits~\cite{Ruster2016} with very long coherence times, such that the error on the identity quantum circuit would be very hard to beat using any small QEC code. Additionally, single-qubit and two-qubit gates with bare qubits have, so far, the smallest achieved in-fidelities in any experimental platform~\cite{16Ballance,PhysRevLett.117.060505}, and it also seems  unlikely that small QEC codes, with their large overhead in complexity, will be capable of beating them. 
We thus believe that alternative   criteria have to  be established, which serve as reasonable guiding principles in the development of future technologies that improve upon existing QEC codes.

\subsection*{Break-even point for useful QEC}

In this section, we introduce the criteria used in our work to judge whether a particular combination of hardware and quantum code can successfully perform QEC,   sustaining thus a logical encoded qubit. As mentioned in the introduction,  a first necessary condition that must be verified by any  implementation of a QEC code is that the effect of a complete round of error detection and correction proves to be {beneficial}.  In order to make it quantitative,  we must define a measure for the quality, or {\it integrity}, of a logical qubit. 

The {\it fidelity} of the logical  encoded state subjected to noise/errors $\tilde{\rho}_L$ with respect to its ideal form $\rho_L=\ket{\psi_L}\bra{\psi_L}$, namely $\mathcal{F}=\langle \psi_{L}|\tilde{\rho
}_L|\psi_L \rangle$, and in particular how it  changes if we perform a QEC cycle on the imperfect $\tilde{\rho}_L$ or not, might first appear as  a natural measure. However, one encounters difficulties. Consider  a logical qubit $\tilde{\rho}_L$ that has completely decohered under the effect of independent depolarizing noise on the constituent $n$ physical qubits. The collective entity no longer contains any information about the  initial logical state $\ket{\Psi_L}$. Its fidelity is $\mathcal{F}=2^{-n}$ since  the system is in an equal mixture of all possible states regardless of the initial encoded state. The problem is that a round of error correction will {\em seem} to improve the quality of this logical qubit: it will map all the states in the mixture  to either the logical zero $\ket{0}_L$, or the logical one $\ket{1}_L$, creating a mixture of these two $\rho_{\rm QEC}=\half(\ket{0}_L\bra{0}_L+\ket{1}_L\bra{1}_L)$. Consequently the fidelity will rise to $\mathcal{F}_{\rm QEC}=\frac{1}{2}$ under an ideal QEC cycle (or  close to this number for imperfect correcting circuits). Thus, if we were to select fidelity as our measure for the quality of a logical qubit, we would be faced with the unsatisfactory feature that a logical qubit that has been completely lost, and is free of any meaningful information, can seem to be partially recovered. 

One might attempt to correct this issue by projecting the $n$-qubit state into the logical subspace, and only then computing the fidelity. Nonetheless, this leads to another unsatisfactory feature: one would find an equality in apparent performance between a device that maintains the logical qubit entirely in that subspace and one which allows a large component of the state to leave the subspace, {\it regardless} of the nature of the part of the state outside the proper subspace. As we will later remark, the approach we  take in this work can be thought of as essentially a more sophisticated variant of this idea. 

\begin{figure}
 \begin{centering}
  \includegraphics[width=1.0\columnwidth]{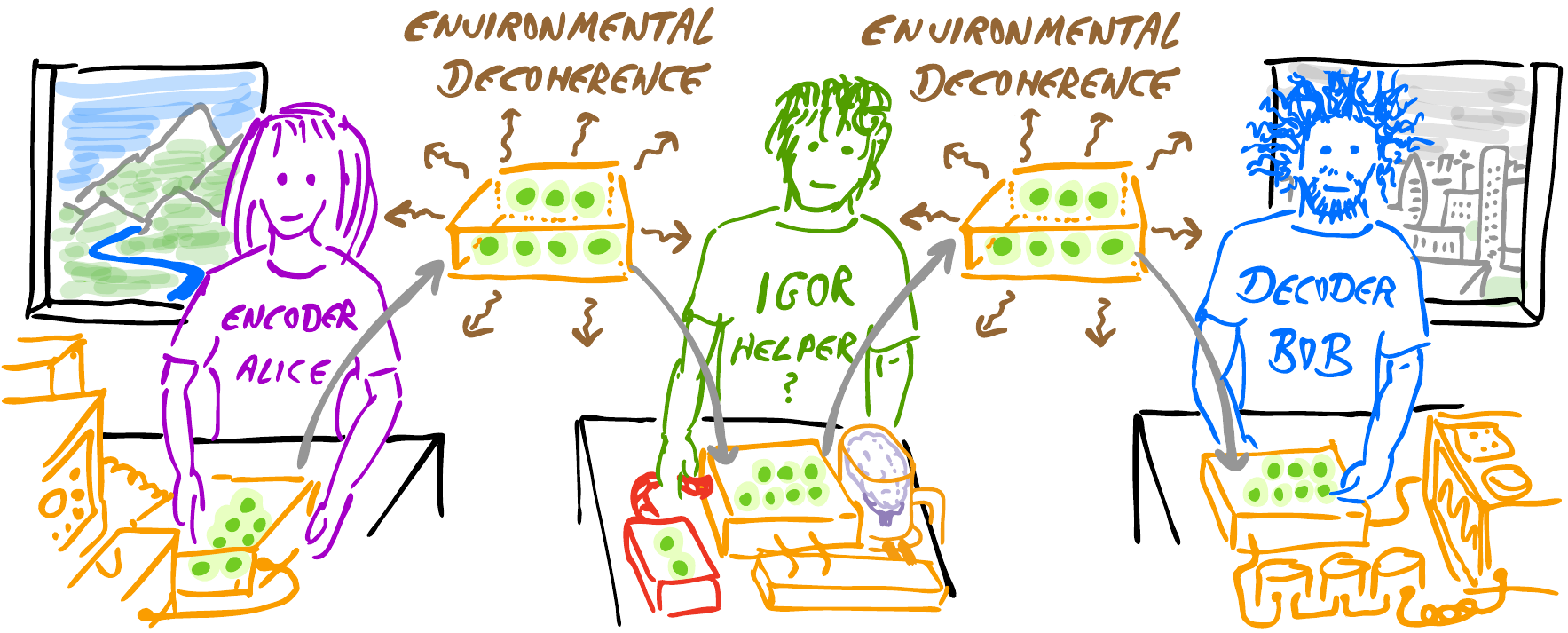}\\
  \caption{\label{Fig:IgorFig} {\bf Cartoon illustration of the protocol for assessing the efficacy of our QEC cycle}. Strictly, Alice and Bob have ideal experimental equipment capable of encoding and decoding a quantum state perfectly, whereas only Igor  has the imperfections of our real laboratory setting. The stages of the protocol are detailed in Table~\ref{table:first_criterion}.}
\end{centering}
\end{figure}

The alternative measure we will employ has a very clear and intuitive operational meaning. It is best illustrated with a quantum-information protocol that separates the role of {\em encoding} the logical qubit, from the task of {\it reading} it. We will use the  labels Alice and Bob for two  entities that have these roles (see Fig.~\ref{Fig:IgorFig}). Now suppose that a random qubit state $\ket{\psi}=\alpha\vert{0}\rangle+\beta\vert{1}\rangle$ is selected, and Alice is instructed to prepare a logical qubit $\ket{\psi_L}=\alpha\vert{0}\rangle_L+\beta\vert{1}\rangle_L$ using the code of her choice. This logical qubit of $n$ physical qubits $\rho_L$ is then   subjected to some noise channel, which may have any form, including correlated noise (e.g. spatially or temporally correlations arising from global fluctuating magnetic fields with a characteristic correlation time) and coherent noise (e.g. resulting from over-rotations in single-qubit gate operations). The now-imperfect logical qubit $\tilde{\rho}_L$ is presented to Bob, along with the following classical information: ``The original state was either $\ket{\psi}=\alpha\vert{0}\rangle+\beta\vert{1}\rangle$ or $\ket{\psi_\bot}=\beta^*\vert{0}\rangle-\alpha^*\vert{1}\rangle$.'' In other words, Bob is given a choice of two states, the true qubit state and the state which is orthogonal to it. Now Bob is challenged to make his best guess as to which state the $n$ physical qubits encode: it is a problem of {\it state discrimination} with the imperfect logical qubit as the resource. He may use any physically allowed process in his analysis, and in particular he can perform error correction, decode to a single physical qubit, and destructively measure it. 

 For simplicity,  we will assume that Alice and Bob are ideal agents in the following, i.e. encoding and analysis of the logical qubit takes place perfectly. Then the probability $\mathcal{P}_{\rm B}(\rho_L,\tilde{\rho}_L)$ that Bob guesses correctly will vary only with the quality of the received qubit: an error-free logical qubit will score $\mathcal{P}_{\rm B,max}=1.0$, since Bob will certainly succeed in his state discrimination task, whereas (for example) a logical qubit which has undergone complete depolarization will score $\mathcal{P}_{\rm B,min}=0.5$ since Bob can only guess randomly.
We define the integrity $\mathcal{I}$ of our memory as simply a scaled probability, 
\begin{equation}
\mathcal{I}=2\mathcal{P}_{\rm B}(\rho_L,\tilde{\rho}_L)-1
\label{eqn:scaledIntegrity}
\end{equation}
The scaling thus provides us with the natural limits of unity for a perfect memory, and zero for a memory which provides Bob with no useful information whatsoever.

For any given decoherence model one can find the probability that Bob will guess correctly given a fully decohered logical qubit. If, as in the following analysis, the decoherence is restricted to a specific channel, then Bob's performance can be higher, i.e. there will be instances in the random selection of the qubit state to be  encoded by Alice that happen to be robust against the specific channel. In the cases we will be concerned with in the following section,  restricting to a purely dephasing environment is an excellent approximation, in which case if we happen to select $\vert{0}\rangle$ or $\vert{1}\rangle$ for Alice to encode,  the effect of decoherence on the encoded qubit will not degrade Bob's capacity to differentiate: he need only measure all qubits in the $z$-basis and determine whether the observed pattern belongs to the set of states associated with $\vert{0}\rangle_L$ or $\vert{1}\rangle_L$. Conversely, if we had happened to select $\vert{+}\rangle$ or $\vert{-}\rangle$ for Alice to encode, then after full dephasing Bob will not be able to gain any value from his analysis and his probability of guessing correctly will be $0.5$. His performance when Alice randomly selects qubit states, sampled uniformly over the Bloch sphere, is found to be $\mathcal{P}_{\rm min}=0.75$ after full dephasing. Therefore when we plot the average $\mathcal{P}$ for any degree of pure dephasing we will find it varies in the range $0.75\leq\mathcal{P}_{\rm B}(\rho_L,\tilde{\rho}_L)\leq1$. 

Armed with this notion of the integrity of a qubit as, essentially, the extent to which its state can be read out by Bob, we now  ask the question of  whether the QEC cycle is beneficial or harmful by allowing  for an imperfect round of error detection and correction prior to Bob's guess $\tilde{\rho}_L\to\rho_{\rm QEC}$. The full protocol for our measure, where the code to be used (e.g. surface code, 2D color code, etc) is to be specified, is described in Table~\ref{table:first_criterion}.   According to our criterion, the round of imperfect error correction is now deemed to be beneficial if Bob's probability of subsequently discriminating the state correctly is higher when we indeed perform a round of QEC, versus simply opting not to do so, and allowing the environment to act for time $\tau$ uninterrupted, namely
\beq
\label{eq:be_pont}
\mathcal{P}_{\rm B}(\rho_L,{\rho}_{\rm QEC})>\mathcal{P}_{\rm B}(\rho_L,\tilde{\rho}_L).
\eeq
The break-even point for a beneficial QEC occurs when Eq.~\eqref{eq:be_pont} is satisfied. For convenience of exposition we may imagine that a third party, besides Alice and Bob, is responsible for the cycle of error correction: since this individual is effectively a flawed assistant for Bob, we use the name Igor after the famous fictional lab assistant (see Fig.~\ref{Fig:IgorFig}). Then, our criterion for successful error correction can be summarized as, {\it ``Is Igor a help or a hindrance to Bob?"}.

It is worth noting that Alice's encoding protocol is predetermined and may not vary with the particular choice of qubit she is instructed to encode. Similarly Igor, who does not have the classical description of the encoded qubit, will always perform the same procedure as he attempts to correct it. Moreover, in all the analysis presented in this paper, we also fix Bob's protocol: he simply performs his own (perfect) round of error correction, then decodes the logical qubit to a single qubit and measures that qubit in  the basis of his choice. His optimal basis choice, for all cases considered here, is simply $\{\ket{\psi},\ket{\psi_\bot}\}$ and Bob makes his `guess' according to the outcome. Thus only the final step, the measurement, depends on the choice of encoded qubit which was issued to Alice. It is interesting to observe that with this choice of Bob's protocol, Bob is effectively mapping the state of the $n$ qubits into the logical subspace (with his round of perfect error correction) and then making a guess with a probability of success given by the fidelity of the corrected logical qubit. Then our concept of {\it integrity}  relates directly to the fidelity after the encoded qubit is mapped into the logical subspace {\it via the process} of error correction.

\begin{table}
  \centering
  \begin{tabular}{|c|l|} \hline\hline
    Step & Action \\ \hline
    1&  We select a qubit state at random.\\ \hline
    2& We require Alice to encode it into the $n$ physical\\  
    &  qubits of the code. She does so perfectly.\\ \hline
    3 & The $n$ physical qubits are subjected to \\
    & environmental noise for a time $\tau/2$.\\ \hline
    4 & Optionally, Igor  is asked to apply a full round of \\
    & {\it imperfect} error correction.\\ \hline
      5& The $n$ physical qubits are subjected to \\
      & environmental noise for a further time $\tau/2$. \\ \hline
      6 & Bob  takes the final state of the $n$  qubits, and performs\\ 
      &   an analysis so as to make his best guess of the state.\\
      & He does so perfectly.
\\ \hline             
  \end{tabular}
  \caption{ \bf Protocol for assessing the beneficial role of QEC.}
  \label{table:first_criterion}
\end{table}

Notice that this protocol naturally generalizes to multiple rounds of error correction, i.e. multiple times when the imperfect Igor can attempt to help. We simply wait a time $\tau/(m+1)$, have Igor perform his cycle, and repeat until $m$ cycles are performed. After a final wait of $\tau/(m+1)$ so that a total time $\tau$ has elapsed, Bob receives the $n$ physical qubits. For a sufficiently high performing Igor, and  a long enough time $\tau$, it will be beneficial to have multiple rounds. Note that in the numerical simulation of the protocols, as discussed at later stages of this paper, we will take into account the finite duration that applications of the QEC cycles require.

Provided that this criterion has been fulfilled, and that QEC is proven beneficial, we can turn to the second desirable property  of QEC, namely  that   encoding, error detection and error correction, are  {\em beneficial} in comparison to the degradation of an unprotected physical qubit $\rho=\ket{\psi}\bra{\psi}\to\tilde{\rho}$. For the particular task at hand, this amounts to proving that
\beq
\label{eq:be_pont_bare}
\mathcal{P}_{\rm B}(\rho_L,{\rho}_{\rm QEC})>\mathcal{P}_{\rm B}(\rho,\tilde{\rho}),
\eeq
and would essentially demonstrate that the encoded logical qubit outperforms the   quantum memory built with a single unprotected physical qubit of the same sort as those used to form the logical qubit.


\section{\bf Trapped-ion experimental toolbox for QEC}
\label{sec:expt_system}

\subsection{Experimental Architecture}
\label{sec:expt_architecture}
The proposed setup consists of a 1D~segmented high-optical-access
(HOA) ion trap fabricated by Sandia National Laboratories~\cite{hoa_source},  and operated in a cryogenic
environment (see Fig.\ref{Fig:ExpHoa}). We consider 
$^{40}$Ca$^{+}$ ions  for hosting the qubits, and $^{88}$Sr$^{+}$  ions for providing the capabilities
for sympathetic cooling, and mixed-species readout for syndrome
extraction. We consider that ions undergoing the quantum logic operations can be separated
and shuttled across the segmented trap array by using high-speed
(diabatic), low-excitation protocols in order to minimize cross-talk
on neighboring qubits. The required  pulsed control of the qubits, system
synchronization, measurement and fast-feedback as required
for QEC, can be achieved by a custom-engineered high-speed controller.

The choice of the trap is motivated by the requirements for the  realization of  a  QEC code, which demand high-fidelity quantum
operations on the order of more than 10 ions. Therefore, a micro-fabricated
segmented ion trap that  enables multiple
trapping zones and  versatile ion crystal reconfigurations is required. This increases the complexity of the trap to a level that can,
to date, only be satisfied by a quasi-planar trap structures,  which
reduce the trapping depth such that   precautions against ion loss
have to be taken. This can be mitigated by lowering
the pressure of the vacuum environment by operating the
experiment at cryogenic temperatures. 

The encoded qubit will be realized in $^{40}$Ca$^{+}$ ions which
allow for an optical as well as a ground state qubit. The chosen species
enables high-fidelity state detection of the optical qubit due to its
simple electronic structure. The optical qubit is formed by the
$4S_{1/2}(m_f=-1/2)$ ground state and $3D_{5/2}(m_f=-1/2)$ metastable excited state.
 The excited state has a
lifetime of 1.1s which sets the upper limit on the qubit storage
time~\cite{schindler-njp-15-123012}. Quantum operations are performed
with a laser nearly resonant to this transition at a wavelength
of about 729nm. It
is also possible to encode a qubit in the two $4S_{1/2}(m_f=\pm1/2)$
Zeeman substates. The coherence of this qubit is only limited by
magnetic field fluctuations, where recent improvements resulted in a
coherence time of more than one second~\cite{Ruster2016}. State
manipulation of this qubit is performed by Raman lasers close to the
$4S_{1/2}$ to $4P_{1/2}$ transition. Due to the complexity of the QEC  algorithm, a second ion
species for sympathetic cooling and stabilizer readout will also be explored. For this, $^{88}$Sr$^{+}$ ions can be used.

The quality of quantum operations is limited by different processes
for the optical and the spin qubit. For the optical qubit, the
absolute phase noise of the laser driving the transition limits the
achievable coherence, whereas the spin qubit is only sensitive to the
differential phase noise in the two Raman laser beams. For the spin
qubit, off-resonant excitation of the $4P_{1/2}$ state is a process
reducing the gate fidelity, which can only be mitigated by increasing simultaneously the
intensity and detuning of the Raman laser beams. However, dynamic control of light at
397nm is more challenging than control at a longer wavelength of 729nm.

It is expected that the encoding and QEC of a single  logical qubit with a low-distance code can be implemented in a
single segmented linear trap with the ion crystal reconfiguration
techniques outlined below. However, multiple logical qubits
will likely require a more capable architecture in which ion
reconfiguration can be performed more efficiently using three- or
four-way junctions. This allows multiple processing regions where
syndrome measurements can be performed in parallel, which is also
crucial for an extensible QEC architecture.


\subsection{Gate Operations}
\label{sec:gate_operations}

\begin{table}
  \centering
  \begin{tabular}{|l |c|c|c|c|} \hline\hline
    Operation & Current & Current & Anticipated & Anticipated  \\
      &  duration & infidelity  &  duration &   Infidelity \\\hline
    Single-qubit gates & 5$\,\mu$s & $5 \cdot 10^{-5}$ & 1$\,\mu$s & $1 \cdot 10^{-5}$ \\ \hline
    Entangling  (2 qubits) & 40$\,\mu$s & $1 \cdot 10^{-2}$ & 15$\,\mu$s & $2 \cdot 10^{-4}$ \\ \hline
    Entangling  (5 qubits) & 60$\,\mu$s & $5 \cdot 10^{-2}$ & 15$\,\mu$s & $1 \cdot 10^{-3}$ \\ \hline
    Dual species & 60\,$\mu$s & $3 \cdot 10^{-2}$ & 15\,$\mu$s & $4 \cdot 10^{-4}$ \\ 
      entangling   (2 qubits) &  &  &  &  \\ \hline
    Dual species & 80\,$\mu$s & $5 \cdot 10^{-2}$ & 15\,$\mu$s & $6 \cdot 10^{-4}$ \\ 
    entangling   (3 qubits) & & &  & \\\hline
    Dual species   & - & - & 15\,$\mu$s & $2 \cdot 10^{-3}$ \\ 
    entangling (5 qubits) &  &  &  & \\ \hline
    Measurement & 400$\,\mu$s & $1 \cdot 10^{-3}$ & 30$\,\mu$s & $1 \cdot 10^{-4}$ \\ \hline
    Re-cooling & 400$\,\mu$s & $\bar{n} < 0.1$ & 100$\,\mu$s & $\bar{n} < 0.1$ \\ \hline
    Qubit reset & 50$\,\mu$s & $5 \cdot 10^{-3}$  & 10$\,\mu$s &$5 \cdot 10^{-3}\ ^\star$ \\ \hline
  \end{tabular}
  \caption{{\bf Current and anticipated gate-operation infidelities and durations}. Single-qubit operations are a 90 degree rotation on the Bloch's sphere, whole entangling operations correspond to fully entangling M{\o}lmer-S{\o}rensen operations (see Sec.~\ref{sec:expt_system}). The reported dual-species operations have been performed in a $^9$Be$^+$-- $^{40}$Ca$^+$ crystal. For the parameter marked by the $\star$ symbol, i.e. the anticipated value of the qubit reset fidelity, numerical simulations were performed both for the value $5 \cdot 10^{-3}$ and the value $1 \cdot 10^{-4}$. }
  \label{tab:summary_gates}
\end{table}

{\it (i) State of the art.--} The experimentally-available set of 
operations considered in this work consists of {\it(i)} global laser-driven M{\o}lmer-S{\o}rensen (MS)
entangling operations~\cite{ PhysRevA.62.022311,molmer-prl-82-1835}, which can be expressed as
\beq
\label{eq:MS}
U_{\rm MS,\phi}(\theta)=\ee^{-\ii\frac{\theta}{4} S^2_\phi},\hspace{1ex}  S_\phi=\sum_{i=1}^n(\cos\phi X_i+\sin\phi Y_i)
\eeq
where $\phi$ is controlled by the laser phase, and $\theta$ by its intensity and pulse duration. Here, $X_i=\sigma_i^x$ and $Y_i=\sigma_i^y$ are Pauli matrices. Additionally,
the global laser beams can also drive {\it(ii)} global rotations around the Bloch sphere of an
individual qubit with a rotation axis in the equatorial plane
\beq
\label{eq:rot_xy}
U_{\rm R,\phi}(\theta)=\ee^{-\ii\frac{\theta}{2}S_\phi},
\eeq
which are also controlled via the phase, intensity, and pulse duration of the laser beam. Finally,
{\it(iii)} addressed ac-Stark shifts result in rotations around the
$z$-axis on the Bloch sphere of an individual qubit
\beq
\label{eq:rot_z}
U_{{\rm R}_j,z}(\theta)=\ee^{-\ii\frac{\theta}{2} Z_j},
\eeq
where $\theta$ is controlled by the intensity of the off-resonant laser beam, its detuning, and the pulse duration. Here, $Z_i=\sigma_i^z$ is one of the Pauli matrices. This gate set is
described in detail in~\cite{schindler-njp-15-123012}, and a numerical method to
find an efficient decomposition of an arbitrary quantum algorithm in a
sequence of these gates is presented in~\cite{Martinez2016}. In sections below, we shall use extensively the following single-qubit operations
\begin{equation} 
X_j(\theta) = \ee^{  - \frac{\ii\theta}{2} X_j}, \hspace{1ex}Y_j(\theta) = \ee^{  - \frac{\ii\theta}{2} Y_j}, \hspace{1ex} Z_j(\theta) =\ee^{  - \frac{\ii\theta}{2} Z_j},
\label{eq:single_qubit_rot}
\end{equation}
which can be obtained either directly from the available set $Z_j(\theta)=U_{{\rm R}_j,z}(\theta),$ or by means of dynamic error suppression sequences (see Sec~\ref{Sec:DES}).  In addition, we shall also use MS gates~\eqref{eq:MS} of $X$-type  or $Y$-type  acting on a pair of ions/qubits $i$ and $j$. These MS gates will be  defined as 
\begin{equation}
\label{2_qubit_MS}
X^2_{i,j}(\theta) = \ee^{ - \frac{\ii\theta}{2} X_iX_j}, \hspace{1ex} Y^2_{i,j}(\theta) = \ee^{  - \frac{\ii\theta}{2} Y_iY_j},
\end{equation}
which are obtained (up to a global phase) from  $U_{\rm MS,\phi}(\theta)$ in Eq.~\eqref{eq:MS} by setting $\phi=0$ and $\phi=\frac{\pi}{2}$ respectively, and can be implemented by using spectroscopic decoupling techniques or ion-crystal reconfigurations steps, such that the MS laser beams only couple to the ion pair $i,j$.
Using this notation, $\theta=\frac{\pi}{2}$ MS gates are "fully-entangling", as they map the computational basis states of $N$ qubits to GHZ states (up to local unitary rotations) of $N$ qubits. For instance, for two ions $X^2_{i,j}(\pi/2) =(\mathbb{I} - \ii X_iX_j)/\sqrt{2}$ and one finds $X^2_{i,j}(\pi/2)\ket{0}_i \ket{0}_j =  (\ket{0}_i \ket{0}_j - \ii \ket{1}_i \ket{1}_j)/\sqrt{2}$. Throughout this manuscript, we will use the term multi-qubit MS gates to refer to MS gates acting on more than 2 qubits.

The entangling MS gates are performed on the
axial center-of-mass mode of the ion string. This has the advantage
that only a single loop in phase-space has to be closed to erase
unwanted spin-motion entanglement. However, this mode cannot be used 
 for longer ion strings as the ion string approaches a zig-zag configuration. 
This scheme allows for operations generating
a maximally entangled GHZ state of up to 8 ions, which can be implemented in
about 50$\mu$s for optical qubits with state fidelities of \{98.6, 95.7, 81.7\}\% for \{2,4,8\} ions \cite{monz-prl-106-130506}.  If the limiting factor 
on the gate quality is  phase-noise on
the laser driving the qubit transition, this could be improved by a
laser with smaller phase noise or by switching to ground-state Zeeman qubits which
are only susceptible to the phase difference of the two Raman
lasers. Recently, high fidelity entangling operations for two ground-state hyperfine qubits
have been demonstrated, reaching infidelities below
$10^{-3}$~\cite{16Ballance,PhysRevLett.117.060505}.

Carrying out high-fidelity  dual-species QIP protocols, such as $^{40}$Ca$^+$ and
$^{88}$Sr$^+$ in our case, is  generally more difficult than
single-species experiments.  Complicating factors include a more
complex motional mode structure and cooling requirements. However,  dual-species
entangling gates have already been achieved~\cite{15Tan, 15Ballance}, showing Bell-state infidelities
of $2 \cdot 10^{-2}$ and $2 \cdot 10^{-3}$. Moreover, we have also achieved preliminary 
dual-species operations for QEC
with $^{40}$Ca$^+$ and $^9$Be$^+$ 
 using global optical beams, including
experimental approaches for handling dual-species crystals, as well as
initial dual-species  gates for $XX$
and $ZZ$ stabilizer readout of a two-ion $^9$Be$^+$ crystal with a
single $^{40}$Ca$^+$ ion. Preliminary infidelity estimates obtained from Bell state preparation are listed in Table
\ref{tab:summary_gates}; SPAM (state preparation and measurement) errors are included in the errors given.

\textit{ (ii) Experimental input for the noise model.--} To assess the performance of a QEC procedure with a reasonably-sized ion register,
one requires a simplified, yet sufficiently realistic, error model. In several  studies, circuit noise is assumed to affect equally the single- and two-qubit gates  of the QEC protocol. However, in many experimental setups, the leading source of noise affecting idle qubits, single-qubit gates and entangling operations can be very different, requiring thus more elaborate noise models. The chosen error model in this work
includes perfect gate operations followed by a {\it depolarizing channel} on
the active qubits and inactive qubits are affected by {\it dephasing
noise}. We use microscopic calculations and experimental results to set the parameters of this error model, which has the advantage that it can be numerically simulated efficiently even for large qubit registers by using parallelizable Monte Carlo techniques, 
providing  quantitative target gate fidelities. More details on this error model can be found in
section~\ref{sec:error_models}. 

\textit{ (iii) Expected performance.--} The required parameters for the
chosen noise model can be fixed by the knowledge of {\it (i)} the state  infidelity for the gates, which describes the
strength of the depolarizing noise, and {\it (ii)} the duration of the gates, which are used
to estimate the effect of dephasing  on idling qubits. Table
\ref{tab:summary_gates} shows a summary of current and anticipated
gate operations for these parameters. The current coherence
time on the ground-state qubit is 2 seconds. By improving the magnetic
field stability it is expected that this can be extended to 10
seconds. The current coherence time on the optical qubit is 200\,ms,
which is anticipated to be improved to the limit given by the
spontaneous decay from the metastable excited state to about 2
seconds.

\subsection{Dynamic Error Suppression}\label{Sec:DES}

In developing a trapped-ion experimental toolbox for QEC we are motivated to explore how the capabilities on hand in the laboratory may be crafted to maximize compatibility with the stringent demands on quantum error correction.  Our primary objectives are twofold: (1) ensure gate errors are suppressed to rates as low as practicable relative to fault-tolerance thresholds; and (2) ensure compatibility of the underlying error model with the mathematical assumptions of fault-tolerance in QEC.  In our experiments we therefore routinely turn to open-loop control protocols applied at the physical layer and designed to improve gate performance in advance of QEC.

The strict requirements of fault-tolerance on qubit error rates have motivated the development of error-suppressing physical-layer control techniques~\cite{BallNPJQuantum2016,Viola1998,Viola1999,Zanardi1999,Vitali1999,Viola2003,Byrd2003,Khodjasteh2005,Yao2007,Uhrig2007,Gordon2008,Khodjasteh2009dcg,Khodjasteh2009,Khodjasteh2010,Liu2010,Biercuk_Filter,CPMG_ions} known as dynamic error suppression (DES).  In these  feedback-free protocols, temporal modulation of the qubit control field is employed in order to effectively average away decoherence induced by environmental fluctuations or control imperfections.  These protocols are considered an important complement to QEC~\cite{Preskill_Layered,Khodjasteh2005, JonesPRX2012}, both because of their potential to improve the resource-efficiency of QEC, and the fact that these protocols work in the presence of noise with strong temporal correlations, a regime which violates most error models underpinning the functionality of QEC. In fact, even in the presence of strong qubit decoherence DES can extend the effective qubit lifetime by decoupling from slowly varying noise sources. We expect that, in general, targeted application of DES will be implemented at the physical level for qubit manipulation.

{\it (i) Relevant control protocols.--}
We consider control protocols with diverse historical origins, but a common framework for implementation.  Physical qubit operations consist of multiple elementary control operations, which are sequentially applied in such a way that the desired target operation (quantum gate) is realized while simultaneously reducing the net sensitivity to error.  We treat control protocols taking the form of an $n$-segment sequence of unitaries, executed over the time period $[0, \tau]$. This implies a partition of the sequence duration $\tau$ into $n$ subintervals $I_l = [t_{l-1},t_l]$, $l\in\{1,...,n\}$, such that the $l$th control unitary has duration $\tau_l = t_l-t_{l-1}$.  The total operation can thus  be expressed as
\beq
\label{PrimitivePulseForm}R(\theta,\phi,\tau) := \prod_{l} U_{\rm R,\phi_l}(\theta_l)= \prod_{l}\ee^{-\ii\frac{\theta_l}{2}S_{\phi_l}},\\
\eeq
where we have used the  rotations in Eq.~\eqref{eq:rot_xy}, and defined $\theta_l \equiv \Omega_l\tau_l$ in terms of the 
   the time-independent Rabi rate $\Omega_l$ during the $l$th time interval $[t_{l-1},t_l]$. The resultant rotation generator  generates a rotation of the Bloch vector through an angle $\theta_l $ about an arbitrary axis $\boldsymbol{l} = \left(\cos(\phi_l),\hspace{0.1cm}\sin(\phi_l),\hspace{0.1cm}0\right)$.

The assignment of the relevant control parameters for each segment, $\{\Omega_{l}, \tau_{l}, \phi_{l}\}$, may be determined through a variety of techniques. ``Composite-pulse'' constructions address a combination of {\em static} pulse-length and off-resonance control errors, and are generally implemented via piecewise-constant phase modulation. Representative sequences include the so-called SK1 and BB1, correcting for pure amplitude 
errors~\cite{Brown2004,Wimperis1994}, CORPSE for pure detuning errors~\cite{Wimperis1994,Cummins2003}, and both reduced CinSK (CORPSE in SK1) 
and reduced CinBB (CORPSE in BB1)~\cite{Bando2013} for simultaneous errors.  Dynamically corrected gate (DCG) protocols are constructed similarly (via different underlying mathematics - recently unified in ~\cite{Chingiz2014, ViolaGeneral, Ball2014}), but employ piecewise-constant amplitude and phase modulation 
of the applied segmented control fields.  Representative approaches include the Walsh family of DCGs~\cite{SoareNatPhys2014, Ball2014}.

\begin{table*}[htbp] 
\begin{tabular}{|l|c|c|c|c|c|c|c|} \hline \hline \hspace{2ex}Composite pulse \hspace{2ex} &  \hspace{2ex}Error model  \hspace{2ex}&   ($\theta_1, \phi_1$) &   ($\theta_2, \phi_2$) &  ($\theta_3, \phi_3$) &  ($\theta_4, \phi_4$) & ($\theta_5, \phi_5$) & ($\theta_6, \phi_6$)   \\
\hline SK1    & $a$ &($\theta$, 0) &  (2$\pi, -\phi_{1}$)  &  (2$\pi, \phi_{1}$)   & - & - & -\\
\hline BB1    & $a$ &($\theta$, 0) &  ($\pi, \phi_{1}$)  &  (2$\pi, 3 \phi_{1} $) & 
($\pi, \phi_{1}$) & - & - \\
\hline CORPSE    & $d$ & \hspace{2ex}($2\pi + \theta/2 - k, 0$) \hspace{2ex} &  ($2\pi -2k, \pi$) & ($\theta/2 - k, 0$)   
& - & - & - \\
\hline WAMF    & $d$ &($\Omega$, $\theta$, 0)  &  \hspace{2ex} ($\Omega/2$, $\theta/2$, 0) \hspace{2ex}  & \hspace{2ex} ($\Omega/2$, $\theta/2$, 0)  \hspace{2ex} &($\Omega$, $\theta$, 0) 
 & - & - \\
\hline Reduced CinSK  & $s$ & ($2\pi + \theta/2 - k, 0$) &  ($2\pi -2k, \pi$) & ($\theta/2 - k, 0$)   &  \hspace{1ex}(2$\pi, -\phi_{1}$) \hspace{1ex} &   \hspace{2ex}(2$\pi, \phi_{1}$) \hspace{2ex} & - \\
\hline Reduced CinBB  & $s$ & ($2\pi + \theta/2 - k, 0$) &  ($2\pi -2k, \pi$) & ($\theta/2 - k, 0$)   & ($\pi, \phi_{1}$)  &  (2$\pi, 3\phi_{1} $)   & ($\pi, \phi_{1}$) \\
\hline \end{tabular}
\caption{\label{tab:comp_pulses}{\bf Single qubit DES protected gates enacting net operation $R(\theta, 0)$}, following~\cite{Chingiz2014}. 
Here, $\phi_1 = \cos^{-1}(-\theta/4\pi)$, $k = \arcsin[\sin(\theta/2)/2]$, $a$: amplitude noise; $d$: detuning noise; $s$: simultaneous amplitude and detuning noise. Unless otherwise noted the rabi rate, $\Omega$ remains fixed during all segments.  The Walsh modulated DCG sequence (WAMF)~\cite{Khodjasteh2010,SoareNatPhys2014, Ball2014} maintains constant segment durations, $\tau_{l}$, and employs amplitude modulation of the Rabi rate as described below. }
\end{table*}

The approach of producing composite sequential operations achieved through modulation of a control field can be extended to the implementation of two-qubit M\o lmer-S\o rensen gates.  Here, one may exploit phase-modulation of the driving field used to generate the effective spin-spin coupling via an intermediate bosonic mode~\cite{GreenPhiM}.  Application of a piecewise-constant, phase-modulation pattern to the driving field permits simultaneous decoupling of ``spectator'' bosonic modes and the suppression of temporal fluctuations in control amplitude without the need to consider nonlinearities in optical instruments associated with amplitude modulation.  The analytic framework in which these gates may be defined rests on a mathematical underpinning similar to that used for the construction of single-qubit DES strategies.  This approach is particularly effective in achieving high-order suppression of residual spin-motional entanglement in the ion chain.

DES protocols adapted for idle periods, and known as dynamic decoupling~\cite{Biercuk_Filter} are also commonly implemented to correct for a variety of error sources.  For instance, we employ spectroscopic decoupling to store idle qubits temporarily in Zeeman
sub-levels that are not affected by the lasers responsible for QEC
gates~\cite{nigg-science-345-302}. This decoupling is achieved by a
sequence composed of $N_{\rm p}=9$ pulses that can be applied to a
 set of $\ell$ idle ions to be hidden/un-hidden, labeled by
$h_1,\cdots h_{\ell}\in\{1,\cdots, N\}$. The composite pulse sequence is  designed in a way which, to lowest order, echoes out addressing errors due to residual light intensity on neighboring ions~\cite{nigg-science-345-302}.

\textit{(ii) Evaluating control performance.--} 
The operational fidelity for an imperfect operation is given by $\mathcal{F}_{av}(\tau)=\frac{1}{4}\langle |\text{Tr}(\tilde{U}(\tau))|^2\rangle$, following~\cite{GreenFF, ToddNJP}, where the error propagator, $\tilde{U}(t)$, captures the influence of noise and approaches the identity in the limit of vanishing errors.  Calculating the fidelity requires the error propagator to be expressed as an infinite series using the so-called Magnus Expansion as $\tilde{U}(\tau) = \exp[-\ii\Phi(\tau)]$, where the effective error operator $\Phi(\tau) = \sum_{\mu = 1}^\infty\Phi_\mu(\tau)$ at the end of the interaction has  expansion terms taking the form of time-ordered integrals over nested commutators of the so-called toggling-frame Hamiltonian.  Considering unitary errors, it is convenient to define the \emph{error vectors} $\boldsymbol{a}_\mu(\tau)$ by re-expressing the operators $\Phi_\mu(\tau) = \boldsymbol{a}_\mu(\tau)\cdot\boldsymbol{\sigma}$ in the basis of Pauli operators~\cite{ToddNJP}.  Then, one can expand the exponential in the error propagator to obtain the fidelity in the small noise limit
\begin{align}
\mathcal{F}_{av}=&1
-\langle a_{1}^{2}\rangle
-\left[\langle a_{2}^{2}\rangle+2\langle \boldsymbol{a}_{1}\boldsymbol{a}^{T}_{3}\rangle-\frac{\langle a_{1}^{4}\rangle}{3}\right]+\sum_{k=3}^{\infty}\mathcal{O}(\xi^{2k})
\end{align}
with $a_\mu^2 := \boldsymbol{a}_\mu(\tau)\boldsymbol{a}_\mu(\tau)^T$ the norm square of the error vector.  This expression contains a collection of terms with equal magnitude arising from \emph{different orders} of the Magnus expansion (e.g. $a_{2}^{2}$ vs $a_{1}^{4}$).  An expression for the leading order fidelity which keeps terms only to $\boldsymbol{a}_{1}(\tau)$ but approximates the full expansion~\cite{SoareNatPhys2014} is given by
\begin{align}
\label{eq:chi_fid}
  \mathcal{F}_{av}(\tau) \approx \mathcal{F}_\chi= \frac{1}{2}\Big\{1+\exp[-\chi(\tau)]\Big\}
\end{align}
where we have defined $\chi(\tau):=\langle a_1^2\rangle$.  We may conveniently move to the Fourier domain via the formalism of  the filter-transfer function  using
\begin{equation}
\label{eq:fidloss}
\chi(\tau)= \frac{1}{2\pi}\int_{-\infty}^{\infty}\frac{d\omega }{\omega^2}
\sum_{i=a,d}S_{i}(\omega)F_{i}(\omega).
\end{equation}
Here, we have introduced the noise power spectral densities in the amplitude ($i=a$) and dephasing ($i=d$) quadratures, $S_{i}(\omega)$, describing the statistical properties of the environmental noise process afflicting the control operations (see the second column of Table~\ref{tab:comp_pulses} for different examples).  

According to this discussion, the key quantities describing the effect of the control modulation are then $F_{i}(\omega)$; these objects characterize the spectral properties of the applied control, and can be calculated analytically for any piecewise-constant sequence~\cite{Biercuk2011,ToddNJP},  thus providing a simple quantitative means to compare control protocols of interest in the presence of generic, multi-axis time-dependent noise.  Because the net infidelity for an operation is given via an overlap integral of the noise power spectral density $S_{i}(\omega)$, and $F_{i}(\omega)$ for the control, we may describe these objects using the language of filter-design and refer to them as {\it filter transfer functions}. 

The filter order characterizes the performance of a filter transfer function by performing a Taylor expansion of the filter-transfer function about $\omega = 0$.  Assuming noise with dominant spectral weight at low frequencies, the approximation $F(\omega)\propto (\omega)^{2p}$ holds for some $p$ associated with the most significant power law expansion term.  The associated control protocol thus defines a high-pass filter with \emph{filter order} $p-1$.  This parameter takes on particular relevance in determining the efficacy of a selected control protocol subject to broadband noise. This filter order must be distinguished from the \emph{Magnus order} of error cancellation associated with quasistatic errors, which can be understood from   the DC limit of our filter-function formalism for constant noise fields.  A pulse sequence for which the Magnus expansion terms fulfill  $\Phi^\text{(DC)}_1 = ...= \Phi^\text{(DC)}_{\mu-1} = 0$ is then said to compensate static errors to Magnus order $(\mu-1)$ (see Refs.~\cite{ToddNJP,Ball2014}). The residual error is then dominated by terms proportional to the $\mu$th power in the magnitude of the error scaling.  This distinction is particularly important when considering more general expressions for the fidelity beyond leading order, in which contributions to the error from multiple Magnus orders appear in the Fidelity, and are captured through the exponentiated form of $\mathcal{F}_\chi$in Eq.~\eqref{eq:chi_fid}.

\textit{(iii) Expected performance and protocol selection.--} The tools outlined above and detailed in publications including references~\cite{GreenFF, ToddNJP, Chingiz2014, SoareNatPhys2014, GreenPhiM} suggest an efficient suppression of gate \emph{error rates} due to noise processes exhibiting strong temporal correlations.  Given realistic error models for dephasing noise and slow control-amplitude drifts, factors of error suppression exceeding $\sim100\times$ are projected using state-of-the-art systems, and substantiated using both numeric simulations and analytic calculations~\cite{Chingiz2014}.  Key implementation challenges relate to the calibration of the requisite control phases and amplitudes, generally achieved through rf modulation protocols such as $IQ$ or $\Phi M$. The addition of time segments to a basic gate operation or complex modulation patterns introduces new paths for error.  Those errors which are systematic may be efficiently suppressed by judicious choice of DES strategy.  Stochastic errors may accumulate as a result of the more complex protocol, but due to their independence they scale only approximately linearly with added gate time under DES (a proxy measure for complexity).  Therefore on balance DES has the potential to provide substantial benefits.

Taking into consideration the discussion above, we determine a critical path to selection of appropriate modulation protocols.  We first observe that \emph{high-order error suppression in the Magnus expansion does not imply high-order time-domain noise filtering} and vice versa. This has been validated using experiments on trapped ions~\cite{SoareNatPhys2014}, and formalized rigorously in Ref.~\cite{ViolaGeneral}.  Given the ``whitening'' effect of DES protocols on low-frequency-dominated noise, it is naively expected that the residual errors under DES will exhibit lower correlations than would otherwise be achieved.  However, the order of error cancellation in the Magnus expansion is the primary determinant of correlations between residual errors that can cause failure of QEC protocols.  Accordingly, the choice of a DES strategy will involve a determination first of the requisite Magnus order of error cancellation to suppress residual error correlations, next a determination of the high-frequency behavior of system noise, and finally consideration of how added complexity in high-order DES strategies introduces new pathways for error due to poor pulse calibration.   Demonstrations of the suppression of residual error correlations using analyses of randomized benchmarking validate this general approach and will be the subject of a forthcoming manuscript.


\subsection{Ion Crystal Reconfiguration Techniques}
\label{sec:ion_reconfiguration}

\textit{ (i) State of the art.--}
Since the seminal proposal for the \textit{Quantum CCD} \cite{Kielpinski2002}, the advent of segmented ion traps and fast multichannel arbitrary waveform generators has enabled the demonstration of ion shuttling operations \cite{ROWE2002}. These operations need to be performed fast on the timescale set by gate operations. This is required to avoid excessive overhead and decoherence from qubit dephasing, as well as anomalous heating of the ion crystal. On the other hand,  motional excitations from shuttling must also be avoided  in order not to compromise the phonon-mediated MS entangling operations. Thus, the required waveform generators have to fulfill the requirements of {\it (i)} analog update rates below typical trap frequencies, {\it (ii)} simultaneous and real-time update of many (10-80) channels, and {\it (iii)} superior signal integrity, i.e. low noise at trap frequencies, low glitch impulse areas and low digital crosstalk. Designs for such devices have been reported in~\cite{BAIG2013,BOWLER2013}.\\
With segmented traps and waveform generators available, inter-segment shuttling of single ions within few trap periods has been reported for $^9$Be$^+$ \cite{BOWLER2012} and for $^{40}$Ca$^+$~\cite{fast_ion_splitting}. While fast separation has also been reported in \cite{BOWLER2012}, the realization for $^{40}$Ca$^+$ from \cite{fast_ion_splitting} has been more challenging due to the low transient minimum trap frequency resulting from the increased mass. Recently, a fast  rotation of two $^{40}$Ca$^+$ ions with low resulting excitation, which can be used  for reordering the qubit register,  has also been demonstrated~\cite{fast_rotation}. These experimental results are summarized in Table \ref{tbl:shuttlingops}.

\textit{ (ii) Role of ion-string length on crystal reconfiguration.--}
The extent to which shuttling operations have to be employed for logical qubit encoding, syndrome readout, error correction and gate operations, depends on the experimental capabilities to store and coherently manipulate  ion crystals of intermediate size. 
For ion strings of increasing size, addressing errors increase, and the presence of more spectator vibrational modes decreases the fidelities of  entangling MS gates~\cite{PhysRevA.62.022311}. Additionally, for segmented micro-traps, the ions are confined in smaller potential wells with increased anharmonicities. Furthermore, the more complex geometry does not always allow for precise micromotion compensation in all spatial directions at a reasonable experimental effort. These two effects can give rise to decreased confinement stability, presumably via parametric resonances. As a consequence, de-crystallization and trap loss occur at increased rates,  such that this can represent a serious obstacle. To our knowledge, these effects have not been thoroughly investigated or  quantitatively characterized.

Therefore, the following particular points have to be addressed by future experimental investigations:
{\it (a)}
the actual extent to which the speed of low-excitation shuttling operations can be increased, see Table \ref{tbl:shuttlingops}.
{\it (b)}
The extensibility of low-excitation separation/merging and reordering operations beyond two ions, and  to mixed-species scenarios.
{\it (c)}
The scaling of the attainable fidelities of entangling gates with the ion register size.
{\it (d)}
The actual decrease of duration/increase of fidelity of laser-addressed hide/unhide operations, which -as shuttling operations- serve the task of selecting a subset of ions for QEC.
{\it (e)}
The impact of decay from the metastable state for hidden qubits on the overall error rates, which is to be determined from simulations.

 \begin{table}
\begin{tabular}{|l | c | c | c|} \hline\hline
Operation & Shuttling (one  & Separation & Rotation\cite{fast_rotation} \\
&  segment)\cite{fast_transport} & /merge\cite{fast_ion_splitting} & \\\hline
Duration  & 3.6$\,\mu$s & 80$\,\mu$s & 42$\,\mu$s \\ \hline
Excitation & $<$0.1 & 6 & $<$0.3  \\ 
axial (phonons) &  &  &   \\ \hline
Excitation  & N/A & $<$0.1 & $<$0.2 \\ 
radial (phonons) & & & \\ \hline
Anticipated duration  & 5$\,\mu$s & 30$\,\mu$s & 20$\,\mu$s\\ \hline
Anticipated excitation  & $<$0.2 & $<$1 & $<$0.2\\
axial (phonons) &  &  & \\ \hline
Anticipated excitation & $<$0.01 & $<$0.1 & $<$0.1\\ 
radial (phonons) &  &  & \\ \hline
\end{tabular}
\caption{{\bf Current and anticipated metrics for different shuttling operations}, carried out with $^{40}$Ca$^+$ ions in a multilayer trap  (see Sec.~\ref{sec:expt_system} and lower inset of Fig.~\ref{Fig:ExpHoa}). The axial trap frequency is about $2\pi\times$1.4~MHz, while the radial frequencies range around $2\pi\times$3~MHz. The shuttling is carried out with one ion, while the other operations are carried out with two ions. Note that the 3.6~$\mu$s duration for low excitation shuttling is obtained with an amplitude- and phase-calibrated de-excitation kick \cite{fast_transport}. Since shuttling duration will not be the bottleneck as compared to other operations, we can  anticipate a slightly longer duration of 5~$\mu$s for similarly low excitation  with a smaller calibration effort.}
 \label{tbl:shuttlingops}
\end{table}

\textit{ (iii) Experimental input for the effective noise model.--}
In order for simulation results to provide guidance towards the best strategy for logical qubit operation, we need to establish a noise model that captures the essential mechanisms how shuttling operations contribute to errors, but keeps the complexity and computational requirements reasonably small. We thus chose the following model: each shuttling operation contributes with a fixed amount of energy to the radial and axial degrees of freedom of each ion involved in the operation. Despite the fact that the energy is mostly contributed in the form of a coherent oscillator displacement, we assume that there is no fixed phase relation between consecutive displacements corresponding to different shuttling operations. Therefore, the shuttling operations lead to momentum kicks, which heat up the ions. We do not distinguish different collective modes and rather keep track of the mean motional energies of each ion. For merging of ion strings, we assume instantaneous thermalization, such that the total energies are equally distributed among the ions. Whenever entangling gate operations are carried out, we take the motional excitation into account  to estimate   gate imperfections according to the infidelity estimates  discussed in Sec.~\ref{sec:error_models},  which consider the excitation on spectator vibrational modes, as well as the excitation of the gate-mediating bus mode.

\textit{ (iv) Expected performance for shuttling operations.--}
 The anticipated improvements are due to ongoing efforts such as {\it (a) filter un-distortion:} the distortion induced by second-order low-pass filters on the segment supply lines are partially undone by correcting for the filter transfer function, at the expense of control voltage amplitude as a resource. This increases the degree of control. {\it (b) Increased control voltage range:} the larger segment voltages generated by a second generation waveform generator will increase the minimum confinement throughout separation/merge operations, and enable crystal reordering at larger radial trap frequencies. {\it (c) Ramp generation:} software for automated voltage ramp generation will find optimized voltage ramps, possibly employing control techniques such as shortcuts-to-adiabaticity or optimal control~\cite{Furst2014}. According to these improvements of an existing setup with $^{40}$Ca$^+$ ions, the anticipated key metrics for the different elementary shuttling operations are  shown in Table~\ref{tbl:shuttlingops}.



\subsection{Readout and Electronic Control}
\label{sec:readout_electronics}

Maintaining a logical qubit via QEC will require
repetitive ancilla readout (see Fig.~\ref{Fig:4qubit_stab_readout}) and reset, feedback on the logical qubits,
and likely sympathetic re-cooling of the  crystal.  Achieving high
single-qubit readout fidelity generally requires a trade-off between
minimizing the dark and bright state histogram overlap, and minimizing
decay from the excited to the ground state (as well as repumping of
dark states in hyperfine qubits).  Additionally,  state
discrimination must be performed in real time; post-processing
techniques cannot be used to enhance fidelities. A control system, the  M-ACTION, has been designed to
address these challenges in the context of maintaining a logical qubit: structured
around a fast CPU communicating to FPGAs, this control system minimizes real-time
processing delay, allows rapid prototyping of algorithms in C++, and
can feed back to hardware with low latency.

\textit{State of the art.--}  Carrying out experiments on most trapped-ion control 
systems has typically involved describing the experiment on a PC in a 
simple domain-specific language, running a simple compiler to produce 
real-time bytecode, and executing this on a peripheral device such as an 
FPGA board or PC card~\cite{schindler-njp-15-123012, 15Mount} running a simple finite-state machine. 
This approach does not support arbitrary feedback requiring non-trivial 
calculation within latencies comparable to other ion-trapping 
operations, i.e. well below 10 $\mu\,$s.

An alternative approach is to design the system to have significant 
\emph{low-level} processing power directly at the FPGA board; this 
allows more complex real-time decisions and calculations without being 
limited by communication bandwidth, and will be essential for QEC and 
other protocols requiring feedback. This design principle has been 
implemented in the M-ACTION system 
\cite{16NegnevitskyPrivateCommunication} used in a number of recent 
experiments on calcium ions \cite{17Leupold, 16Kienzler}. The system 
uses a chip consisting of an FPGA tightly coupled to two physical ARM 
Central Processing Units, which allows standard C++ to be compiled. Thus 
the numerical libraries of C++ can be fully utilized in decision 
processes, allowing low-latency decisions during experimental sequences 
\cite{17Leupold}.

Control electronics including synthesizers generating qubit drive fields must be linked to a stable master clock serving both for synchronization of distributed control electronics, and provision of a stable phase reference against which qubit coherence is measured.  This is vital because the common decoherence mechanism of dephasing represents a relative measure of the phase coherence of two effective oscillators, as outlined in~\cite{BallNPJQuantum2016}.  Common approaches to the provision of stable references include the use of an atomic frequency standard with good long-term stability, such as commercial Rubidium and Cesium clocks, followed by a quartz oscillator providing superior broadband phase noise.  Both long-term stability and short-term phase noise represent critical sources of error; analyses have demonstrated that the use of common lab-grade synthesizers serving as system master clocks can produce error rates nearing the percent level in less than 100$\,\mu$s.  Such error rates are easily suppressed by more than four orders of magnitude through appropriate selection of the master clock.  In future systems with multiple master clocks it will be essential to ensure that slowly varying drifts between clocks are minimized to maintain a fixed laboratory reference frame for operations~\cite{BallNPJQuantum2016}.

\textit{Building blocks and expected performance.--}
In the planned QEC scheme, a common step is to map a
syndrome onto an ancilla qubit, read out its state, and re-initialize
the ancilla along with cooling the ion chain. The dominant source of readout infidelity in both  $^{40}$Ca$^+$ and
$^{88}$Sr$^+$ optical qubits will likely  be background counts for short
readout times, which increase the dark and bright histogram overlap. This
can be counteracted by increasing the photon collection time or
efficiency, such that more photons are collected and the dark and
bright  histograms become more  separated. Spontaneous
decay from the D states is another source of infidelity for longer
readout times, exceeding $10^{-4}$ after roughly 100\, $\mu$s for
$^{40}$Ca$^+$, and dominating for very large  detection times.

Assuming a reasonable collection efficiency of 0.6\,\%, a
background count rate of 10$^4$/s, and  considering the possible beam intensities  similar to those available for $^{40}$Ca$^+$, the delay incurred in reading
out the $^{88}$Sr$^+$ ancilla will be 100--300\,$\mu$s with an infidelity
of below 10$^{-3}$. A readout infidelity of 10$^{-4}$ in 150\,$\mu$s for the optical transition in $^{40}$Ca$^+$ has been achieved using
Bayesian schemes that incorporate photon arrival times in the state
estimate, and attempt to identify spontaneous decays
\cite{08Myerson}. By increasing the photon collection
efficiency and thus reducing the detection time to below 20\,$\mu$s;
this will be attainable with a background count rate of $2\times 10^3$/s and
collection efficiency of 3.5\,\%.

After readout, the ion chain can be re-cooled using EIT cooling on the
radial modes\,\cite{16Lechner} and sideband cooling on the bus mode used
for multi-qubit gates.  EIT cooling takes several hundred microseconds
depending on the geometry, ion level structure and initial
temperature. The initial temperature depends strongly on the
fluorescence lasers: on resonance they will cause significant heating,
whereas by red-detuning and by weakening them, Doppler cooling will occur
at the expense of photon counts. Thus an optimum may be found between
readout time, readout fidelity and heating, such that the total
readout and cooling time is minimized.

After EIT cooling, the mean phonon number will already be below
$\bar{n} = 1$, thus few sideband-cooling pulses are required.  
Since cooling times will be at least several hundred microseconds,
there is significant time available for classical computation
(determining if/where an error has occurred) and feedback latency
(preparing the error correction pulse/pulses) in M-ACTION.  If the
chain were cooled only once per several readouts, however, these
classical delays could become the bottleneck.  Computation for a
7-qubit code, even when using Bayesian readout, should take
\emph{5}\,$\mu$s, and feedback latency is around 50\,$\mu$s. Ongoing efforts to reduce this will lead to a latency of 1.5
$\mu$s.

The corrective operation when an error is detected, which involves a single-qubit
rotation on a processing qubit, is an optional step that can be
avoided by altering future gates on that qubit to take the error into
account. This requires more classical computation, however we do not
anticipate the computation time being a problem for a single logical
qubit.  It could, however, result in an increasingly broad tree of
sequences. These must either be pre-computed and pre-loaded onto the
FPGA hardware, or loaded onto the hardware in real-time.  Pre-loading
the sequences will require more memory on the hardware and is
infeasible beyond approximately  5--10 feedback cycles, whereas real-time
loading will take up to 1\,ms for tens of pulses. A scalable solution
will be real-time loading using either a more efficient encoding
scheme or a high-bandwidth communication link.

\section{\bf Effective error models for elementary QEC operations in trapped ions}
\label{sec:error_models}

In Sec.~\ref{sec_qec_protocols}, we will introduce specific trapped-ion protocols
to assess the progress of QEC. In this section, we build on our previous discussion of the state-of-the-art, and future developments in trapped-ion technology, to model the noise on the elementary operations of these QEC protocols by certain quantum channels. As already noted previously in this manuscript, several works on QEC use circuit error models with a unique quantum channel affecting equally all the elementary operations of the QEC cycle. In this work, we go beyond these assumptions, and develop a more involved   model with several distinct channels, the parameters of which can be set by microscopic calculations and/or experimental measurements.   This model contains certain simplifications/limitations, which we comment upon in due course.

\subsection{Dephasing channel for idle qubits} 
During the QEC cycles, there
are several operations where the internal states of a subset of  qubits is not affected. More specifically, these operations
are {\it (i)} crystal reconfiguration, leaving all the qubit states  
unchanged, {\it (ii)} single-qubit rotations and MS entangling gates, which leave the spectator qubits  unchanged, and
{\it (iii)} ancillary qubit measurement and re-cooling where the
data qubits  remain idle. In all of these processes, the idling qubits suffer
mainly dephasing due to their coupling to the environment,
e.g. fluctuating magnetic fields, which can be modeled by the
identity  followed by a
dephasing channel acting on the particular subset of $m$ idle qubits $i_1,i_2,\cdots i_m\in\{1,\cdots,N\}$.  To simplify the model, we
will assume that the noise channel fulfills the i.i.d. criterion,
i.e. it is temporally and spatially uncorrelated. This leads to the
usual {\it dephasing channel} as described in~\cite{nielsen-book}, but applied to the set of idle qubits  ${\varepsilon_{\mathrm{d}}}(\rho)=\varepsilon^{\rm d}_{i_1}\circ\varepsilon_{i_2}^{\rm d}\circ\cdots\circ\varepsilon^{\rm d}_{i_m}(\rho)$, where 
\begin{equation}
\label{eq:dephaisng_channel}
\varepsilon_i^{\rm d}(\rho)=(1-p_{\mathrm{d}})\rho+p_{\mathrm{d}}\sigma_i^z\rho \sigma_i^z,
\end{equation}
is a Kraus map and $p_{\mathrm{d}}$ is the probability for a single phase flip. It would also be interesting to study  spatially-correlated dephasing, which does not necessarily
imply a faster decoherence as occurs for GHZ
states~\cite{monz-prl-106-130506}, but also enables  almost decoherence-free
subspaces in certain  codes~\cite{nigg-science-345-302}.

One can easily estimate the phase-flip probability by calculating the
time-evolution of a single qubit subjected to a fluctuating shift of
the transition frequency, which is modeled by a stochastic
process. Assuming a Markovian regime, one finds
$p_{\mathrm{d}}=\frac{1}{2}(1-\mathrm{e}^{-\Gamma_{\mathrm{d}}
  t_{\mathrm{i}}})\approx \frac{\Gamma_{\mathrm{d}}}{2} t_{\mathrm{I}}$,
where $t_{\mathrm{I}}$ is the time interval where the qubit remains
idle, and $\Gamma_{\mathrm{d}}$ is the rate of dephasing. This leads to a
dephasing time $T_2=1/\Gamma_{\mathrm{d}}$, as measured in
Ramsey-interferometry experiments where $\langle X_i(t_{\rm I})\rangle=\langle X_i(0)\rangle{\rm e}^{-t_{\rm I}/T_2}$ (see
Sec.~\ref{sec:gate_operations}), which yields $p_{\mathrm{d}}= t_{\mathrm{I}}/2 T_2 $.

\subsection{Depolarizing channel for   stabilizer mappings}

\label{subsec:Depol_models_MS_gate}

During the QEC cycles, the
stabilizer readout is accomplished by mapping the syndrome information
of the data qubits onto  ancillary qubits. As described below, this can be
accomplished by the combination of two multi-qubit MS gates, or by a sequence of two-qubit MS gates. We will model the leading error of this mapping  using a
{\it depolarizing channel}, as described for instance in~\cite{nielsen-book},
after each stabilizer mapping in the QEC protocol. We have explored three types of depolarizing channels
affecting $n$ active qubits  involved
in the MS gates (e.g. $n=5$ active qubits for QEC using multi-qubit MS gates, formed by 4 data and 1 ancillary
qubits labeled by $j_1,j_2,j_3,j_4,j_5\in\{1,\cdots,N\}$): 

{\it (i) Independent depolarizing noise}: The first error model consists of independent
depolarizing channels ${\varepsilon_{\rm MS}}(\rho)=\varepsilon^{\rm MS}_{j_1}\circ\varepsilon^{\rm MS}_{j_2}\circ\cdots \circ\varepsilon^{\rm MS}_{j_n}(\rho)$ acting on each of the active qubits
\begin{equation}
\label{depolarising_channel_iid}
\varepsilon^{\rm MS}_j(\rho)=(1-p_{\rm MS})\rho+\frac{p_{\rm MS}}{3}\sum_{\alpha\in\Lambda_\alpha}\sigma_{j}^\alpha\rho\sigma_j^\alpha,
\end{equation}
where $p_{\rm MS}$ is the probability for a MS depolarizing error, and $\Lambda_\alpha=\{x,y,z\}$. We note that this error model underestimates the occurrence of
multiple-qubit errors during the entangling gate, and can thus
overestimate the correcting power of the QEC.  Therefore, we have also explored other
 channels.

{\it (ii) Two-qubit depolarizing noise}: 
Provided that the $N$-ion
MS  gate~\eqref{eq:MS} can be understood as an all-to-all interaction
between qubit pairs, and is thus local-unitary equivalent to
applying CNOTs between all $N(N-1)/2$ ion pairs, an error model that
considers single- and two-qubit errors  with the same error probability may be
more realistic. This will certainly be the case for the QEC schemes based on sequences of 5-qubit MS gates, where the noise is described by the quantum operation
\begin{equation}
\label{depolarising_channel_1_2}
\begin{split}
{\varepsilon_{\rm MS}}(\rho)&=(1-p_{\rm MS})\rho+\frac{p_{\rm MS}}{105}\sum_{i\in\Lambda_{\rm a}}\sum_{\alpha\in\Lambda_{\alpha}}\sigma_{i}^\alpha\rho\sigma_{i}^\alpha\\
&+\frac{p_{\rm MS}}{105}\sum_{j_1, j_2\in\Lambda_{\rm a}}\sum_{\alpha,\beta\in\Lambda_\alpha}\sigma_{j_1}^\alpha\sigma_{j_2}^\beta\hspace{0.2ex}\rho\hspace{0.2ex}\sigma_{j_1}^\alpha\sigma_{j_2}^\beta,
\end{split}
\end{equation}
where $p_{\rm MS}$ is the probability for a MS depolarizing error, we have introduced the set of $n$ indexes for the active ions $\Lambda_{\rm a}$, and the sum over multiple ion indexes excludes coincidences of the pair of indexes. For the 5-ion MS gate, local-unitary equivalent to 10 CNOTs, with each pair of ions potentially undergoing 15 possible Pauli errors (6 single-qubit and 9 two-qubit Pauli operators), this results in the pre-factor 1/105.

{\it (iii) Multi-qubit depolarizing noise:} As a worst-case scenario for the schemes based on multi-qubit MS gates, we have also explored a model where any 5-qubit error can occur with the same error probability due to a faulty 5-ion MS gate. This can be described by the 
quantum operation

\begin{equation}
\label{depolarising_channel_any}
\begin{split}
{\varepsilon_{\rm MS}}(\rho)&=(1-p_{\rm MS})\rho\\
&+\frac{p_{\rm MS}}{1023}\sum_{\boldsymbol{j}\in\Lambda_{\rm a}}\sum_{\boldsymbol{\alpha}\in\tilde{\Lambda}_\alpha}\sigma_{j_1}^\alpha\sigma_{j_2}^\beta\sigma_{j_3}^\gamma\sigma_{j_4}^\kappa\sigma_{j_5}^\zeta\hspace{0.2ex} \rho \hspace{0.2ex}\sigma_{j_1}^\alpha\sigma_{j_2}^\beta\sigma_{j_3}^\gamma\sigma_{j_4}^\kappa\sigma_{j_5}^\zeta,
\end{split}
\end{equation}
where $p_{\rm MS}$ is the probability for a MS depolarizing error to occur, and we have introduced $\boldsymbol{j}=(j_1,j_2,j_3,j_4,j_5)$, and  $\boldsymbol{\alpha}=(\alpha,\beta,\gamma,\kappa,\zeta)$ in $\tilde{\Lambda}_\alpha=\{0,x,y,z\}$. Here,  the sum over multiple ion indices excludes coincidences of any indexes,  we have introduced $\sigma^0=\mathbb{I}_2$ as the identity operation, and the sum over possible Pauli errors excludes  the global identity $\alpha=\beta=\gamma=\kappa=\zeta=0$ (i.e. no error), thus giving rise to a total of $4^5-1 = 1023$ possible Pauli error configurations.

To estimate how the error probability of the above depolarizing channels
$p_{\rm {MS}}$ depends on the different experimental parameters, we will calculate the state fidelity of the ideal MS gate followed by each of the depolarizing channels in Eqs.(\ref{depolarising_channel_iid}) and (\ref{depolarising_channel_1_2}),  $\mathcal{F}=\langle{\Psi_{\rm t}}|\epsilon_{\rm MS}\left(U_{\rm MS}|{\Psi_0}\rangle\langle {\Psi_0}|U_{\rm MS}^{\dagger}\right)|{\Psi_{\rm t}}\rangle$, where the ideal  gate produces a GHZ-type state $|{\Psi_{\rm t}}\rangle=U_{\rm MS,\phi}(\pi/2)|{\Psi_0}\rangle$. 

For the {\it (i)} independent  depolarizing channel in Eq~(\ref{depolarising_channel_iid}), one obtains
$\mathcal{F}=(1-p_{\rm MS})^5+\frac{1}{30}(1-p_{\rm MS})^3p_{\rm
  MS}^2+\frac{1}{15}(1-p_{\rm MS})p_{\rm MS}^4\approx 1-5p_{\rm MS}$
for $p_{\rm MS}\ll 1$. In this case, only the processes where no error
occurs contribute to the fidelity at the lowest order in $p_{\rm MS}$, such that the
error probability is simply $p_{\rm MS}=(1-\mathcal{F})/5$. For the {\it (ii)} two-qubit depolarizing channel  that includes one- and
two-qubit errors with the same probability~(\ref{depolarising_channel_1_2}),
one finds $\mathcal{F}=1-\frac{95}{105}p_{\rm MS}$, such that
$p_{\rm MS}=105(1-\mathcal{F})/95$. In this case, processes with no error and
two Z-type errors leave the GHZ state invariant, and
contribute with the same order in $p_{\rm MS}$. Finally, for the  {\it (iii)} multi-qubit depolarizing channel  that includes all 5-qubit errors with the same probability~(\ref{depolarising_channel_any}),
one finds $\mathcal{F}=1-\frac{1008}{1023}p_{\rm MS}$, such that
$p_{\rm MS}=1023(1-\mathcal{F})/1008$. In this case, processes with no error, and
two/four $Z$-type errors leave the GHZ state invariant, and
contribute with the same order in $p_{\rm MS}$.

\begin{table}
  \FontsizeTables
  \centering
  \begin{tabular}{|c|c|c|c|c|c|c|}
    \hline
    \multicolumn{1}{|c|}{} & \multicolumn{4}{c|}{Ion crystal reconfiguration} & \multicolumn{2}{c|}{Spectroscopic } \\
        \multicolumn{1}{|c|}{} & \multicolumn{4}{c|}{} & \multicolumn{2}{c|}{ decoupling} \\
     \multicolumn{1}{|c|}{Approach} & \multicolumn{2}{c}{} & \multicolumn{2}{c|}{} & \multicolumn{2}{c|}{} \\
     \multicolumn{1}{|c|}{} & \multicolumn{2}{c|}{1 species} & \multicolumn{2}{c|}{2 species} & \multicolumn{2}{c|}{2 species} \\
     \multicolumn{1}{|c|}{} & \multicolumn{2}{c|}{without re-cooling} & \multicolumn{2}{c|}{with re-cooling} & \multicolumn{2}{c|}{with re-cooling} \\
    \hhline{*7{=}}
    5-ion MS  & current & anticipated & current & anticipated & current & anticipated \\
     infidelity& value & value & value & value & value & value  \\
     \hline
    map $S_1^x$  &$2.1\%$  & $0.29 \%$ & $4.9\%$ & $0.21\%$ & $4.9\%$ & $0.21\%$ \\
    \hline
    map $S_1^z$  &  $5.2\%$ & $0.46\%$ & $4.9\%$ & $0.21\%$ & $4.9\%$ & $0.21\%$  \\
    \hline
    map $S_2^x$  & $-$ & $0.85\%$ & $4.9\%$ & $0.21\%$ & $4.9\%$ & $0.21\%$  \\
    \hline
    map $S_2^z$  & $-$ & $1.5 \%$ & $4.9\%$ & $0.21\%$ & $4.9\%$ & $0.21\%$  \\
    \hline
    map $S_3^x$  & $-$ & $2.4 \%$ & $4.9\%$ & $0.21\%$ & $4.9\%$ & $0.21\%$  \\
    \hline
    map $S_3^z$  & $-$ & $3.2 \%$ & $4.9\%$ & $0.21\%$ & $4.9\%$ & $0.21\%$  \\
    \hline
  \end{tabular}
  \label{tab:MS_gate_errors} 
    \caption{{\bf MS gate infidelities in different QEC cycle  scenarios:} We use our  model of the MS gate $\epsilon=\epsilon_{\rm m}+\epsilon_{\rm d}+\epsilon_{\rm I}$ in Eqs.~(\ref{eq:thermal_error}),~(\ref{eq:dephasing_error}), and (\ref{eq:intensity_error}), to reproduce the current performance of 5-ion MS gates in Table~\ref{tab:summary_gates}. Then, we use the model to
     predict the  infidelities for the stabilizer mappings for the  scenarios of QEC with a warmer ion background, and those with the expected improved conditions. }
     \label{table_errors:MS}
\end{table}

The  probability of the depolarizing channel can then be
extracted by comparing to the GHZ infidelity $\epsilon=1-\mathcal{F}$ obtained by a microscopic Hamiltonian modeling the evolution of the trapped-ion MS gate~\cite{PhysRevA.62.022311}. In this way, one can include possible sources of noise and experimental imperfections that lead to evolutions that  depart from the ideal MS gate \cite{PhysRevA.62.022311, molmer-prl-82-1835}. We now discuss three different sources of infidelity.

{\it (a) Gate infidelity due to the motional excitations.--} The employed MS gate utilizes a bi-chromatic laser-ion interaction
that drives simultaneously the blue and red motional sidebands
corresponding to the centre-of-mass (CoM) axial mode. This acts as a bus
mode that mediates an all-to-all qubit-qubit interaction~\eqref{eq:MS} capable of
generating the aforementioned GHZ states. The motional excitation of
this mode, as well as the presence of additional vibrational modes of
the ion chain, lead to an infidelity caused by two main sources: {\it
  (i)} off-resonant couplings to the sidebands of spectator modes, and
{\it (ii)} fluctuation of the effective Rabi frequency of the
laser-ion coupling due to the motional excitations of bus and
spectator modes, i.e. Debye-Waller factors. If one assumes, as argued
in Sec.~\ref{sec:ion_reconfiguration}, that there is a fast equilibration after  ion-reconfiguration
operations, resulting in a thermal vibrational state with a mean phonon
number that increases after each particular crystal reconfiguration according to
Table~\ref{tbl:shuttlingops}, it is possible to estimate the
infidelity of the $N$-ion fully-entangling MS gate  $\mathcal{F}=1-\epsilon_{\rm m}$ as
\begin{equation}
\label{eq:thermal_error}
\epsilon_{\rm m}\approx  \frac{\pi N(\delta-\omega_z)}{2\omega_z^2 t_{\rm g}}0.8\left(\bar{n}+1\right)+  \frac{\pi^2 N(N-1)\eta^4}{8N^2}\!\left(1.2 \bar{n}^2+1.4\bar{n}\right)\!,
\end{equation}
where $\omega_z$  is the CoM axial mode frequency with mean phonon number $\bar{n}$, $\delta$ is the symmetric detuning of the bi-chromatic laser beams with respect to the electronic transition, $t_{\rm g}$ is the gate time, and $\eta= k_{\rm L}/\sqrt{2m\omega_z}$ is the single-ion axial Lamb-Dicke parameter. The first term in this equation represents the infidelity due to unclosed phase-space trajectories of the spectator modes, whereas the second one is due to the  decrease of the Rabi frequency due to the thermal background of all modes, i.e. Debye-Waller factor. Note that, although all the modes of the ion crystal participate in the infidelity, the error can be bounded with quantities that are characterized by the mean phonon number of the CoM mode.

In order to apply this error estimate~(\ref{eq:thermal_error}) to the different steps of the QEC cycle based on crystal reconfiguration, we   assume  that the gate time $t_{\rm g}(\bar{n}_f)$ after a number of crystal reconfigurations that increase the mean phonon number to $\bar{n}_f$ is modified with respect to the optimized gate time $t_{\rm g}(\bar{n}_0)$ in Table~\ref{tab:summary_gates}, where $\bar{n}_0\approx 0$ after laser cooling. This is accomplished by modifying the laser detuning, such that phase-space trajectories of the CoM mode are still closed for the modified  time
\begin{equation}
\label{eq:thermal_gate_time}
t_{\rm g}(\bar{n}_f)=t_{\rm g}(0)\left(1+\frac{\eta^2(2\bar{n}+1)}{N}\right).
\end{equation}
Therefore, the gates become slightly slower the higher the mean phonon number is. Note that this delay will not result in an appreciable change in the dephasing~(\ref{eq:dephaisng_channel}) that idle qubits suffer for each stabilizer readout, and one can assume that $2t_{\rm g}(0)$ is the  dephasing time for idle qubits during each $N$-ion stabilizer mapping (see Eq.~(\ref{eq:dephaisng_channel}) above).  
 
 This motional infidelity $\epsilon_{\rm m}$ can become the leading source of error in the QEC protocol where the ancillary (readout) and physical ions are of the
same species, and are shuttled/merged/split/rotated during the QEC
cycle to extract the syndrome. Accordingly, re-cooling of the motion
of the ion crystal via the ancillary ion cannot be exploited, as the
scattered light would perturb the quantum state encoded in the
data qubits. In this situation,  the motional excitation of the ion string resulting from the
different reconfiguration steps can become very large (see Table~\ref{tbl:shuttlingops}), yielding a motional infidelity that overcomes other possible sources of noise.

{\it (b) Gate infidelity due to magnetic-field and laser-intensity fluctuations.--} Another possible source of noise in the MS gate is caused by {\it (i)}
fluctuations between the qubit frequency and the laser frequency {\it
  (ii)} and laser-intensity fluctuations.  We model these two sources
of noise by stochastic processes that yield fluctuations of the qubit
frequencies, and of the laser coupling strengths to the motional
sidebands of the MS scheme, respectively. If one assumes that the time
correlations of these processes decay much faster than the gate time
(i.e. Markovian assumptions), then the gate infidelity can be
expressed as
$\mathcal{F}\approx 1-(\epsilon_{\rm d}+\epsilon_{\rm I})$, where
$\epsilon_{\rm I}$ is the error due to intensity fluctuations, and
$\epsilon_{\rm d}$ is the dephasing error due to e.g. fluctuating magnetic
fields experienced by the qubits during the
gate. Such an error can be approximated by
\begin{equation}
\label{eq:dephasing_error}
\epsilon_{\rm d}\approx  2 \Gamma_{\rm d} t_{\rm g}\sum_{i,j}{\rm e}^{-|z_i^0-z_j^0|/\xi_{\rm c}},
\end{equation}
where $\Gamma_{\rm d}$ is the rate of dephasing leading to a dephasing time $T_2=1/\Gamma_{\rm d}$, $z_i^0$ are the positions of the ions in the string, and $\xi_{\rm c}$ a typical length scale over which magnetic-field fluctuations are correlated, i.e. for $\xi_{\rm c}=0$, we have local noise and $\epsilon_{\rm d}\approx  2t_{\rm g}N/T_2$, whereas for $\xi_{\rm c}\gg |z_1^0-z_N^0|$, we have global magnetic-field fluctuations and $\epsilon_{\rm d}\approx 2t_{\rm g}N^2/T_2$ would be the dephasing rate affecting a GHZ state. We note that,  for the different mappings of the syndrome information into the ancilla qubits, the actual dephasing error will lie between these two limits, and its particular value will depend on the  collective state of the qubits at the instants of time where the MS gates are applied. To simplify the description, we consider a conservative, worst-case scenario, and  use an error rate $\epsilon_{\rm d}\approx  2 t_{\rm g}N^2/T_2$ consistent with the values reported in  Table~\ref{tab:summary_gates}.

Finally, intensity fluctuations during the gate will have two
effects. On the one hand, the time-dependence of the fluctuations can
lead to a residual spin-motion entanglement due to imperfect closure
of the phase-space trajectories of the bus CoM mode. On the other
hand, the acquired phase that depends on the area of the enclosed
trajectory may also differ from the one required to generate the
desired GHZ state. These two sources of error are accounted for, in
corresponding order, by means of the following expression
\begin{equation}
\label{eq:intensity_error}
\epsilon_{\rm In}\approx\Gamma_{\rm In} t_{\rm g}\eta^2\left(\bar{n}+\frac{1}{2}\right)+\Gamma_{\rm In} t_{\rm g}\frac{\eta^2(N-1)}{4},
\end{equation}
where $\Gamma_{\rm In}$ is the rate of intensity fluctuations, obtained
through its zero-frequency power spectral density, and sets a typical
time-scale for the effects of intensity noise
$T_{\rm In}=1/\Gamma_{\rm In}$. We will adjust this parameter to be consistent with the fidelities reported in experiments (see  Table~\ref{tab:summary_gates}).

In the QEC protocols based on two species, we use a
different ion species for the ancillary and data qubits. Thus, we can 
exploit the ancillary ion for intermediate sympathetic re-cooling of the ion
crystal. In any case, the population of vibrational modes remains small prior to the stabilizer mapping via MS gates. Hence, the {\it
  (i)} error due to thermal motional excitation in Eq.~(\ref{eq:thermal_error}) will not be leading, but
instead will contribute together with other sources of error. We
consider also {\it (ii)} dephasing and {\it (iii)} intensity
fluctuations during the gate as additional sources of gate infidelity following Eqs.~(\ref{eq:dephasing_error}) and~(\ref{eq:intensity_error}). Therefore, the MS-gate infidelity that can be used to extract the probability of the depolarizing channels corresponds to $\epsilon=\epsilon_{\rm m}+\epsilon_{\rm d}+\epsilon_{\rm I}$.

\subsection{Depolarizing channel} 
\label{subsec:hiding_error}

Due to the use of spectroscopic decoupling protocols (\ref{Sec:DES}), in the noise model used in this study, we do not consider residual errors on the neighboring ions, and will model the error in
this decoupling process by independent single-qubit {\it depolarizing channels} acting on each of
the qubits  being hidden or unhidden ${\varepsilon_{\mathrm{h}}}(\rho)=\varepsilon_{h_1}\circ\varepsilon_{h_2}\circ\cdots\circ\varepsilon_{h_m}(\rho)$, where 
 \begin{equation}
\label{eq:hidding_channel}\varepsilon_h(\rho)=(1-p_{\mathrm{h}})\rho+\frac{p_{\mathrm{h}}}{3}\sum_{\alpha\in\Lambda_{\alpha}}\sigma_h^\alpha\rho\sigma_h^\alpha,
\end{equation}
and $p_{\rm{h}}$ is the error probability. Taking the current and
expected single-qubit gate infidelities $\epsilon_1$ reported in
Table~\ref{tab:summary_gates} into account, we can  estimate
the probability of the depolarizing channel for spectroscopic
decoupling as $p_{\rm h}=9\epsilon_1$. 

\subsection{Bit-flip channel for measurement and  reset }

The QEC cycles also require a measurement and reset of the ancillary qubit. Faulty measurement/reset can be modeled by a {\it bit-flip channel}, as introduced in~\cite{nielsen-book}, that acts on the set of $n_a$ ancillary qubits $\{a_1,a_2,\cdots a_{n_a}\}\in\{1,\cdots,N\}$, namely ${\varepsilon_{\mathrm{b}}}(\rho)=\varepsilon_{a_1}\circ\varepsilon_{a_2}\circ\cdots\circ\varepsilon_{a_{n_a}}(\rho)$, where 
 \begin{equation}
\label{eq:bit_flip_channel}
{\varepsilon_{a}}(\rho)=(1-p_{\mathrm{b}})\rho+p_{\mathrm{b}}\sigma_a^x\rho\sigma_a^x,
\end{equation}
where $p_{\rm b}$ is the probability for a bit-flip error to occur. We estimate the value of this probability through the  infidelities for measurement $p_{\rm b}=\epsilon_{\rm meas}$ and qubit reset $p_{\rm b}=\epsilon_{\rm res}$  reported in Table~\ref{tab:summary_gates}.


\section{\bf Trapped-ion topological QEC and    fault tolerance }
\label{sec:cnot_alternative}

\subsection{Basic properties of the 7-qubit color code}

We will focus on the development  of trapped-ion QEC protocols to implement  a logical qubit based on the 7-qubit color code, assessing their potential to be useful for QEC by the operational measure descried in Sec.~\ref{sec:QEC_criterion}. The 7-qubit code constitutes an enabling building block of two  main routes towards fault-tolerant quantum computation (FTQC). On the one hand, it is equivalent (up to local unitaries) to the 7-qubit Steane code \cite{steane-prl-77-793, nielsen-book} and can, as such, be used as an  elementary unit to achieve more and more robust logical qubits by means of concatenation. On the other hand, it constitutes the smallest, though functionally complete, representative of the class of 2D topological color codes \cite{bombin-prl-97-180501}, for which logical qubits of increasing robustness can be achieved by using codes defined in  lattices  of increasing size. We note that topological codes  typically display higher error thresholds in comparison to concatenated ones, offering thus a practical and very promising route towards large-scale QEC.

One of the goals of our study is the identification of the accuracy requirements of the experimental building blocks used to realize complete QEC cycles on the logical qubit. For instance, a series of limitations on the experimental approach used in \cite{nigg-science-345-302}, such as e.g.~a large overhead in spectroscopic decoupling operations, has already been identified. By a numerical analysis of the operational measures in Eqs.~\eqref{eq:be_pont}-\eqref{eq:be_pont_bare}, we aim at deriving quantitative estimates on the experimental  requirements to make a QEC based on this protocol beneficial. Moreover, the measure will also allow us to benchmark the performance of other protocols that avoid spectroscopic decoupling. In this sense, exploring a variety of protocols for this code is an ideal testbed for the development of key tools, which would be equally required in the implementation of other small- and medium-size quantum  codes, such as e.g.~the 9-qubit Bacon-Shor-code (see e.g.~\cite{RevModPhys-Terhal-87-307-2015}) or the rotated 9-qubit surface code of distance 3 \cite{Horsman-njp-2012}.

\begin{figure}[t]
 \begin{centering}
  \includegraphics[width=0.8\columnwidth]{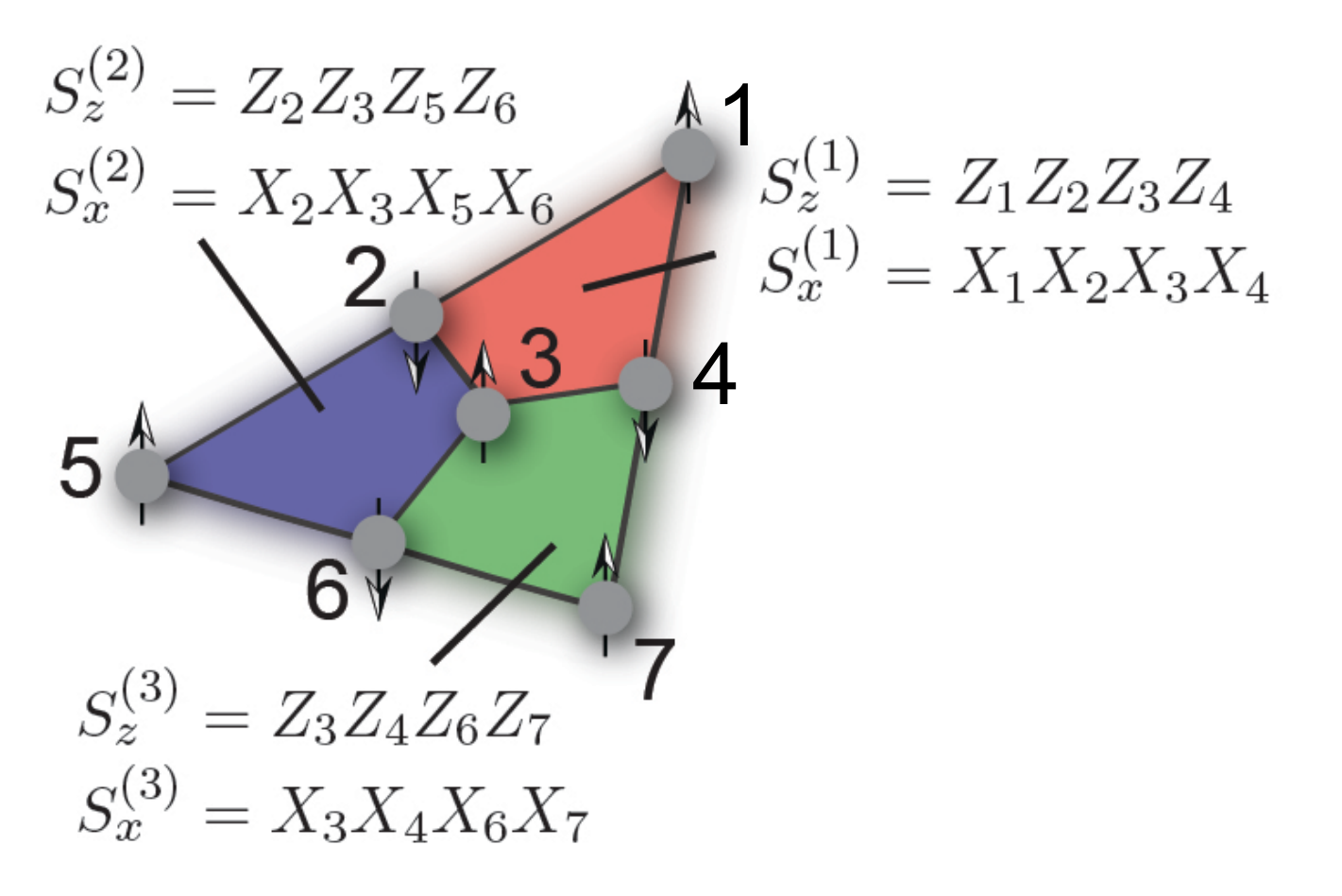}\\
  \caption{\label{Fig:7qubitCode} {\bf Scheme of the 7-qubit color code:} One logical qubit is embedded in 7 data qubits forming a 2D triangular color code structure of three plaquettes. The code space is defined via $S_i^x$ and $S_i^z$ stabilizer operators (generators), each acting on a plaquette that involves four physical qubits.  Logical operators such as $Z_L = \prod_{i} Z_i$, and similarly the other logical single-qubit Clifford gate generators $X_L:=\prod_iX_i$, $H_L:=\prod_{i} H_i=\prod_{i} \frac{1}{\sqrt{2}}(X_i+Z_i)$ and $S_L:=\prod_{i} S_i^\dagger=\prod_i\ee^{-\ii\frac{\pi}{4}(1-Z_i)}$   can be realized in a transversal, i.e.~bit-wise way.}
\end{centering}
\end{figure}

Let us thus briefly summarize a few central properties of the 7-qubit color code, which are relevant for the QEC protocols developed and studied below. This code allows one to store and manipulate $k=1$ logical qubit, which is redundantly encoded in entangled states distributed over $n=7$ physical qubits. The code has a logical distance of $d=3$, and thus allows one to detect and correct at least $t=1$ arbitrary error (phase and/or bit flip error) on any of the 7 physical  qubits. The code belongs to the family of CSS codes \cite{calderbank-pra-54-1098,steane-prl-77-793}, and thus allows one to independently detect and correct bit- and phase-flip errors. Errors are identified by measuring the corresponding error syndrome, which is deduced from the sets of three four-qubit Z and X-type stabilizer operators associated to the three plaquettes of the code (see Fig.~\ref{Fig:7qubitCode}).   If we denote the set of Pauli matrices of the physical qubits as $\{X_i=\sigma_i^x,Y_i=\sigma_i^y,Z_i=\sigma_i^z\}_{i=1}^n$, the stabilizers are
\beq
\label{eq:stabilizers}
\begin{split}
&S^{(1)}_x=X_1X_2X_3X_4,\hspace{1ex}S^{(2)}_x=X_2X_3X_5X_6,\hspace{1ex}S^{(3)}_x=X_3X_4X_6X_7,\\
&S^{(1)}_z=Z_1Z_2Z_3Z_4,\hspace{1ex}S^{(2)}_z=Z_2Z_3Z_5Z_6,\hspace{1ex}S^{(3)}_z=Z_3Z_4Z_6Z_7.
\end{split}
\eeq
Logical states are encoded in the code space, which is defined as the simultaneous eigenspace of eigenvalue +1 of the set of all $s=6$ stabilizer generators~\eqref{eq:stabilizers}, such that $k=n-s=1$ coincides with  the number of encoded qubits. Logical qubits employing larger  distance codes, and thus allowing for the correction of multiple errors, can be constructed by encoding a logical qubit in larger lattice structures involving more physical qubits \cite{bombin-prl-97-180501}. An interesting representative of this procedure is the distance-5 color code (of 4-8-8-type \cite{bombin-prl-97-180501}), which encodes a single logical qubit in 17 physical qubits arranged on a 2D lattice structure of 8 plaquettes.

A distinguishing feature, as compared e.g.~to Kitaev's surface code \cite{dennis-j-mat-phys-43-4452, raussendorf-prl-98-190504}, is that the color code \cite{bombin-prl-97-180501} permits a transversal realization of the entire group of Clifford gate operations~\cite{nielsen-book}. Thus, the realization of a logical Clifford gate  on the logical qubit amounts to a bit-wise application of the corresponding gates  on all physical qubits $X_L=\prod_{i} X_i,Z_L = \prod_{i} Z_i$, and similarly the other logical single-qubit Clifford gate operations, such as the Hadamard $H_L$ and K-gate $K_L$. This property not only does facilitate the practical implementation of logical gate operations but, more crucially, also prevents an uncontrolled propagation of errors through the quantum hardware -- a central requirement to ultimately reach the FTQC regime. A universal set of logical gate operations can be achieved by complementing the Clifford operations with a single non-Clifford gate. For 2D color codes such an additional gate operation, e.g. the T-gate \cite{nielsen-book} by magic-state injection \cite{bravyi-pra-71-022316} involves a quantum state teleportation process between the register of system qubits and an ancilla qubit.

\subsection{Trapped-ion alternatives to CNOT-based QEC}

In this section,  we develop trapped-ion alternatives to CNOT-based  schemes for the readout of the 4-qubit stabilizer operators of the color code~\eqref{eq:stabilizers}. This is the essential operation in a QEC cycle, which consists of measuring  all X- and Z-type stabilisers, and performing conditional operations on the physical qubits.  These readout schemes are also essential to encode a particular qubit state:  starting from $\otimes_i\ket{0}$, one would measure all of the X-type stabilisers, perform conditional operations to project onto the code subspace, and  apply  a  single-qubit rotation at the logical level followed by the required QEC cycles. 

 In this section, we start by describing rules for the propagation of errors in circuits involving M\o lmer-S\o rensen (MS) gates. These rules are used to understand the properties of schemes that  work with a single ancillary ion (i.e. non-fault-tolerant schemes) and use either  5-ion~\eqref{eq:MS} or 2-ion~\eqref{2_qubit_MS} MS entangling gates. The  motivation to study these schemes  is to gain  insight on how important it is to avoid the direct occurrence of multi-qubit errors  in QEC protocols~\cite{nigg-science-345-302}.  By using 5 or 4 ancillary qubits and  sequences of 2-ion MS gates, it is also possible to implement a trapped-ion version of the CNOT-based schemes for fault-tolerant stabilizer readout by DiVincenzo-Shor (DVS)~\cite{shor_ft_qec}, and DiVincenzo-Aliferis (DVA)~\cite{aliferis_ft_qec},  respectively. The main goal of exploring these  schemes is to assess under which conditions, and in which experimental parameter regimes, such fault-tolerant protocols might offer advantages in  reaching the break-even point for useful QEC. 
Let us remark that the techniques hereby presented can be easily generalized to any other stabilizer of a different QEC code. Therefore, they will be an essential ingredient of future trapped-ion efforts for QEC.

\begin{figure}
\center
\includegraphics[angle=0,width=1 \columnwidth]{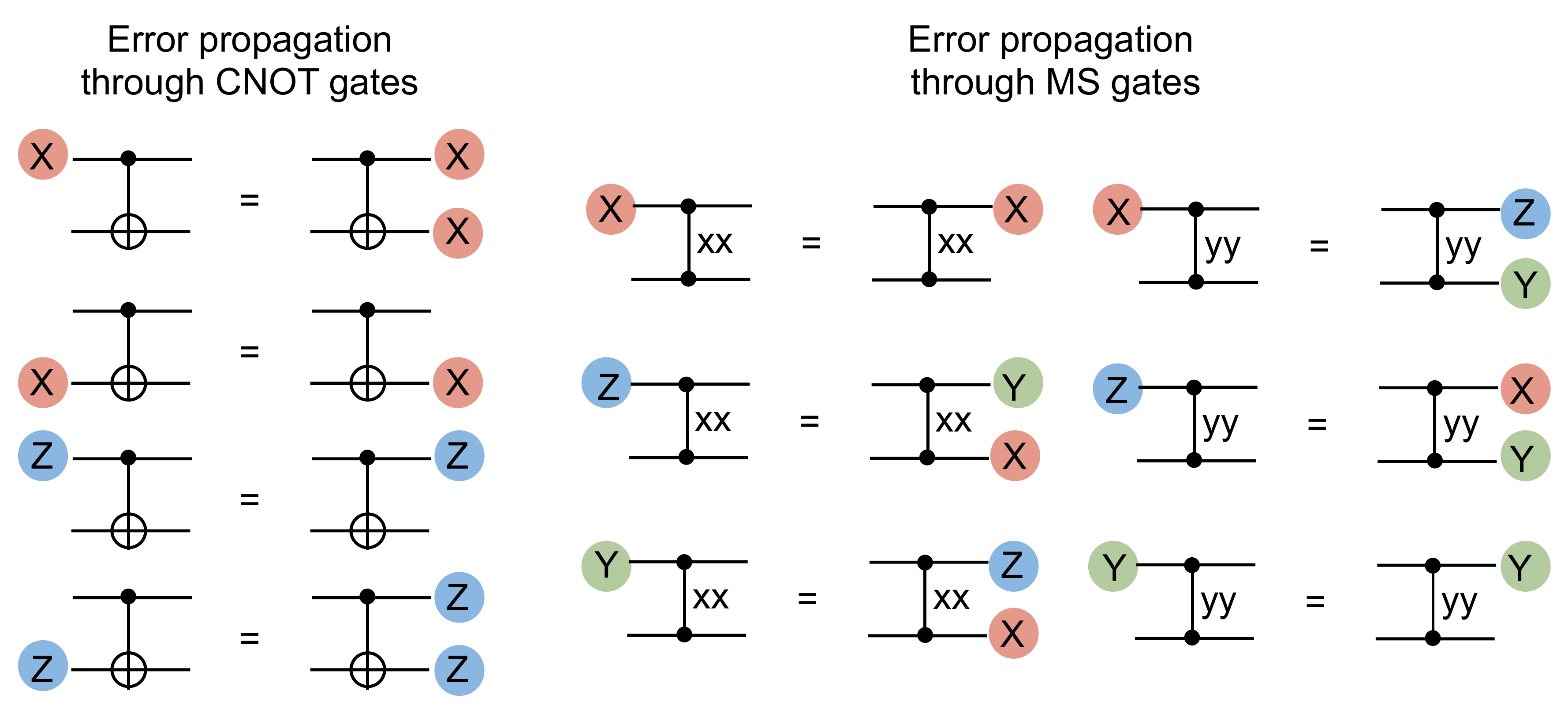} 
\caption{{\bf Error propagation through CNOT and MS gates}: (left panel) An incoming Pauli error of the $X$ ($Z$) type at the control (target) qubit propagates onto a target (control)  error of the same $X$ ($Z$) type after a CNOT gate. Conversely, an incoming Pauli error of the $Z$ ($X$) type at the control (target) qubit does not propagate into the target (control) qubit. (right panel) Error propagation through an MS gate up to phase factors irrelevant for fault-tolerant arguments. The left column corresponds to $XX$-type MS gates $X_{i,j}^2$, whereas the right column describes $YY$-type MS gates $Y_{i,j}^2$. Pauli errors that anticommute with the MS-gate basis are rotated, and propagated to the other qubit. Pauli errors in  the same MS-gate basis do not propagate.}
\label{fig:Error-propagation}
\end{figure}

\subsubsection{ M\o lmer-S\o rensen (MS) gate error propagation}

For the construction of fault-tolerant quantum circuits, it is important to understand how errors  propagate from one qubit to others by means of the entangling gate operations. The circuit identities in the left panel of Fig.~\ref{fig:Error-propagation} show the well-known propagation of $X$ and $Z$-type single qubit errors through CNOT gates.  Analogous error propagation properties can be derived for fully-entangling 2-ion $X^2_{i,j}(\pi/2)$ and $Y^2_{i,j}(\pi/2)$ MS gates~\eqref{2_qubit_MS} (see right panel of Fig.~\ref{fig:Error-propagation}). 
Errors of the same type as the basis of the MS gate commute with the gate operation, e.g.~$X^2_{i,j}(\pi/2) X_i = X_i X^2_{i,j}(\pi/2)$, and thus do not propagate to the second qubit. On the contrary, errors of a  type different from the MS-gate basis, e.g.~a $Z_i$ occurring before a MS gate $X^2_{i,j}(\pi/2)$, are converted into an error of the type that is complementary to the error type, i.e.~into a $Y_i$ error in this case. In addition, this  triggers the creation of an error on the second qubit of the type of the gate, i.e.~here an $X_j$ error. This can be seen from the  identities
$
X^2_{i,j}(\pi/2) Z_i = \frac{1}{\sqrt{2}}(\mathbb{I} - \ii X_iX_j) Z_i = 
  \ii Z_i X_i X_j \frac{1}{\sqrt{2}}(\mathbb{I} - \ii X_iX_j) = -Y_i X_j X^2_{i,j}(\pi/2)
$
which,  in an analogous fashion, also hold for $Y$-type MS gates and the other types of errors. Note that, in contrast to the CNOT gate operation, the MS gate is symmetric in the indices of the two qubits, i.e.~these propagation rules hold equally for errors arriving on the second qubit of the gate, e.g.~$Y^2_{i,j}(\pi/2) Z_j = Y_i X_j Y^2_{i,j}(\pi/2)$.

\subsubsection{ Non-fault-tolerant stabilizer readout}
\label{sec:non-FT-2-ion-scheme}

In this section, we start by reviewing  the  scheme for stabilizer readout of the 7-qubit code~\cite{nigg-science-345-302}, which uses  a single ancilla ion and multi-qubit MS gates. We then extend this scheme to  a protocol that is based on 2-ion MS gates, which will be used   to explore how important it is to avoid the direct occurrence of multi-qubit errors from the  multi-ion MS gates. 

\begin{figure}[t]
 \begin{centering}
  \includegraphics[width=1\columnwidth]{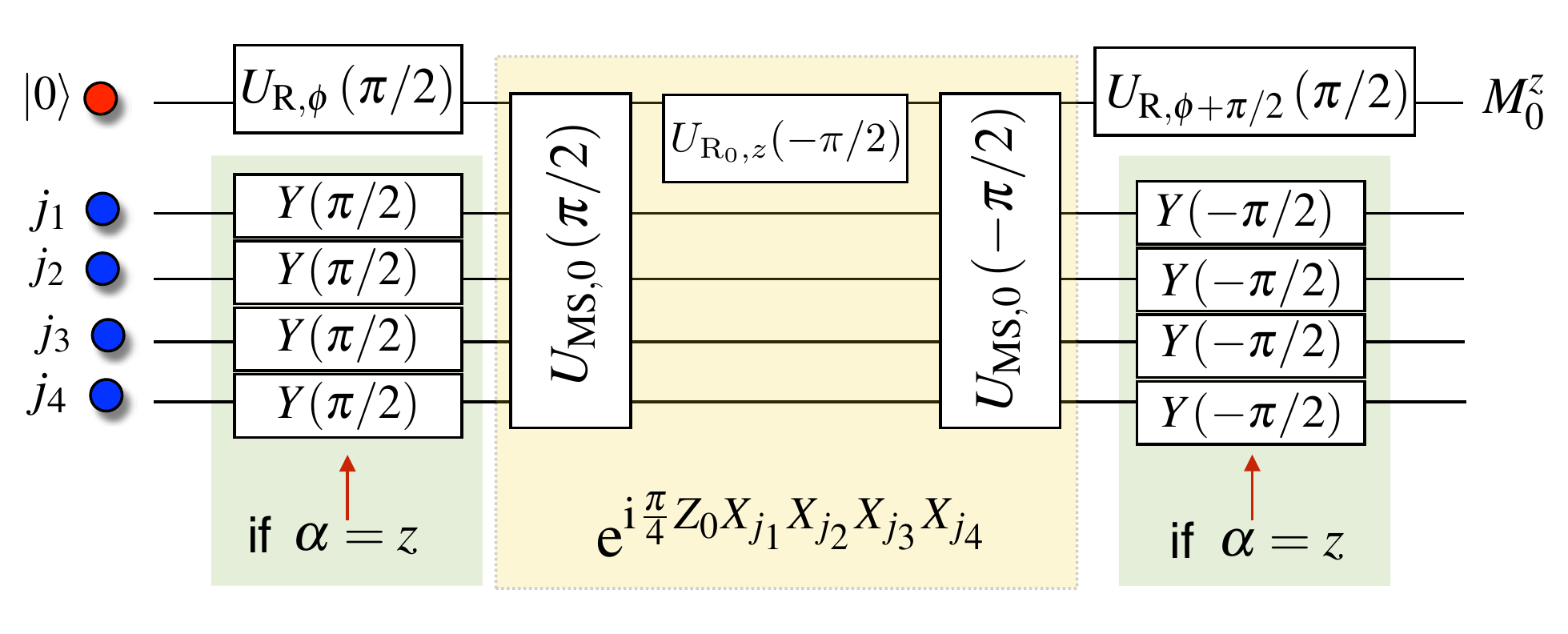}\\
  \caption{\label{Fig:4qubit_stab_readout} {\bf Stabilizer readout based on multi-qubit MS gates}: The yellow box contains two 5-ion MS gates, interspersed by a single-qubit rotation of the ancillary qubit around the $Z$-axis. This sequence realizes a coherent mapping of the $+1$/$-1$ eigenvalue information of  $X$-type  stabilizers~\eqref{eq:stabilizers} onto two orthogonal states of the ancilla qubit (red dot), initially prepared in a superposition state on the equator of the Bloch sphere. Note that the basis of the initial  rotation $U_{\rm R,\phi}(\pi/2)$ required to prepare the ancilla qubit in this way,  specified by the angle $\phi$,  can be arbitrary as long as the final   rotation $U_{\rm R,\phi+	\pi/2}(\pi/2)$ of the ancilla is shifted by $\pi/2$. In other words, the relative phase between these two single-qubit pulses on the ancilla needs to be well-defined, but there does not need to be a fixed phase reference between the addressed laser,  driving resonant single-qubit rotations, and the lasers driving global rotations and the entangling MS gate. 
For the readout of $Z$-type stabilizers~\eqref{eq:stabilizers},  the data qubits must be rotated via $Y(\pm \pi/2)$ at the beginning and at the end of the mapping, to effectively switch between $X$- and $Z$-type of stabilizers.}
\end{centering}
\end{figure}

{\it (i) Multi-qubit MS stabilizer readout.--} The   readout of the stabilizers~\eqref{eq:stabilizers} can be accomplished by mapping the syndrome information
of the data qubits onto a single ancillary qubit using two global $5$-ion MS gates~\eqref{eq:MS},
and an intermediate single-qubit rotation via a local ac-Stark
shift~\eqref{eq:rot_z}, since 
\beq
U_{\rm MS,0}(-\pi/2)U_{{\rm R}_ j,z}(-\pi/2)U_{\rm MS,0}(\pi/2)=\ee^{\ii\frac{\pi}{4}Z_{ j}\Pi_{i\neq j}X_i},
\eeq
as shown in~\cite{mueller-njp-13-085007}. By applying this  sequence to the ancillary ion and a
particular subset of four  qubits $\{j_1,j_2,j_3,j_4\}$  belonging  to a particular plaquette
stabilizer~\eqref{eq:stabilizers}, one can map all stabilizers $\big\{S_\alpha^{(1)},S_\alpha^{(2)},S_\alpha^{(3)}\big\}_{\alpha=x,z}$ onto the ancillary qubit. For instance, for the first $X$-type stabilizer, one finds  $U_{\rm MS,0}(-\pi/2)U_{{\rm R}_{ 0},z}(-\pi/2)U_{\rm MS,0}(\pi/2)={\rm exp}(\ii\frac{\pi}{4}Z_{ 0}S_x^{(1)})$. The stabilizer information can then be measured by performing a Ramsey-type sequence on the ancillary qubit (see Fig.~\ref{Fig:4qubit_stab_readout}). The ancilla qubit is initialized in $\ket{0}_0$,  and after the  pulse $U_{{\rm R}_0,\phi}=U_{\rm R,\phi}(\theta/2)U_{{\rm R}_0,z}(\pi)U_{\rm R,\phi}(-\theta/2)U_{{\rm R}_0,z}(-\pi)$, one maps the stabiliser information into the ancilla using the above scheme. Finally, after applying the pulse on the ancilla qubit $U_{{\rm R}_0,\phi+\pi/2}=U_{\rm R,\phi+\pi/2}(\theta/2)U_{{\rm R}_0,z}(\pi)U_{\rm R,\phi+\pi/2}(-\theta/2)U_{{\rm R}_0,z}(-\pi)$, one measures it in the computational basis $M_0^z$, obtaining $\pm1$ when the ancilla qubit is found in state $\ket{0}_0$ or $\ket{1}_0$, respectively. These outcomes correspond to the $\pm 1$ eigenvalue information of the corresponding stabiliser.

\begin{figure}
\center
\includegraphics[angle=0,width=0.9 \columnwidth]{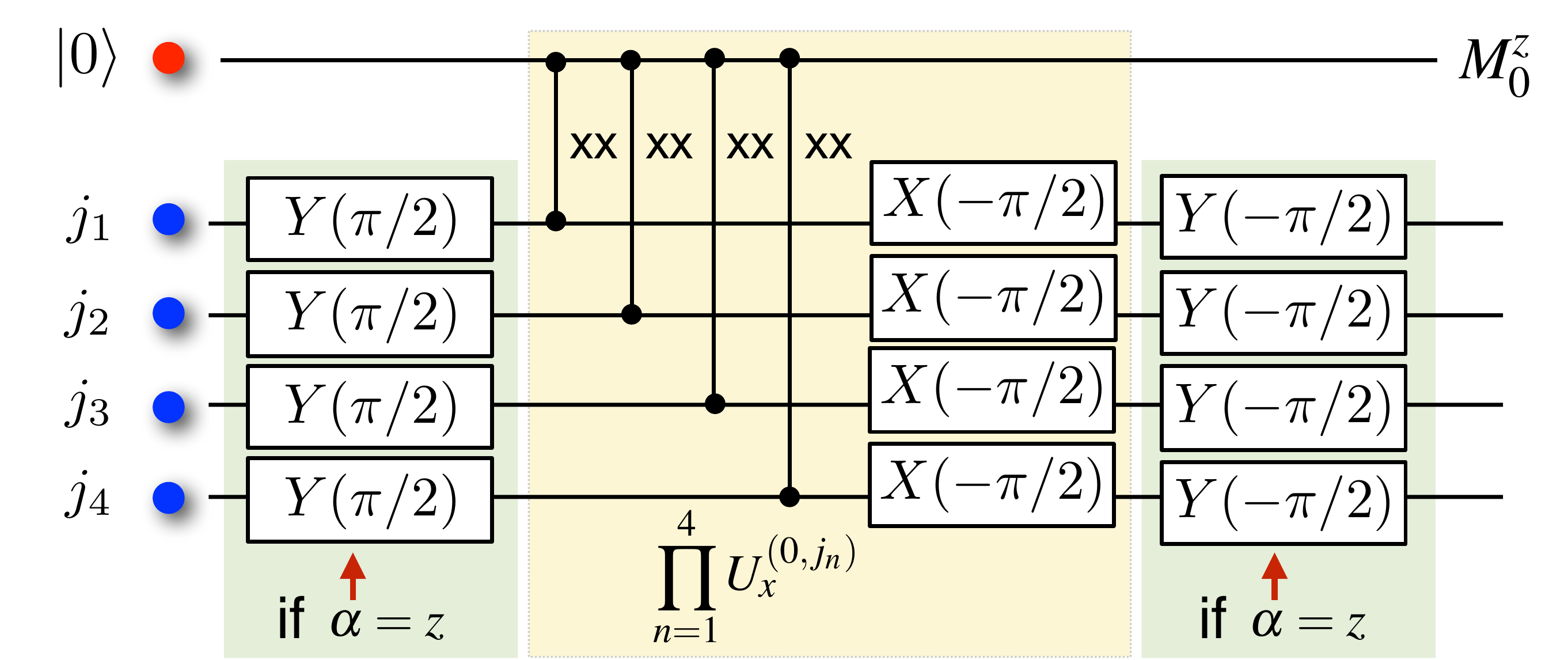} 
\caption{{\bf Stabilizer readout based on sequential  2-qubit MS gates:}  $X$-type ($\alpha=x$) and $Z$-type ($\alpha=z$) stabiliser readout circuits, using sequential 2-ion MS $X_{0,j}^2$ gates~\eqref{2_qubit_MS}, depicted by solid lines with black dots between the single  ancilla qubit and each of the the data qubits involved in a particular stabilizer, and with an $XX$-label that defines the basis of the entangling gate~\eqref{eq:MS}. The MS gates have to be combined with single-qubit rotations $X_j(\pi/2), Y_j(\pm\pi/2)$ on the data qubits~\eqref{eq:cnot_like} to achieve the stabilizer mapping~\eqref{eq:stab_mapping}. The measurement of the ancillary qubit in the computational basis is denoted as $M_0^z$.}
\label{fig:4-1-readout-with-2-ion-MS-gates}
\end{figure}

{\it (ii) Two-qubit MS stabiliser readout.--} Let us now introduce a circuit for the readout of a 4-qubit stabiliser based on 2-ion MS gates.  Such a circuit could be constructed by compiling the known circuits based on 4 CNOTs, using the fact that a 2-ion MS gate is equivalent to a 2-qubit CNOT, up to local unitary operations~\cite{Nebendahl-PRA-2009}. Alternatively, one can construct such circuits minimizing the building blocks by  noticing that 
\beq
\label{eq:cnot_like}
\begin{split}
U_x^{(i,j)} \!&:= 
 \ket{x_+} \bra{x_+}_i \otimes \mathbb{I}_j + \ket{x_-} \bra{x_-}_i \otimes \ii X_j= X_j(\textstyle{\frac{-\pi}{2}}) X^2_{i,j}(\textstyle{\frac{\pi}{2}}),\\
 U_y^{(i,j)} \!&:= 
 \ket{y_+} \bra{y_+}_i \otimes \mathbb{I}_j + \ket{y_-} \bra{y_-}_i \otimes \ii Y_j= Y_j(\textstyle{\frac{-\pi}{2}}) Y^2_{i,j}(\textstyle{\frac{\pi}{2}}),
 \end{split}
\eeq
where  $\ket{x_\pm}=(\ket{0}\pm\ket{1})/\sqrt{2}$, and $\ket{y_\pm}=(\ket{0}\pm\ii\ket{1})/\sqrt{2}$. This identity shows that a combination of a two-qubit MS gate between an ancillary-data qubit pair $(0,j)$, and a single-qubit $\pi/2$-pulse on the data  qubit $j$, acts essentially as a rotated version of a CNOT gate.
Thus, applying sequentially this pair of operations between the ancilla qubit  and the four plaquette  qubits, realizes the mapping of certain stabiliser $S^{(m)}_x = X_{j_1} X_{j_2} X_{j_3} X_{j_4}$ onto the ancilla qubit. Considering that the ancilla qubit is initially in  $\ket{0}_0$, while the data qubits are in an arbitrary  state $\ket{\psi}$, we find
\begin{align}
\label{eq:stab_mapping}
\prod_{n=1}^4 U_x^{(0,j_n)}\!\! \ket{0}_0\! \ket{\psi} 
 = \ket{0}_0\! \half (1 + S^{(m)}_x)\! \ket{\psi} + \ket{1}_0\! \half (1 - S^{(m)}_x)\! \ket{\psi}.
\end{align}
Hence, the two possible values of the stabilizer $\pm 1$  can be directly inferred by measuring the ancilla qubit $M_0^z$ in the $Z$-basis, such that  the initial and final Ramsey pulses of the multi-qubit readout of Fig.~\ref{Fig:4qubit_stab_readout} are not required any longer. Similarly, $Z$-type stabiliser operators can be measured by interchanging the basis of the data (plaquette) qubits from $X$ to $Z$ before and after applying the above sequence of gates, $Y_j(\pi/2) X_j Y_j(-\pi/2) =- Z_j$. The resulting circuits for measuring $X$- and $Z$-type stabiliser operators are shown in Fig.~\ref{fig:4-1-readout-with-2-ion-MS-gates}.

\begin{figure}
\center
\includegraphics[angle=0,width=1\columnwidth]{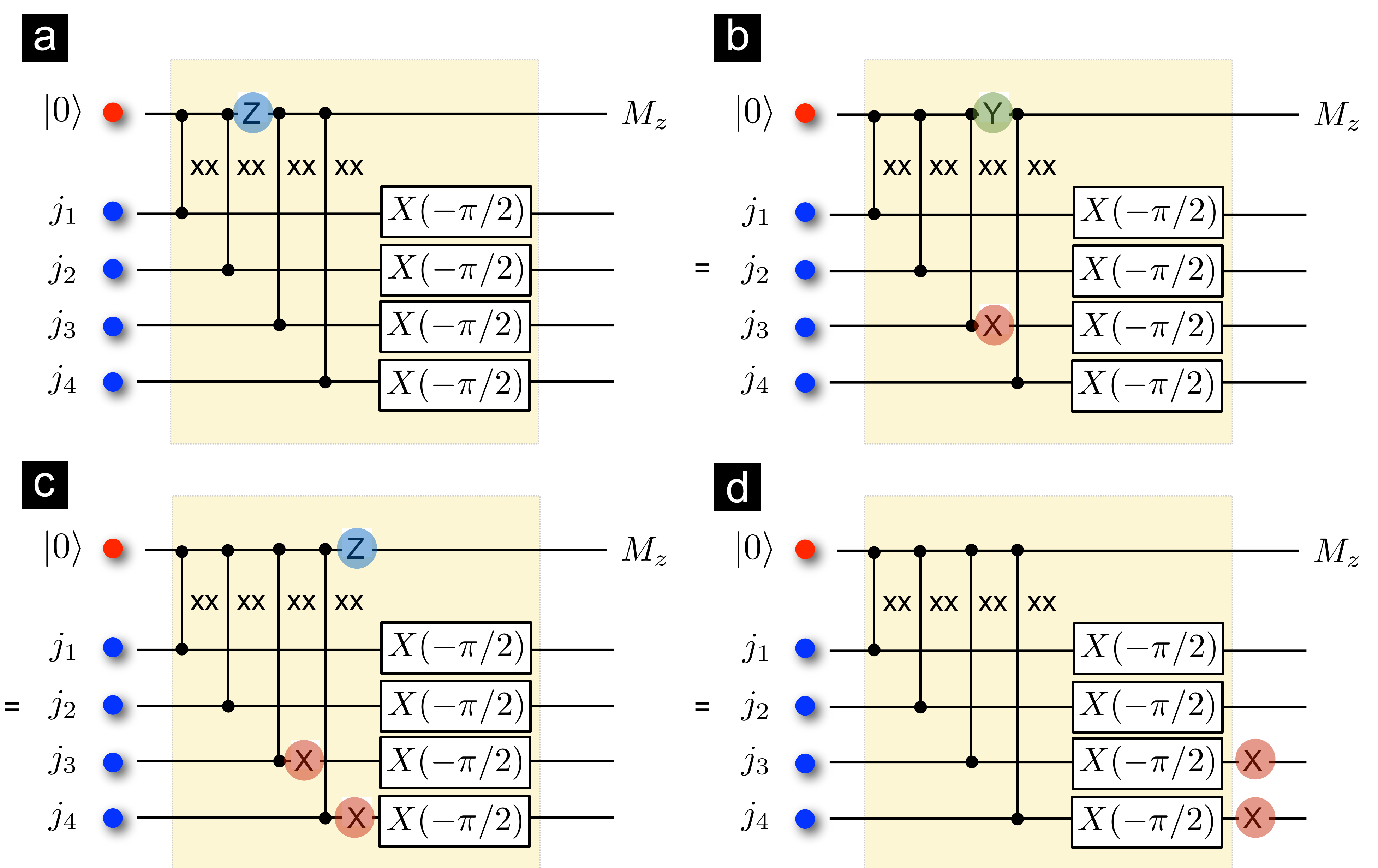} 
\caption{{\bf Non-fault-tolerance of single-ancilla MS-based stabilizer readout}: Illustration of an error event, in which a single phase flip error $Z$ on the ancilla qubit occurs between the second and the third MS gates (a). This error  propagates into the data qubit layer (b) after the third MS gate, where it results in two bit single-qubit flip errors on the data qubits (c). These two errors correspond ultimately to a logical bit flip error (d) that cannot be corrected by the  code. }
\label{fig:Error_propagation_into_data_layer}
\end{figure}

Note, however, that using a single ancillary qubit does not allow for a fault-tolerant measurement of the stabilizers for the 7-qubit code. The reason is that a single-qubit error can propagate to the set of data qubits and lead to two errors therein. These will then, in the subsequent round of QEC, lead to a logical error (see Fig.~\ref{fig:Error_propagation_into_data_layer} for details). Similar effects would occur for the previous scheme based on 5-ion MS gates, also forbidding a fault-tolerant behavior.

\subsubsection{ Fault-tolerant DiVincenzo-Shor stabilizer readout with 2-ion MS gates}
\label{sec:Shor-QEC}

\begin{figure}
\center
\includegraphics[angle=0,width=0.9\columnwidth]{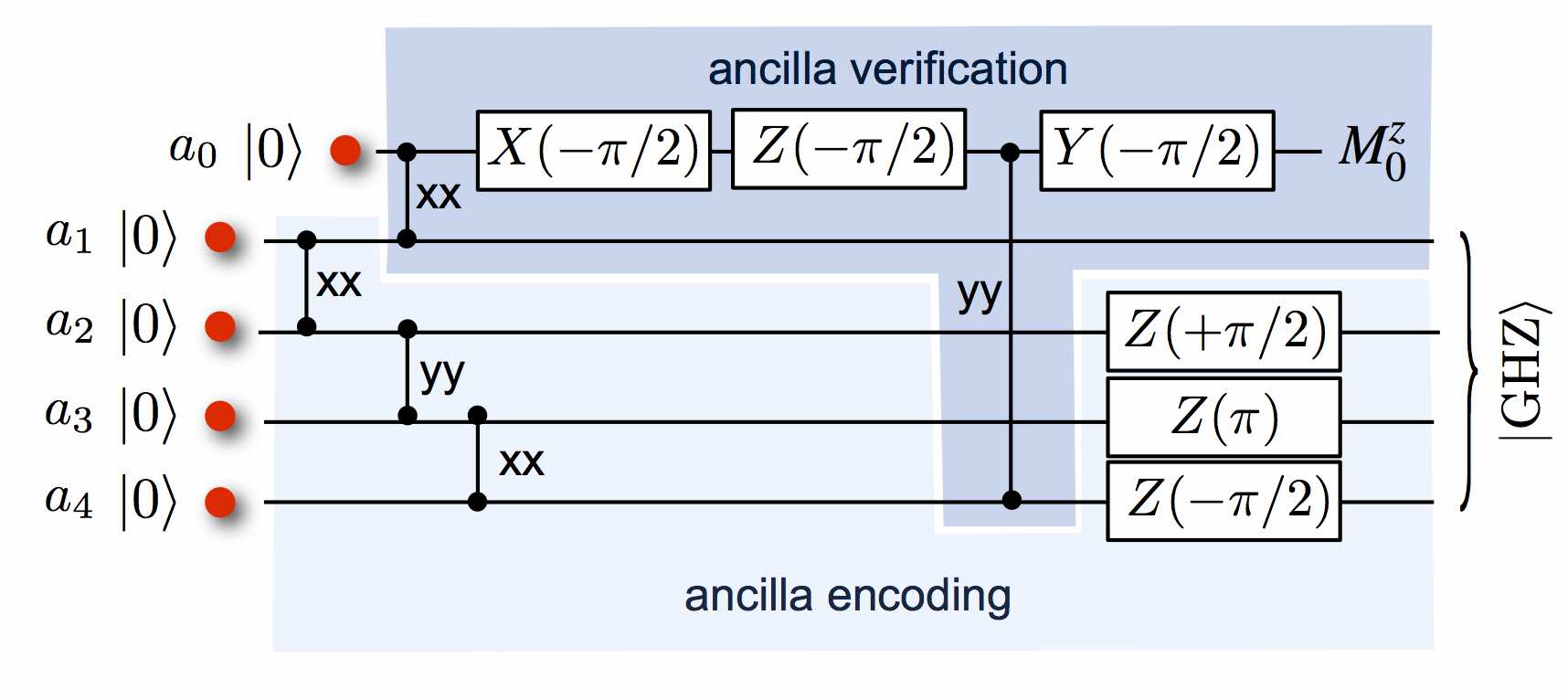} 
\caption{{\bf MS-based circuit for preparation and verification of a GHZ state in the DiVincenzo-Shor scheme}:  The first three MS gates create a 4-qubit GHZ state which is, up to single-qubit $Z$-rotations, equivalent to the desired GHZ state $\frac{1}{\sqrt{2}}(\ket{x_+,x_+,x_+,x_+} + \ket{x_-,x_-,x_-,x_-})$. The state is verified by coupling it via two additional MS gates involving the first and the fourth ancilla qubit to a verification qubit. Note that the $Z$-type rotation on the ancilla verification qubit is incorporated to remove a relative phase in the GHZ state. Note that some parts of the circuit can be executed in parallel, such as e.g. part of the verification circuit, while the GHZ state is still being built up.}
\label{fig:Shor-QEC-state-preparation}
\end{figure}

\begin{figure}
\center
\includegraphics[angle=0,width=0.85\columnwidth]{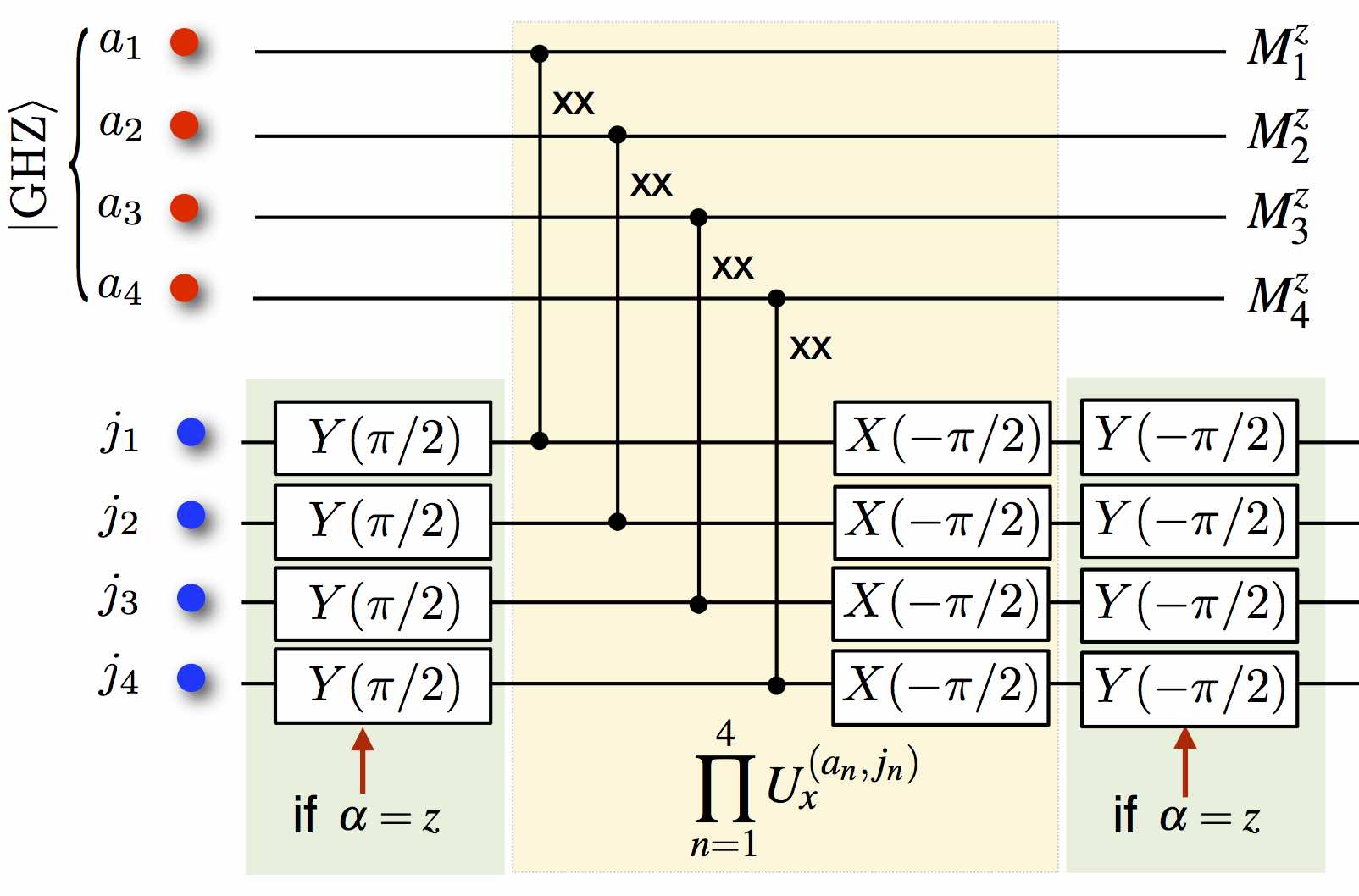} 
\caption{{\bf MS-based circuit for transversal coupling   in the DiVincenzo-Shor scheme}: The stabiliser $S_\alpha^{(n)}$  information is encoded in the relative phase of the  ancilla GHZ state $\frac{1}{\sqrt{2}}(\ket{x_+x_+x_+x_+} \pm \ket{x_-x_-x_-x_-})$ by a sequence of two-ion MS gates and local rotations~\eqref{eq:cnot_like}. This relative phase is revealed by the even or odd parity of the   $Z$-basis measurements of the four ancilla qubits.}
\label{fig:Transversal-stab-type-readout}
\end{figure}

In this section, we develop a fault-tolerant version of the stabiliser readout using 2-ion MS gates. Let us consider the  CNOT-based approach by DiVincenzo-Shor (DVS)~\cite{shor_ft_qec}, in which a 4-qubit ancilla GHZ-type state ("cat state") is prepared, verified, and subsequently coupled transversally, and thus fault-tolerantly, to the respective four data qubits. This scheme requires, besides the 7 data qubits of the color code, 5 additional ancilla qubits: 4 for the ancilla GHZ state (indices $a_1$, $a_2$, $a_3$ and $a_4$) and one extra ancilla qubit ($a_0$) for  verifying the GHZ state in a measurement. The preparation, verification, transversal coupling, and decoding for the readout of a single stabiliser can be accomplished by 12 CNOT gates, and a couple of Hadamards. Since the CNOT for two ion qubits in a larger register can be constructed using a single MS-gate and 4 single-qubit rotations $Y_{a_n}\big(\frac{-\pi}{2}\big)X_{a_n}\big(\frac{-\pi}{2}\big)X_{j_n}\big(\frac{-\pi}{2}\big)X_{a_n,j_n}^2\big(\frac{\pi}{2}\big)Y_{a_n}\big(\frac{\pi}{2}\big)=\ee^{\ii\pi/4}(\ket{0}\bra{0}_{a_n}\otimes\mathbb{I}_{j_n}+\ket{1}\bra{1}_{a_n}\otimes X_{j_n})$, the straightforward translation of the DVS protocol onto a trapped-ion hardware would require 12 MS gates, and 50 single-qubit rotations.  Let us now discuss, step by step, an alternative MS-based approach that reduces the total number of gates (see Figs.~\ref{fig:Shor-QEC-state-preparation} and~\ref{fig:Transversal-stab-type-readout}).

\textit{ (i) Preparation of the ancilla GHZ state:} The 4-qubit GHZ state can be prepared by a sequence of 2-ion $XX$ and $YY$ MS gates acting on the ancilla qubits  initially prepared in $\ket{\psi_0}=\ket{0_{a_1},0_{a_2},0_{a_3},0_{a_4}}$. A single $XX$ MS gate leads to a Bell state of the ancilla qubits $a_1$,$a_2$, rewritten as   $
X^2_{a_1,a_2}\big(\frac{\pi}{2}\big) \ket{\psi_0} = \frac{1}{\sqrt{2}} \big( \ket{x_+}_{a_1} \! \ket{y_-}_{a_2} + \ket{x_-}_{a_1}\!  \ket{y_+}_{a_2}\big)\ket{0_{a_3},0_{a_4}}=:\ket{\psi_1}$. A subsequent $YY$ entangling gate applied to  $a_2$ and $a_3$ extends this state into  a 3-qubit GHZ-type state, namely  $
Y^2_{a_2,a_3}\big(\frac{\pi}{2}\big)  \ket{\psi_1}
= \frac{1}{\sqrt{2}} \big( \ket{x_+}_{a_1}\!  \ket{y_-}_{a_2} \!\ket{x_-}_{a_3} + \ket{x_-}_{a_1}\! \ket{y_+}_{a_2}\! \ket{x_+}_{a_3}\big)\ket{0_{a_4}}=: \ket{\psi_2}.
$
 Finally, a $X$-type MS gate on $a_3$ and $a_4$ produces a  GHZ-type state
$
 X^2_{a_3,a_4}\big(\frac{\pi}{2}\big) \ket{\psi_2}=:\ket{\psi_3}$, where we have introduced
 \beq
 \label{eq:ghz_like}
\ket{\psi_3}\!=\!  \textstyle{\frac{1}{\sqrt{2}}\! \big( \ket{{x_+}}_{a_1}\! \!\ket{y_-}_{a_2}\!\! \ket{x_-}_{a_3}\! \!\ket{y_+}_{a_4}\!\!\! + \ket{x_-}_{a_1}\!\! \ket{y_+}_{a_2}\!\! \ket{x_+}_{a_3} \!\!\ket{y_-}_{a_4}\!\!\big)},
\eeq 
which can be converted into a standard GHZ-type state after its verification.

\textit{(ii) Verification of the GHZ state:} This GHZ-type state $\ket{\psi_3}$~\eqref{eq:ghz_like} can be verified by coupling the first $a_1$ and  fourth $a_4$ ancilla qubits to the verification ancilla qubit $a_0$ via  $X$- and  $Y$-type MS gates  which, together with local unitary rotations on the verification qubit, yield the operations $U_x^{(a_1,a_0)}$ and   $U_y^{(a_4,a_0)}$ in Eq.~\eqref{eq:cnot_like}, respectively.  The expressions~\eqref{eq:cnot_like} show that, under these two operations, the ancilla verification qubit initially prepared in $\ket{0}_{a_0}$ is either not flipped  at all (i.e. first component of the GHZ state $\ket{\psi_3}$), or flipped twice  (i.e. second component of the GHZ state $\ket{\psi_3}$), gaining an additional phase shift that can be compensated with a local Z-rotation. Therefore, for a perfect preparation of the GHZ-type state $U_y^{(a_4,a_0)}Z_{a_0}\big(\frac{-\pi}{2}\big)U_x^{(a_1,a_0)}\ket{0_{a_0}}\ket{\psi_3}=\ket{0_{a_0}}\ket{\psi_3}$, and the verification qubit should ideally end up in $\ket{0}_{a_0}$ which can be checked by measuring $M_0^z=+1$ in the computational basis. 

The state  $\ket{\psi_3}$ is finally converted by  local $Z$-rotations into the desired  GHZ state 
$
 Z_{a_2}\big(\frac{\pi}{2}\big)  Z_{a_3}(\pi) Z_{a_4}\big(\frac{-\pi}{2}\big)  \ket{\psi_3} =\ket{\text{GHZ}}$, where we have introduced
 \beq
 \label{eq:ghz}
\ket{\text{GHZ}}=  \textstyle{\frac{1}{\sqrt{2}} \big( \ket{x_+,x_+,x_+,x_+} + \ket{x_-,x_-,x_-,x_-}\big)}, 
  \eeq
which will be coupled transversally to the data qubits. The  circuit for the MS-based scheme used for the preparation as well as verification of the ancilla GHZ state is shown in Fig.~\ref{fig:Shor-QEC-state-preparation}.

\textit{ (iii) Coupling of the ancilla GHZ state to the data qubits:} To realize the readout of an $X$-type stabiliser operator, the verified 4-qubit ancilla GHZ state is then coupled transversally to the corresponding four data qubits. Again, using the operators in Eq.~\eqref{eq:cnot_like}, 
it can be shown that the sequence of pairwise unitaries $U_x^{(4)}=\prod_{n=1}^4 U_x^{(a_n,j_n)}$  leads to
$
\ket{\psi_4}=U_x^{(4)}\ket{\text{GHZ}} \ket{\psi} =  \frac{1}{\sqrt{2}}(\ket{x_+,x_+,x_+,x_+} \ket{\psi} + \ket{x_-,x_-,x_-,x_-} S_x^{(m)} \ket{\psi}),
$ where we have introduced an arbitrary basis state of the four data qubits $\ket{\psi}$. 
Accordingly, these sequential operations map the $\pm1$ eigenvalue information of a stabiliser $S^{(m)}_x = X_{j_1}X_{j_2}X_{j_3}X_{j_4}$, and thus part of the error syndrome,  onto the relative phase of the  ancillas   $\ket{\psi_4}=\frac{1}{\sqrt{2}}\big(\ket{x_+,x_+,x_+,x_+}  \pm \ket{x_-,x_-,x_-,x_-}\big)\ket{\psi}$.
This relative phase of $+1$ or $-1$ will result in even/odd parity of the combined outcome $(M_1^z,M_2^z,M_3^z,M_4^z)$ of a subsequent measurement of the four ancilla qubits in the $Z$-basis as shown in Fig.~\ref{fig:Transversal-stab-type-readout}. In this figure, we also show the additional local rotations that must be performed for the measurement of $Z$-type stabilizers.
As customary for stabilizer codes, the same circuit can be used as a building block to prepare the  state of the encoded qubit.  

To demonstrate the fault-tolerant nature of the constructed circuit, we must show that, if at most a single error (single-qubit error, two-qubit gate error, or measurement error) occurs anywhere in the circuit, it  will not result in an uncorrectable error on the data qubits that would yield  a logical error. Note that $X$ errors resulting at the output of the circuit are not dangerous as they can only result in a wrong relative sign of the GHZ state, which is later coupled to the data qubits. This can ultimately result in a stabiliser measurement error. Such stabiliser measurement errors are taken care of by applying several rounds (up to three) of readout to reliably decode the error syndrome. With respect to the fault-tolerance of the circuit, the key point is that such $X$ errors never propagate to data qubits since they commute with the $XX$ MS gates  used during the coupling stage, both for $X$- and $Z$-type stabiliser readout. 

In contrast, undetected phase flip errors ($Z$) generated during the preparation and verification of the GHZ state in Fig.~\ref{fig:Shor-QEC-state-preparation} will propagate to the data qubit layer during the coupling step in Fig.~\ref{fig:Transversal-stab-type-readout}, resulting   in bit flip errors ($X$). If two bit flip errors are introduced into the code, this will result in an uncorrectable logical error. However, 
the preparation and verification circuit is constructed in such a way that any combination of two phase flip errors is detectable, as it  will necessarily result in a $M_0^z=-1$ measurement of the ancilla verification qubit $a_0$. If this is the case, the GHZ state must be discarded, and another attempt at preparing and verifying the required ancilla GHZ state is started. It is tedious but straightforward to show that all dangerous two-qubit phase flip errors that can   affect the data qubits are equivalent to a $Z_3Z_4$ error, and will be detected in the verification step through the outcome $M_0^z=-1$. Note that a $Z_1Z_2$ error is equivalent to a $Z_3Z_4$ error, as the resulting two bit flip errors in the code  are equivalent up to an $S_x$-stabiliser. Three-qubit phase flip errors, e.g. $Z_2Z_3Z_4$, are equivalent to a single phase flip error $Z_1$  using the same argument, which propagates to the data layer, but only results in a single $X_1$ bit flip error. This is an allowed process, as single-qubit errors will be picked up and corrected  for in the next QEC round to  leading order.

\subsubsection{Fault-tolerant  DiVincenzo-Aliferis stabilizer readout with 2-ion MS gates}
\label{sec:Aliferis-diVincenzo-QEC}

\begin{figure}
\center
\includegraphics[angle=0,width=0.75\columnwidth]{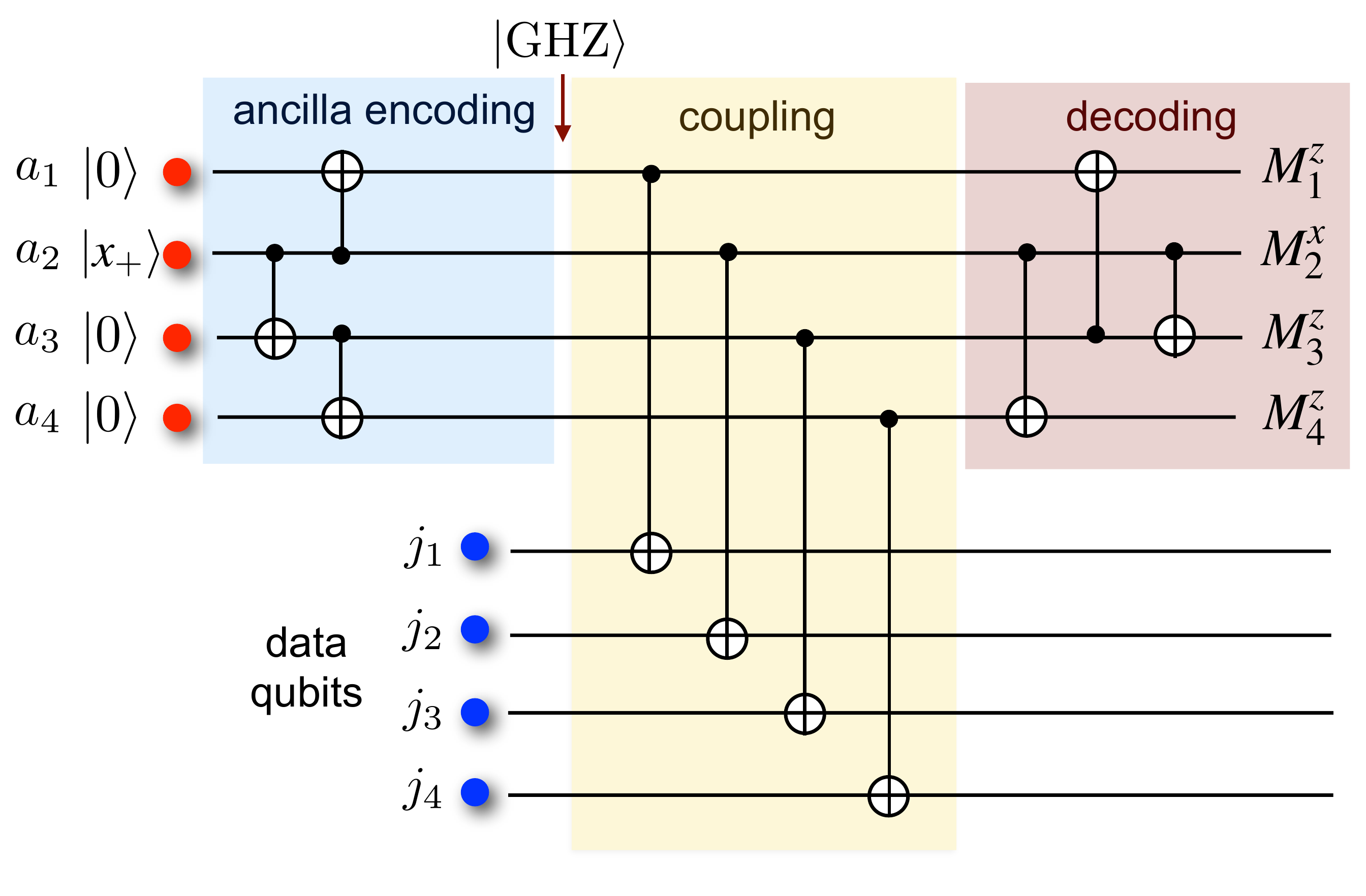} 
\caption{{\bf  CNOT-based circuit for DiVincenzo-Aliferis-type QEC}: This circuit  involves the preparation of an ancilla 4-qubit GHZ state by CNOT gates, its transversal coupling to the data qubit layer,  the ancilla state decoding,  and the  measurements of the ancillas.}
\label{fig:CNOT-DiVincenzo-Aliferis-QEC}
\end{figure}
In this section, we develop a fault-tolerant MS-based version of  the  DiVicenzo-Aliferis (DVA)  approach~\cite{aliferis_ft_qec}, which was originally introduced in  terms of CNOT gates.  Similar to the DVS  scheme discussed above,  an ancilla GHZ state~\eqref{eq:ghz} is initially prepared by a sequence of MS gates. The main difference of the DVA protocol is that its verification  is  postponed  until the end of the readout step. Hence, the GHZ state is  coupled transversally to the data qubits, after which  the ancilla state  is decoded and  measured to obtain the  stabiliser  information.  Importantly, the decoding circuit is constructed in such a way that it also allows one  to  detect unambiguously the occurrence of two single-qubit errors that  have  propagated to the data qubits  potentially causing a logical error. If such a situation is detected, the corresponding two-qubit error correction operation is either physically applied to the data qubits, or used on a software-level to update the Pauli frame.  

DiVincenzo and Aliferis  argue that, by postponing the verification step involving measurements to the end, this scheme can be highly beneficial and avoid bottlenecks when the measurements of qubits are much slower than gate operations~\cite{aliferis_ft_qec}, which is typically the case of  trapped-ion hardware (see Tables~\ref{tab:summary_gates} and~\ref{tbl:shuttlingops}). Furthermore, only 4 ancilla qubits are needed, as compared to the 5 needed for the DVS scheme of Fig.~\ref{fig:Shor-QEC-state-preparation}. Another nice feature is that the verification is not of a stochastic nature, which can simplify considerably the time control and synchronization in a larger QEC protocol. Previous studies have aimed at a comparison of DVS and DVA CNOT-based schemes for the Steane code in a trapped-ion architecture \cite{Tomita-Gutierrez-PRA-2013}. Here, we develop similar schemes based on 2-ion MS gates, and take into account the specific architectural constraints of our experimental system.

\begin{figure}
\center
\includegraphics[angle=0,width=1\columnwidth]{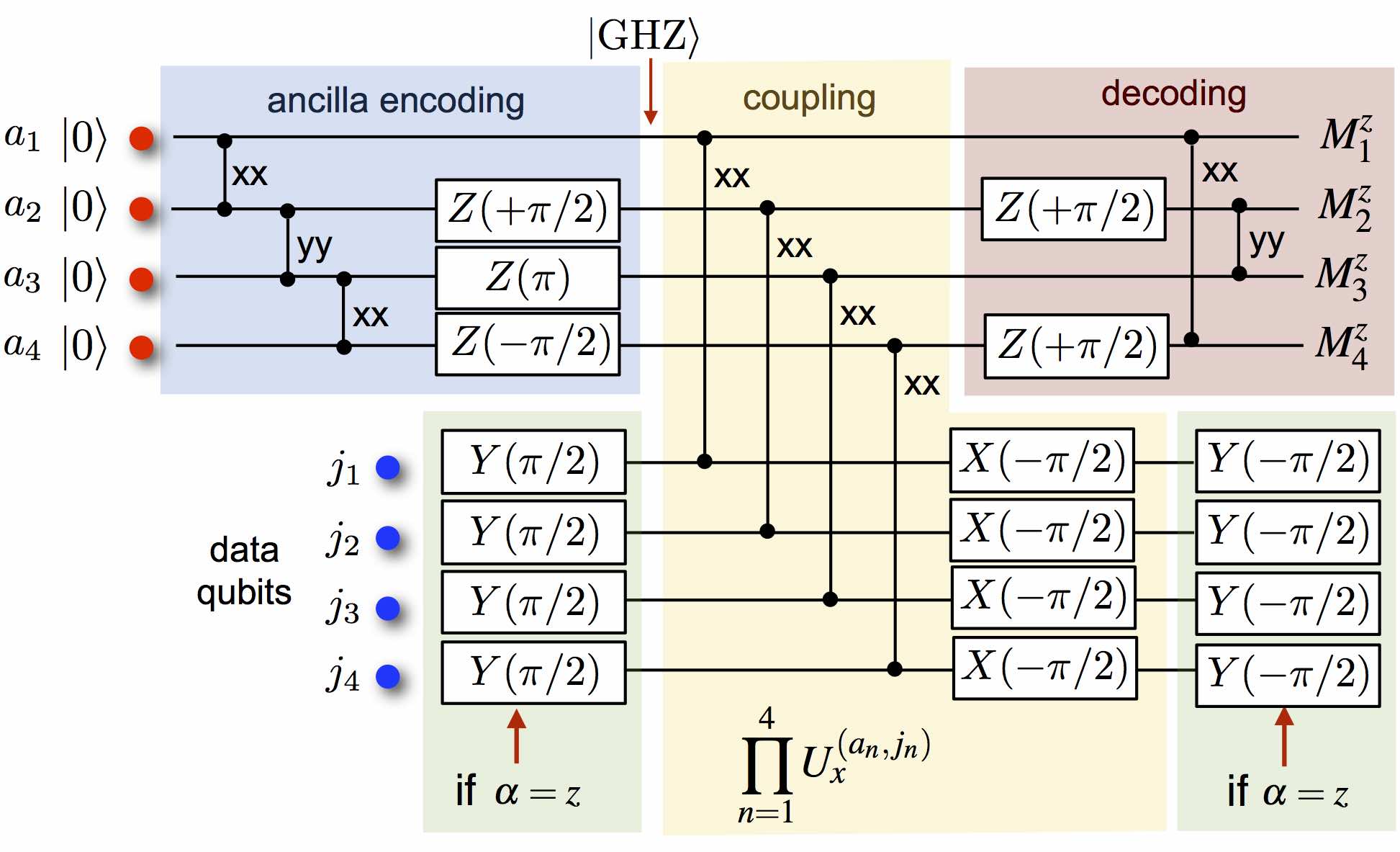} 
\caption{{\bf MS-based  circuit for DiVincenzo-Aliferis-type QEC}: Initially, a sequence of three 2-ion MS gates and local $Z$-rotations is used to prepare the ancilla $\ket{\text{GHZ}}$ (see Eq.~\eqref{eq:ghz}). This state is subsequently coupled transversally by four $X$-type MS gates to the data qubits, thereby mapping the $S_x$ stabiliser eigenvalue information onto the relative phase of the GHZ state. The decoding circuit, formed by  local $Z$-rotations and two 2-ion MS gates, is constructed in such a way that a harmful two qubit error, which has propagated to the code layer, is unambiguously signaled by the $M_3^z=+1, M_4^z=-1$ measurement outcome of the third and fourth ancilla qubit. The  measurement of $Z$-type stabilizers ($\alpha=z$) is almost identical, introducing also  $Y$-rotations  of the four data qubits before and after applying the four $X$-type MS gates of the coupling step.}
\label{fig:DiVincenzo-Aliferis-2ion-MS-gates}
\end{figure}

\vspace{1ex}
{\it (i) DiVincenzo-Aliferis QEC based on CNOT gates.--}Figure~\ref{fig:CNOT-DiVincenzo-Aliferis-QEC} shows the standard circuit based on CNOT gates  for the fault-tolerant measurement  of a four-qubit stabiliser. If no error  occurs at all,  the measurement $M_2^x$  in the $X$ basis of the second ancilla qubit, which was initially prepared in $\ket{x_+}_{a_2}$, will reveal the desired $\pm1$ stabiliser information  after decoding. Additionally,  the remaining  ancilla qubits ("check qubits") will end up in the state $\ket{0}_{a_n}$,  and yield a $M_1^z=+1$, $M_3^z=+1$ and $M_4^z=+1$ outcome. For this circuit, it can be shown that all dangerous two-qubit errors on the ancilla qubits that propagate to the data qubits, and would  induce a logical error, are equivalent to an $X_{a_3}X_{a_4}$ error. This error, which could result e.g.~from a bit-flip error  before  the CNOT gate involving  ancilla qubits $a_3$ and $a_4$ during the GHZ-state preparation (encoding), would propagate through two of the CNOT gates in the coupling step to the $j_3$ and fourth $j_4$ qubits. Note, however, that the  circuit is constructed in such a way that these two errors, $X_{a_3}X_{a_4}$, propagate among the ancilla qubits during the decoding stage causing a bit flip on all three ancilla "check" qubits,  and thus yielding $M_1^z=-1$, $M_3^z=-1$ and $M_4^z=-1$. One  can finally show that this outcome on the "check" ancillas is uniquely associated to the occurrence of such an $X_{a_3}X_{a_4}$ error. In other words, any single error on a data qubit occurring during the decoding circuit cannot cause the same  outcome, and cannot  be thus confused with the $X_{a_3}X_{a_4}$ error. Thus, in case  the $(-1,-1,-1)$ outcome is obtained,  one can  correct the two errors by applying $X_{j_3}X_{j_4}$  to the data qubits, be it physically or on a software book-keeping level.

Note that an additional useful feature is the fact that whenever any single or two of the ancilla "check" qubits yields an outcome $M_{a_n}^z=-1$, this also signals that something has gone wrong, and the stabiliser information obtained from measuring the second ancilla qubit $M_2^x$ is better discarded. 

\begin{figure}
\center
\includegraphics[angle=0,width=0.9\columnwidth]{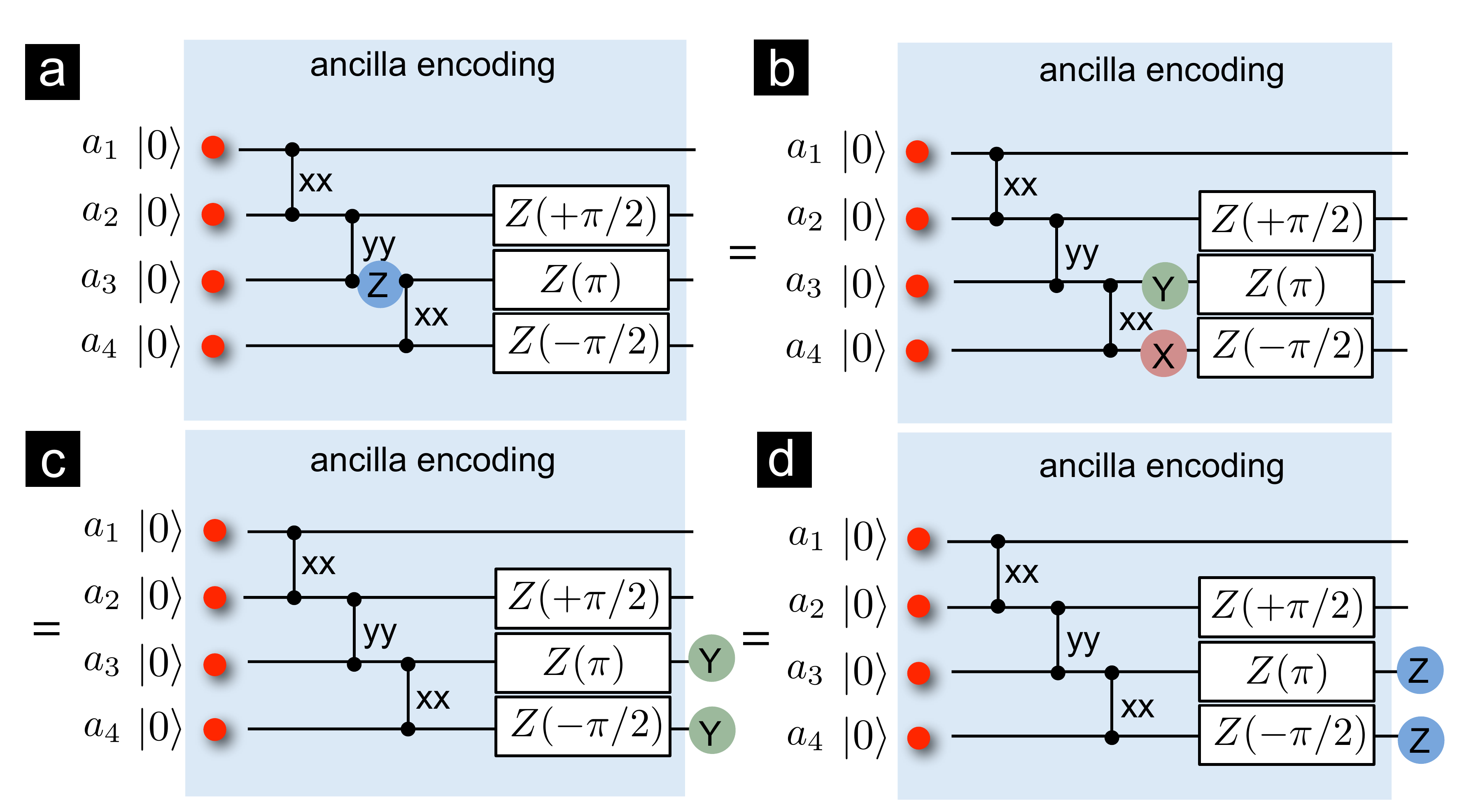} 
\caption{{\bf Dangerous error propagation during the  encoding step of the DiVincenzo-Aliferis-type scheme}: Subfigures (a) to (d) illustrate how a single phase flip error $Z_{a_3}$, occurring after the $YY$ MS gate of the encoding step, results in two phase flip errors $Z_{a_3}Z_{a_4}$ that feed into the data qubits. Note that in the last step (from (c) to (d)) we have made use of the fact that a two-qubit $Y_{a_3}Y_{a_4}$ error  is equivalent to $Z_{a_3}Z_{a_4}$. This can be seen, e.g.,~by recognizing that the $Y_{a_3}Y_{a_4}$ error corresponds to a simultaneous bit- and  phase-flip errors on both qubits, $Y_{a_3}Y_{a_4} \equiv Z_{a_3} Z_{a_4} X_{a_3} X_{a_4}$, and that $X_{a_3}X_{a_4}$ leaves the  ancilla GHZ state $\frac{1}{\sqrt{2}}(\ket{x_+x_+x_+x_+} + \ket{x_-x_-x_-x_-})$ invariant.}
\label{DiVincenzo-Aliferis-error-propagation_1}
\end{figure}

\vspace{1ex}
{\it (ii) DiVincenzo-Aliferis   QEC based on MS gates.--}
Figure~\ref{fig:DiVincenzo-Aliferis-2ion-MS-gates} displays the detailed circuit for a fault-tolerant DiVincenzo-Aliferis-type stabiliser measurement based on 2-ion MS gates. The encoding of the four ancilla qubits, i.e.~the preparation of the four-qubit GHZ state (see Eq.~(\ref{eq:ghz}))  by three 2-ion MS gates followed by three local $Z$-type rotations, occurs in  the same way as for DVS scheme as discussed in Sec.~\ref{sec:Shor-QEC}. Similarly, the coupling to the data qubits takes place transversally by a series of four $XX$ MS gates and four local $X$-rotations on the data qubits. Before the decoding step, the four ancilla qubits are ideally (if no error has occurred) in the state $\frac{1}{\sqrt{2}}(\ket{x_+x_+x_+x_+} \pm \ket{x_-x_-x_-x_-})$, depending on whether the   data qubits are in a $+1$ or $-1$ eigenstate of the measured stabiliser. Under the  two $Z$-rotations of Fig.~\ref{fig:DiVincenzo-Aliferis-2ion-MS-gates},  this  state is transformed into $
\textstyle{Z_{a_2}\big(\frac{\pi}{2}\big) Z_{a_4}\big(\frac{\pi}{2}\big)}  \frac{1}{\sqrt{2}}(\ket{x_+}_{a_1}\!\!\ket{x_+}_{a_2}\!\!\ket{x_+}_{a_3}\!\!\ket{x_+}_{a_4} \pm \ket{x_-}_{a_1}\!\!\ket{x_-}_{a_2}\!\!\ket{x_-}_{a_3}\!\!\ket{x_-}_{a_4})=:\ket{\tilde{\psi}_3}$,  which is a locally equivalent GHZ-type state
\beq
\ket{\tilde{\psi}_3}\!= \!\textstyle{\frac{1}{\sqrt{2}}\!\big(\!\ket{x_+}_{a_1}\!\!\ket{y_+}_{a_2}\!\!\ket{x_+}_{a_3}\!\!\ket{y_+}_{a_4} \pm \ket{x_-}_{a_1}\!\!\ket{y_-}_{a_2}\!\!\ket{x_-}_{a_3}\!\!\ket{y_-}_{a_4}\!\!\big)}.
\eeq
From this state, it can be readily seen that the subsequent $X$-type MS gate on ancillas $a_1$, $a_4$ and the $Y$-type MS gate on ancillas $a_2$, $a_3$ decouple the third and fourth qubit, leaving the first two ancilla qubits in one of two possible Bell-type states
$
\textstyle{Y^2_{a_2, a_3}\big(\frac{\pi}{2}\big) X^2_{a_1, a_4}\big(\frac{\pi}{2}\big) \ket{\tilde{\psi}_3}=\ket{{\rm B}_{\pm}}\otimes \ket{1}_{a_3} \otimes \ket{0}_{a_4}}$, where
\beq
\label{eq:bell}
\ket{{\rm B}_{\pm}}=\frac{1}{\sqrt{2}}\left(\ket{x_+}_{a_1}\ket{y_+}_{a_2} \mp \ket{x_-}_{a_1}\ket{y_-}_{a_2}\right).
\eeq

\begin{figure}
\center
\includegraphics[angle=0,width=0.9\columnwidth]{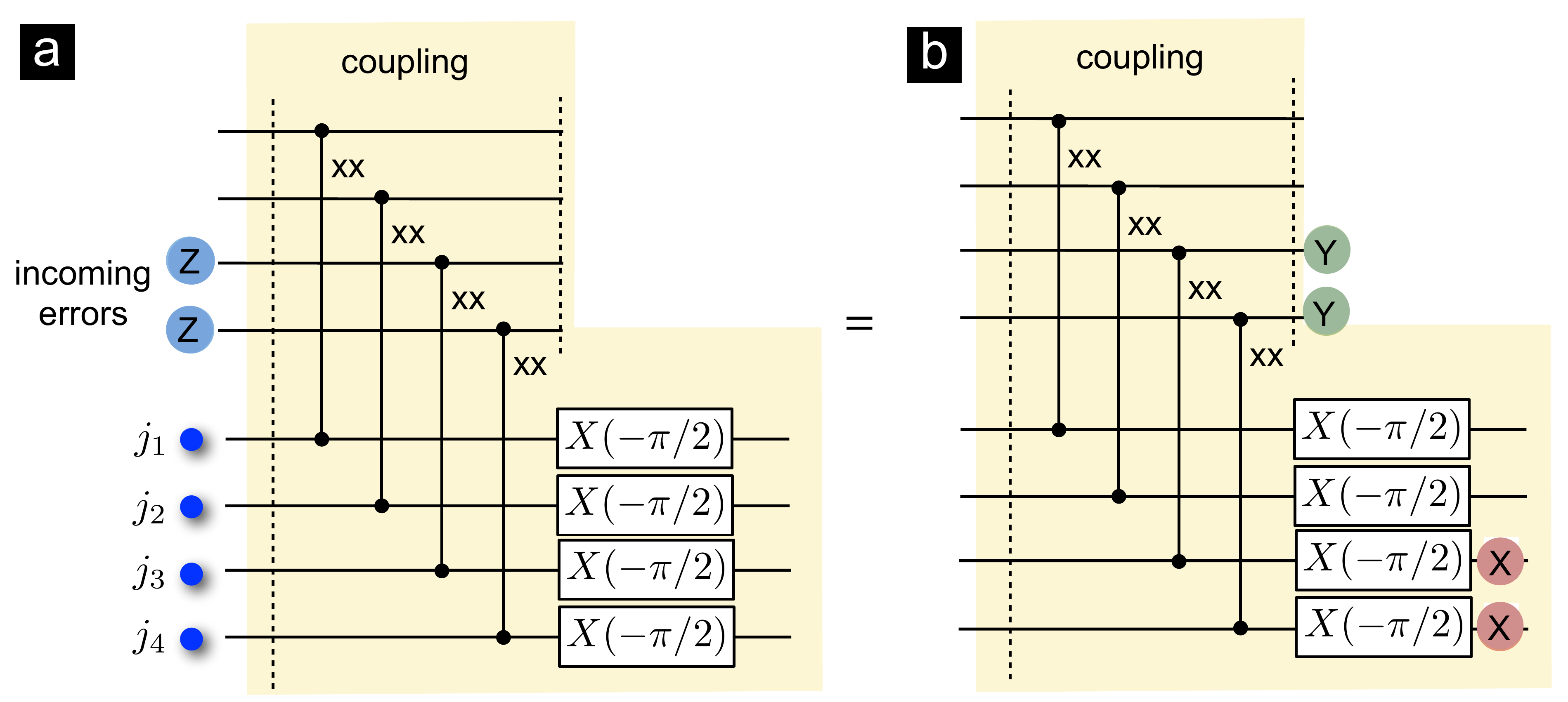} 
\caption{{\bf Dangerous error propagation during the coupling step of the DiVincenzo-Aliferis-type scheme:} The circuits show how an incoming $Z_{a_3}Z_{a_4}$ error, resulting e.g.~from a single phase-flip error in the encoding step in Fig.~\ref{DiVincenzo-Aliferis-error-propagation_1}, is converted into a pair of $Y$-type errors on the ancilla qubits, and furthermore propagate into the data qubit layer, where they result in two bit flip $X_{j_3}X_{j_4}$ errors.}
\label{DiVincenzo-Aliferis-error-propagation_2}
\end{figure}

\begin{figure}
	\center
	\includegraphics[angle=0,width=1\columnwidth]{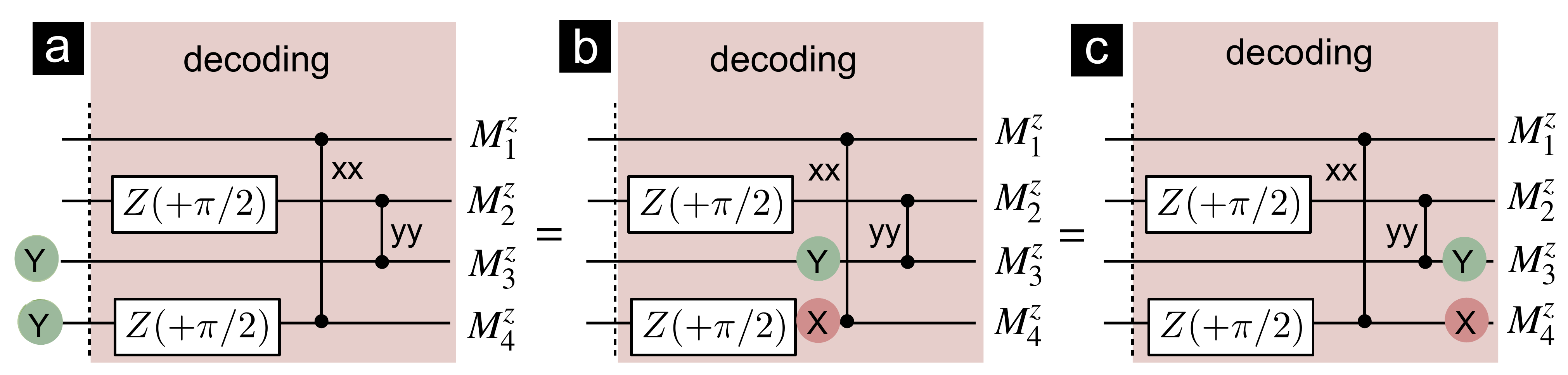} 
	\caption{{\bf Detection of the dangerous two-qubit errors in the DiVincenzo-Aliferis-type scheme}: The circuits show how an incoming $Y_{a_3}Y_{a_4}$ error, resulting from the transversal-coupling step in Fig.~\ref{DiVincenzo-Aliferis-error-propagation_2}), results in bit flips ($Y_{a_3}$ and $X_{a_4}$) on the third and fourth ancilla qubit. Only in this case, the third and fourth qubit end up in $\ket{0}_{a_3}$ and $\ket{1}_{a_4}$, respectively, which results in a $M_3^z=+1, M_4^z=-1$ measurement outcome.}
	\label{DiVincenzo-Aliferis-error-propagation_3}
\end{figure}

Let us now discuss how to access the stabiliser eigenvalue information, simultaneously tracking possible dangerous errors. The first two ancilla qubits being in the Bell-type state~\eqref{eq:bell} are  measured,  revealing the  $\pm1$ stabiliser eigenvalue  by the odd/even parity of the outcome $(M^z_1,M^z_2)$. The third and fourth ancilla qubits act as checks, and are expected to yield $M_3^z = -1$ and $M_4^z = +1$  in the ideal case without any errors. Similar to the case of the DVS scheme discussed above, it can be shown that all dangerous two-qubit errors in  the circuit of Fig.~\ref{fig:DiVincenzo-Aliferis-2ion-MS-gates} are equivalent to a  $Z_{a_3}Z_{a_4}$ error at the end of the encoding circuit. One example of such dangerous error histories is shown in Figs.~\ref{DiVincenzo-Aliferis-error-propagation_1} to \ref{DiVincenzo-Aliferis-error-propagation_3} for the encoding, transversal coupling, and decoding steps of the DVA-type scheme, respectively. As shown in these figures, this type of $Z_{a_3}Z_{a_4}$ error (see Fig.~\ref{DiVincenzo-Aliferis-error-propagation_1}), which has propagated to two $X_{j_3}X_{j_4}$ errors in the code layer (see Fig.~\ref{DiVincenzo-Aliferis-error-propagation_2}), is unambiguously signaled by the $M_3^z=+1, M_4^z=-1$  outcome of the third and fourth ancilla qubit (see Fig.~\ref{DiVincenzo-Aliferis-error-propagation_3}).

\section{\bf Trapped-ion  protocols to assess the break-even point for beneficial QEC}
\label{sec_qec_protocols}

Based on the  description of the experimental capabilities in Sec.~\ref{sec:expt_system},  the development of an effective error model composed of different quantum channels in Sec.~\ref{sec:error_models}, and the trapped-ion toolbox for QEC in Sec.~\ref{sec:cnot_alternative}, we can now present the different trapped-ion  protocols to asses the break-even point for beneficial QEC~\eqref{eq:be_pont}-\eqref{eq:be_pont_bare}  in the 7-qubit color code.  This code, being equivalent to the 7-qubit Steane code~\cite{steane-prl-77-793}, has been subject to a series of studies assessing its QEC performance and error thresholds (we refer the reader to Ref.~\cite{Terhal2009} for a comparative study of this code and other small-scale QEC codes). Depending on the modelling of the noise, typical error thresholds lie around the $10^{-4}$ level~\cite{Terhal2009,Aho2006}.  Additionally, we would like to remark that previous work has also explored the performance of this code in a trapped-ion architecture~\cite{Tomita-Gutierrez-PRA-2013}, considering CNOT-gate and shuttling-based protocols in a two-dimensional array of coupled traps, paying particular attention to the influence of available ancilla resources.

As advanced in the introduction, the QEC  protocols that we explore in this work exploit 
 optimised MS-gate-based circuitry, and are embedded in a single linear segmented trap. The various protocols we investigate are based on either on {\it (i) spectroscopic decoupling}, i.e. storing
idling qubits temporarily in Zeeman sub-levels that do not belong to
the computational subspace,  or on {\it (ii) ion reconfiguration},
i.e. combination of shuttling, splitting, merging, and rotations of
the ion crystal to physically move idle qubits to the storage
region. For all the different protocols within these two types, we have developed three conceptual layers: 
\begin{enumerate}
\item
\textit{Real-space representation:} A sketch of the sequence of operations in real space and time, which is particularly useful to visualize the effect of ion crystal reconfigurations in the shuttling-based protocols. 
\item 
\textit{Circuit representation:} Circuit diagrams which show the entirety of elementary operations that should be applied in the real experiment. Besides the gate operations, measurements, etc., these contain a list of the ion crystal reconfiguration building blocks (splitting, shuttling and merging operations).
\item
\textit{Quantum channel representation:} Circuit diagrams  containing information about the sequence of quantum channels, which reflect the effective noise models used to describe imperfect operations in Sec.~\ref{sec:error_models}, as well as the dependences of channel parameters on the experimentally relevant metrics (e.g. gate  times and infidelities, coherence times). This underlies the  numerical simulations of the following section.
\end{enumerate}

In the following, we shall describe in detail these three layers for the first shuttling-based protocol, and restrict to the real-space  representation for the remaining ones. 

\subsection{Non-fault-tolerant trapped-ion  QEC protocols }

In this section, we discuss the protocols for a QEC cycle with an  8- or 9-ion  crystal, where $7$  data qubits are used to implement the aforementioned 7-qubit color code,  and 1 additional ancillary qubit  is exploited for  syndrome extraction by measuring the 6 stabilizers in Eq.~\eqref{eq:stabilizers} (see Fig.~\ref{Fig:7qubitCode}). For one of the protocols, we will require an extra ion to implement laser cooling of the ion crystal.  By using a single ancillary qubit, this QEC protocol cannot be made fault-tolerant (see Sec.~\ref{sec:non-FT-2-ion-scheme}). However, even in this case, it is very instructive to use it to benchmark the experimental progress towards fault-tolerant quantum computation according to the criteria in Eqs.~\eqref{eq:be_pont}-\eqref{eq:be_pont_bare}. As shown in Sec.~\ref{sec:numerical_studies}, one could already prove the non-trivial beneficial character of our  QEC protocol by improving the experimental hardware according to Tables~\ref{tab:summary_gates} and~\ref{tbl:shuttlingops}. Moreover, this is the less-demanding possible scenario for future trapped-ion experiments along the lines of~\cite{nigg-science-345-302}.

We consider both shuttling- and hiding-based approaches, as well as stabiliser mappings based on either  multi-qubit MS gates (cf. Fig.~\ref{Fig:4qubit_stab_readout}) or on a sequence of 2-qubit MS gates (cf. Fig.~\ref{fig:4-1-readout-with-2-ion-MS-gates}). The criteria for beneficial QEC exposed in Eqs.~\eqref{eq:be_pont}-\eqref{eq:be_pont_bare} of Sec.~\ref{sec:QEC_criterion}, can be assessed through the quantum-information protocols detailed in Table~\ref{table:first_criterion}. In order to implement this protocol numerically, we now present the explicit  scheduling of different  shuttling- and hiding-based approaches for a single QEC cycle in a trapped-ion  7-qubit color code.

\subsubsection{ Shuttling-based, single-species, multi-qubit gate protocol}

\begin{figure}
 \begin{centering}
  \includegraphics[width=1\columnwidth]{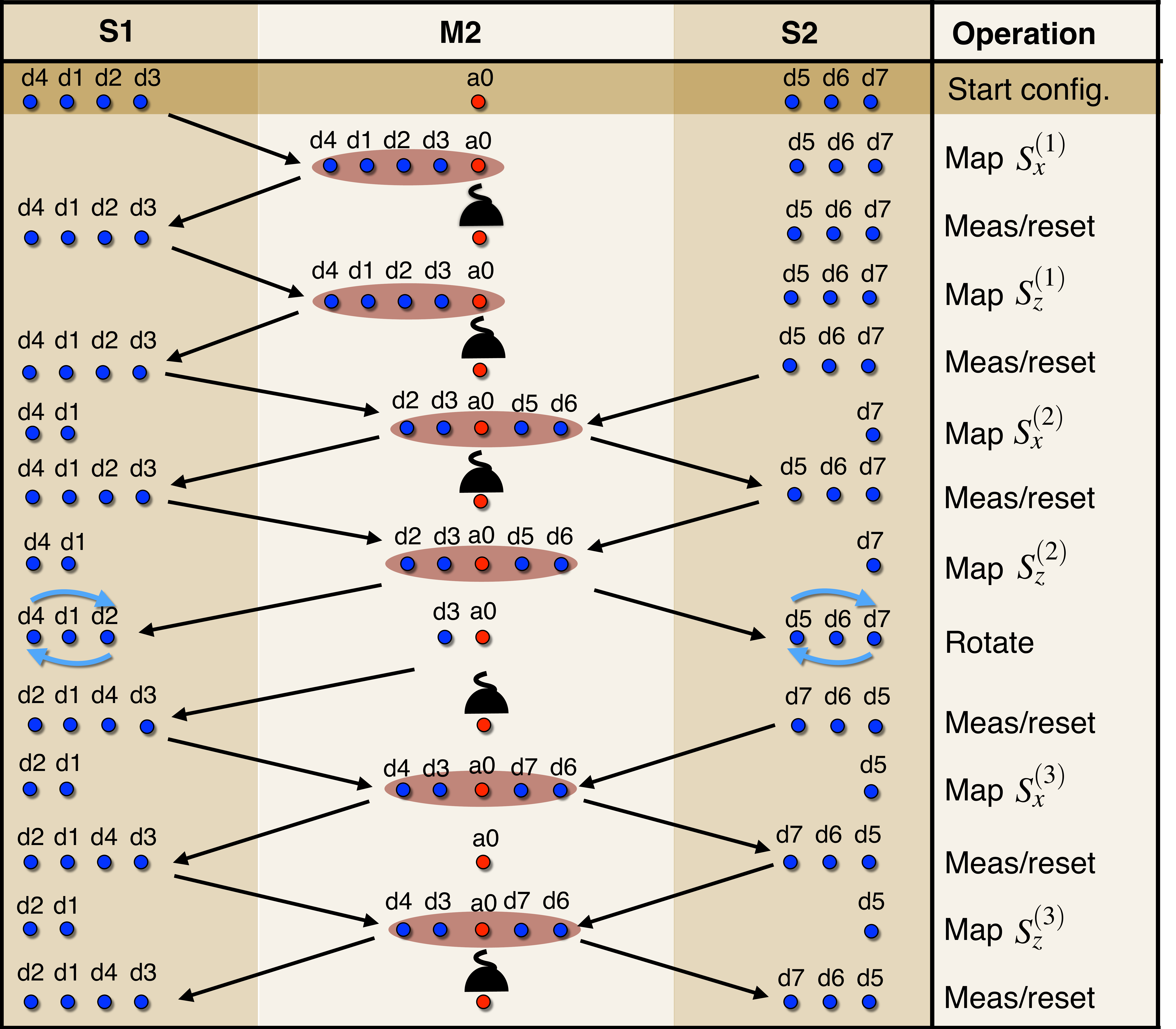}\\
  \caption{\label{Fig:Shuttling_1species_overview} {\bf Real-space representation of  shuttling-based one-species QEC cycle with multi-qubit MS gates:} Data and ancillary ions are represented by blue and red dots, respectively, and are distributed within the storage and manipulation regions of the first row. The black arrows pointing onto the storage zone represent splitting operations followed by a shuttling of a subset of physical ions onto the processing zone, after which they all merge with the static ancillary ion. The black arrows towards the storage zones represent splitting operations, followed by a shuttling of the physical ions back to the storage zone, where they are merged with any physical ions that could already be present there. The blue arrows within the storage zone represent crystal rotations that reorder the physical ions. On the right column, we specify the   times where the stabilizer mappings  $ Map\, S_\alpha^{(m)}$, implemented by multi-qubit entangling gates and represented by red ellipses, and ancillary measurements/reset ($  Meas/reset$), represented by a black detector, are applied.}
\end{centering}
\end{figure}

Let us start by  considering a shuttling-based approach to the  multi-qubit mapping of each of the data-qubit stabilizers   onto the ancillary qubit  (see Fig.~\ref{Fig:4qubit_stab_readout}), both of which belong to the same atomic species. Since all the ions in this protocol are of the same  species, and the lasers responsible for the entangling MS gates~\eqref{eq:MS} act globally on the ion chain, one needs to combine  storage and processing/manipulation zones in the trap, and reconfigure the ion crystal in order to apply sequentially the readout of all six  stabilizers of the code~\eqref{eq:stabilizers}. 

\begin{figure*}
 \begin{centering}
  \includegraphics[width=2\columnwidth]{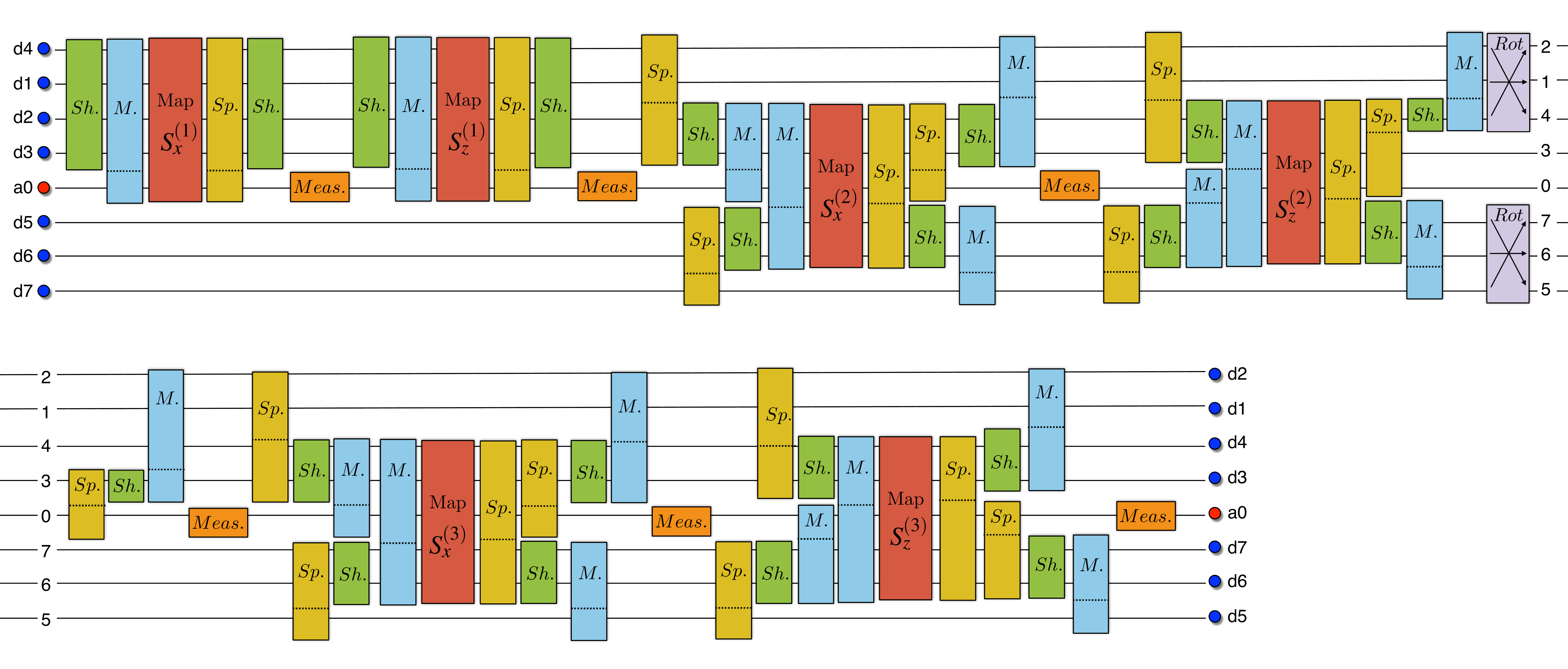}\\
  \caption{\label{Fig:Shuttling_scheme_1species_noRecooling} {\bf Circuit representation of shuttling-based one-species QEC cycle with multi-qubit MS gates:} The data $\{d_1,\cdots, d_7\}$ and ancillary $a_0$ ions are arranged vertically, and a set of boxes represents the elementary operations taking place at a particular time step: {\it Sh.} stands for shuttling of the ions within the green boxes, {\it M.} stands for the merging of the two sets of ions separated by a dotted line within the blue boxes, {\it Sp.} stands for the splitting of the crystal into two sets of ions separated by a dotted line within the yellow boxes,  {\it Meas.} stands for the ancillary ion measurement in the orange boxes, and {\it Rot.} stands for the rotation of the ion crystal within the purple boxes, with arrows representing the corresponding crystal reordering. Note that some crystal reconfiguration operations, such as merging/splitting and shuttling, can be operated simultaneously in different trap zones. Finally, {\it Map $S_{\alpha}^{(m)}$} stands for the mapping of a particular stabilizer $S_{\alpha}^{(m)}$ involving the data  and   ancillary ions within the red boxes. These red boxes contain the sequence of quantum gates in Fig.~\ref{Fig:4qubit_stab_readout}.} 
\end{centering}
\end{figure*}

\begin{figure*}
 \begin{centering}
  \includegraphics[width=2\columnwidth]{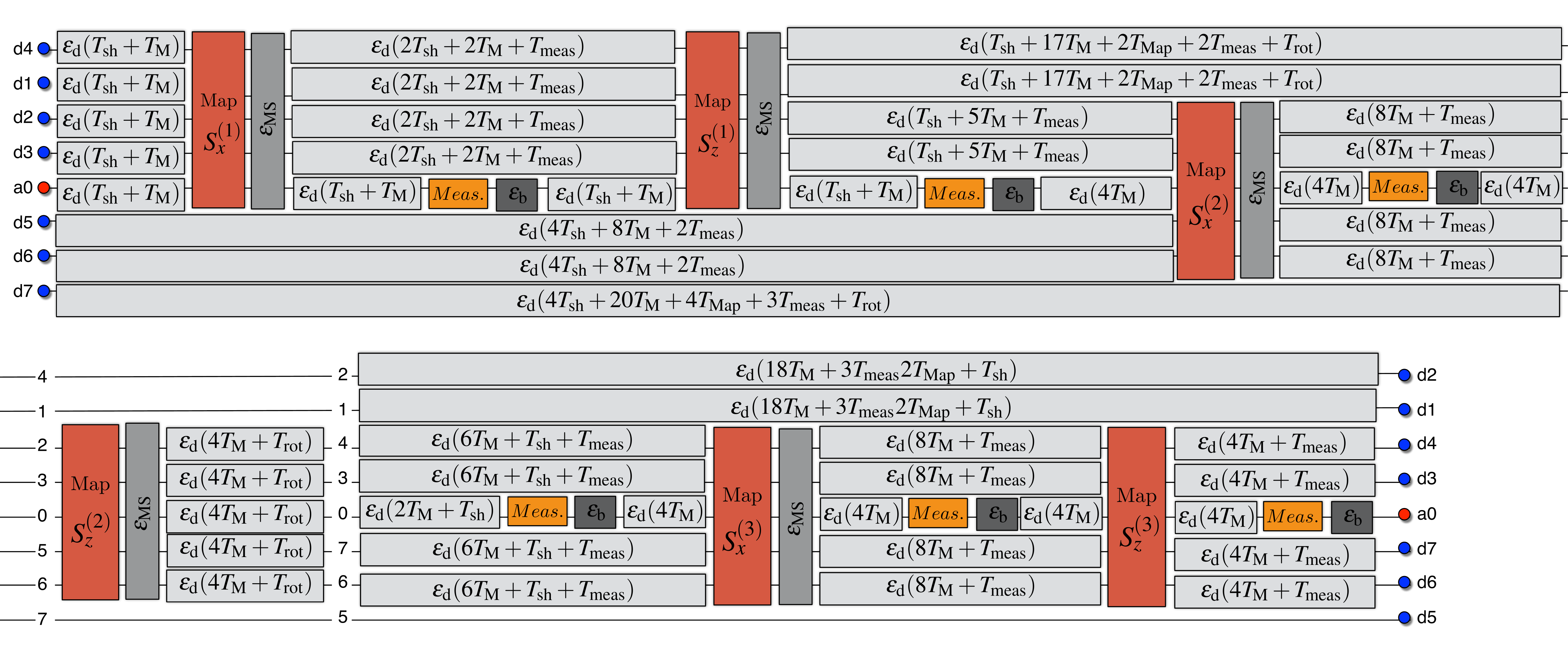}\\
  \caption{\label{Fig:Shuttling_circuit_1species_noRecooling} {\bf Quantum channel representation of shuttling-based one-species QEC cycle with multi-qubit MS gates:} The ion reconfiguration steps of Fig.~\ref{Fig:Shuttling_scheme_1species_noRecooling} lead to an increase of the phonon populations, and a dephasing  of the idle qubits during the time required for these reconfigurations to take place. Additionally, idle ions also dephase during the time lapse of the stabilizer mapping. These time intervals are $T_{\rm sh}$ for shuttling, $T_{\rm M}$ for merging/splitting, $T_{\rm Map}$ for stabilizer mapping, $T_{\rm meas}$ for measuring,  and $T_{\rm rot}$ for rotations. The dephasing channel applied during a certain period is  depicted by  light grey boxes with the channel  $\epsilon_{\rm d}(n_{sh}T_{sh}+n_{M}T_{M}+n_{Map}T_{Map}+n_{meas}T_{meas}+n_{rot}T_{rot})$  acting on a particular qubit, where $n_o$ is the number of operations of the type $o$ that occur within that period.  The actual channel corresponds to Eq.~(\ref{eq:dephaisng_channel}) with a probability $p_{\rm d}=\sum_on_oT_o/2T_2$, with $T_2$ being the coherence time of the qubits. The ancilla readout and reset $Meas$ is modeled by a bit-flip channel~(\ref{eq:bit_flip_channel}) acting on the ancillary qubit $\epsilon_{\rm b}$ with a probability given by the sum of the measurement and reset errors  $p_{b}=\epsilon_{meas}+\epsilon_{res}$, as reported in Table~\ref{tab:summary_gates}, and represented by dark grey boxes after each measurement. Finally,  $Map$  is modeled by an ideal stabilizer mapping acting on the particular set of qubits, followed by a depolarizing channel $\epsilon_{\rm MS}$ (i.e. grey box) given by one of the three channels of Eqs.~(\ref{depolarising_channel_iid})-(\ref{depolarising_channel_any}) with an error probability $p_{\rm d}=\epsilon_{\rm m}+\epsilon_{\rm d}+\epsilon_{\rm I}$ that depends on the current phonon number through  Eq.(\ref{eq:thermal_error}), and the gate time via Eqs.~(\ref{eq:dephasing_error})-(\ref{eq:intensity_error}). }
\end{centering}
\end{figure*}

In Fig.~\ref{Fig:Shuttling_1species_overview},   we depict the real-space representation of this protocol which utilizes a single arm of the segmented ion trap of Fig.~\ref{Fig:ExpHoa}. In this case, we will only make use of  two storage zones ${ S}_{1}$ and ${ S}_{2}$, surrounding a central manipulation region $M_2$, within one arm of the trap. Initially,  Alice encodes an arbitrary state $\ket{\psi}=\alpha\vert{0}\rangle+\beta\vert{1}\rangle$ in the 7 data qubits, and   hands them to Bob, who can ask his assistant Igor  to perform an imperfect round of QEC before he decides if the original state was either $\ket{\psi}$ or $\ket{\psi_\bot}=\beta^*\vert{0}\rangle-\alpha^*\vert{1}\rangle$. In this protocol, Igor has an additional ancillary ion of the same species, which can be used for the readout of the stabilizer information. He distributes the   7 data qubits  within the two storage zones forming two separate ion crystals, and locates the ancillary ion in the manipulation zone (see Fig.~\ref{Fig:Shuttling_1species_overview}).
The steps of such a QEC cycle are described by different operations in the figure  (time evolution occurs downwards), in which these crystals are split, such that the data qubits of  a given stabilizer can be shuttled to the manipulation region and merged with the ancillary ion. At this stage, Igor  applies the stabilizer mapping  in Fig.~\ref{Fig:4qubit_stab_readout} by shining the corresponding lasers onto the ions of the manipulation zone. After splitting the chain, and shuttling the data qubits back  to the storage zone, the ancillary ion can be measured by state-dependent fluorescence, such that Igor can collect the syndrome information. Note that  the scattered photons do not affect the information stored in the physical qubits, as these  have been shuttled to the distant storage regions.

These steps must be repeated for each of the stabilizers~\eqref{eq:stabilizers}  of the 7-qubit color code. As depicted in Fig.~\ref{Fig:Shuttling_1species_overview}, the last two stabilizers mappings  require a reordering of the ion crystals in the storage zones, which can be accomplished by rotating the  crystal, which effectively implements a mirror image about the symmetry axis of the code, such that the roles of plaquettes 2 and 3 are interchanged (See Fig.~\ref{Fig:7qubitCode}). In this way, one can repeat the same crystal-reconfiguration operations of the second set of stabilizers, and finalize  the syndrome extraction.  We note that similar combinations of rotations and split/shuttle/merge operations will be a crucial building block of all shuttling-based QEC protocols in the following sections.

In Fig.~\ref{Fig:Shuttling_scheme_1species_noRecooling}, we describe the circuit representation of this protocol, where one can keep track of the sequential order of the different elementary operations (time evolving from left to right in this figure). At each time-step, a given box describes which operation takes place on which set of ions (see the caption of Fig.~\ref{Fig:Shuttling_scheme_1species_noRecooling} for a detailed account).

 In order to explore the performance of Igor's QEC cycle under realistic experimental errors and sources of noise, we need to {\it (i)} consider the time intervals during which these operations take place, as idle qubits will be subjected to a dephasing noise~\eqref{eq:dephaisng_channel} that degrades the information stored in the code. Moreover, we also need  to {\it (ii)}  consider that the ion reconfigurations excite the motional modes of the ion crystal, affecting the fidelity of  the MS gates~\eqref{eq:thermal_error}   in subsequent stabilizer mappings, which enter in the depolarizing channels of Eqs.~(\ref{depolarising_channel_iid})-(\ref{depolarising_channel_any}). This goes beyond standard noise models in the literature, all of which assume that the gate errors are non-increasing with the depth of the circuit. Finally, we need to {\it (iii)} take into account  that the measurement and reset of the ancillary ion are also faulty, which is accounted by the bit-flip channel~\eqref{eq:bit_flip_channel}.  Accordingly, the circuit representation is translated onto the  quantum channel description of Fig.~\ref{Fig:Shuttling_circuit_1species_noRecooling}, where the above errors are represented as particular Markovian error channels (see Sec.~\ref{sec:error_models}) with error probabilities that depend on the history of previous operations, and on the total time required for each of the steps (see the caption of Fig.~\ref{Fig:Shuttling_circuit_1species_noRecooling} for the details). This circuit of ideal gates interspersed with dephasing, depolarizing, and bit-flip channels, is the one that is numerically simulated in Sec.~\ref{sec:numerical_studies} to estimate  the break-even point that determines when the  QEC cycle becomes useful.

\begin{table*}[htbp] 
	\begin{tabular}{|l||c|c|c|c|c|c|c||c|c|}  \hline 
		\hspace{2ex} \hspace{2ex} &  2-ion MS  &   5-ion MS  &  Single-qubit  &  Meas. &  Re-cooling & Split, shuttle  & Rotation & Total time & Total time  \\
	
	 \hspace{2ex}  &  gate &   gate &   gate &   &   & and merge &  &  (current)  &  (anticipated)  \\
	  \hspace{2ex}  &   &    &    &   &   &  &  &   (ms) &  (ms) \\
		\hline

		\multicolumn{10}{|l|}{\bf Non-fault-tolerant trapped-ion QEC protocols}   \\
						\hline
	Shuttling-based, single-species  &  -  &  12  &   42 & 6  &  - & 20 & 2 &  6.7 & 1.7   \\
		multi-qubit gate (A.1.)  &   &    &    &   &   &  &  &   &   \\
		\hline
			Shuttling-based, two-species  & -  &   12 &   42 &  6 &   6& 6 & 2 & 6.8  & 1.4  \\
		multi-qubit gate (A.2.)  &   &    &    &   &   &  &  &   &   \\
		\hline
			Shuttling-based, two-species  & 24  & -   & 48    & 6  &  24 & 54 & 36 &  23.6  & 7.2   \\
		two-qubit gate (A.3.)  &   &    &    &   &   &  &  &   &   \\
		\hline
			Hiding-based, two-species  &  - & 12   &  150  & 6  & 6  & - & - &  6.3 & 1.1   \\
		multi-qubit gate (A.4.)  &   &    &    &   &   &  &  &   &   \\
		\hline
		\multicolumn{10}{|l|}{\bf Fault-tolerant trapped-ion QEC protocols}   \\
		\hline
			Shuttling-based, two-species  &  54 &  -  &  84  &  24  & 54  & 190  & 150  &  71.2 & 22.4   \\
		DiVincenzo-Shor (B.1.)  &   &    &    &   &   &   &   &    &   \\
		\hline
		Shuttling-based, two-species  &  54 &   - &   78  & 24  & 54  & 190  & 144  & 71.0  &  22.2  \\
		DiVincenzo-Aliferis (B.2.)  &   &    &    &   &   &  &  &   &   \\
		\hline
	 \end{tabular}
	\caption{ {\bf Resource overview for the different QEC protocols:} We describe the number of various operations that conform each QEC cycle (i.e. measurement of all the six stabilisers~\eqref{eq:stabilizers}) within the different trapped-ion protocols introduced in Sec.~\ref{sec_qec_protocols}. The number of basic operations can be obtained from the different schemes presented in the Figs.~\ref{Fig:Shuttling_1species_overview}-\ref{Fig:DVA_real_space_scheme}.  In the two rightmost columns, we present the total  time required for each QEC cycle according to the current and anticipated values given in Tables~\ref{tab:summary_gates} and~\ref{tbl:shuttlingops}. Note, however, that  some of the local single qubit gates require the use of refocusing techniques that combine various single-qubit gates, and that some operations can be done in parallel in different manipulation/processing regions of the ion trap. The latter can lead to a minimization of the overall time of the QEC cycle (see table~\ref{tab:timeCosts} for the optimised QEC cycle times).     }
\end{table*}

\subsubsection{ Shuttling-based,  two-species, multi-qubit gate protocol}

\begin{figure}
 \begin{centering}
  \includegraphics[width=1\columnwidth]{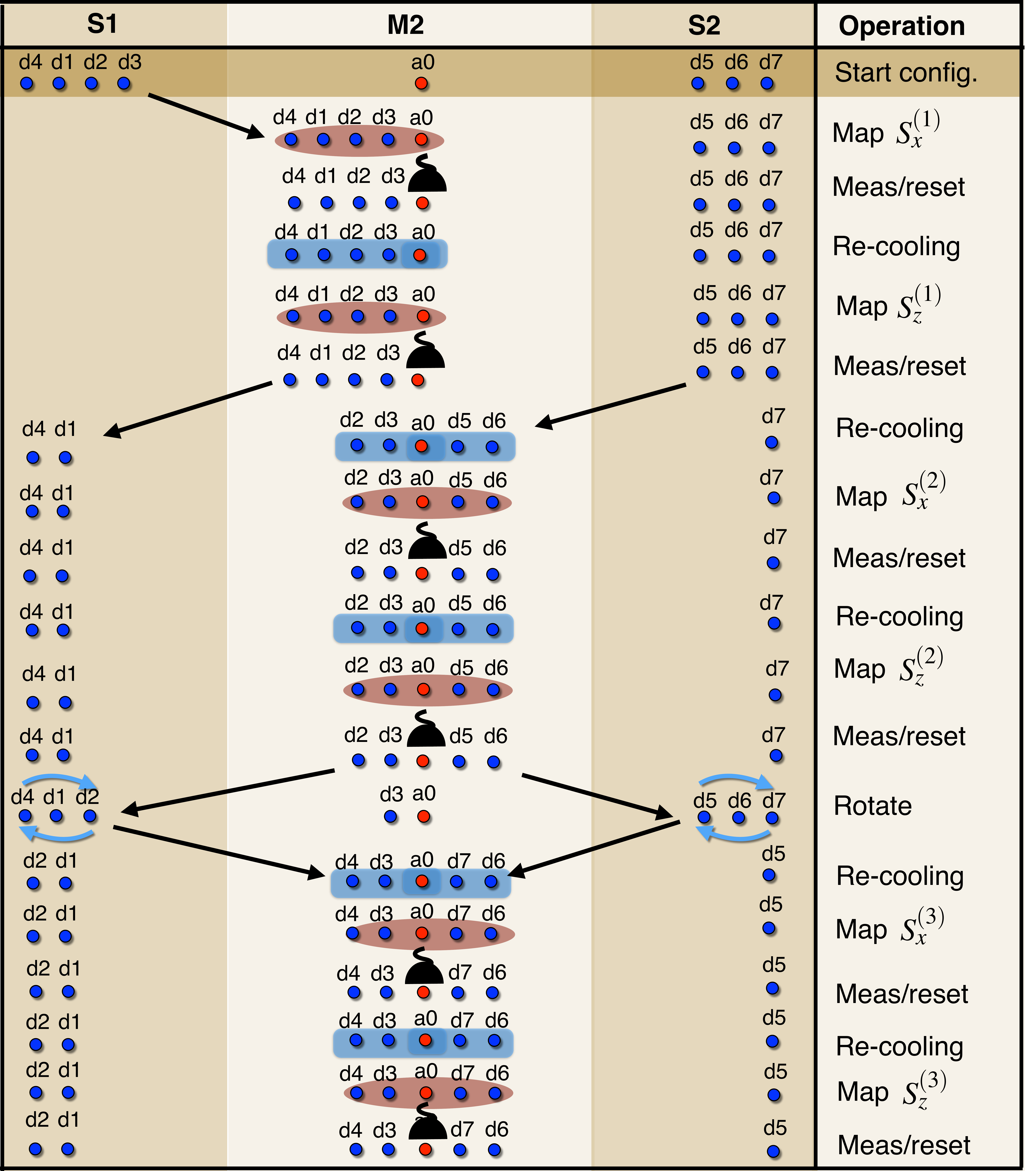}\\
  \caption{\label{Fig:Shuttling_2species_overview}  {\bf Real-space representation of  shuttling-based two-species QEC cycle with multi-qubit MS gates:}  We use the same conventions as in Fig.~\ref{Fig:Shuttling_1species_overview}. As announced in the main text, the number of crystal reconfigurations is reduced with respect to the shuttling-based one-species protocol.  In the rightmost column, we specify the  times when crystal rotations,  stabilizer mappings,  ancillary measurements, and the new sympathetic  ($ Re$-$cooling$)  represented by blue rectangles, are applied.}
\end{centering}
\end{figure}

\begin{figure}
 \begin{centering}
  \includegraphics[width=1.\columnwidth]{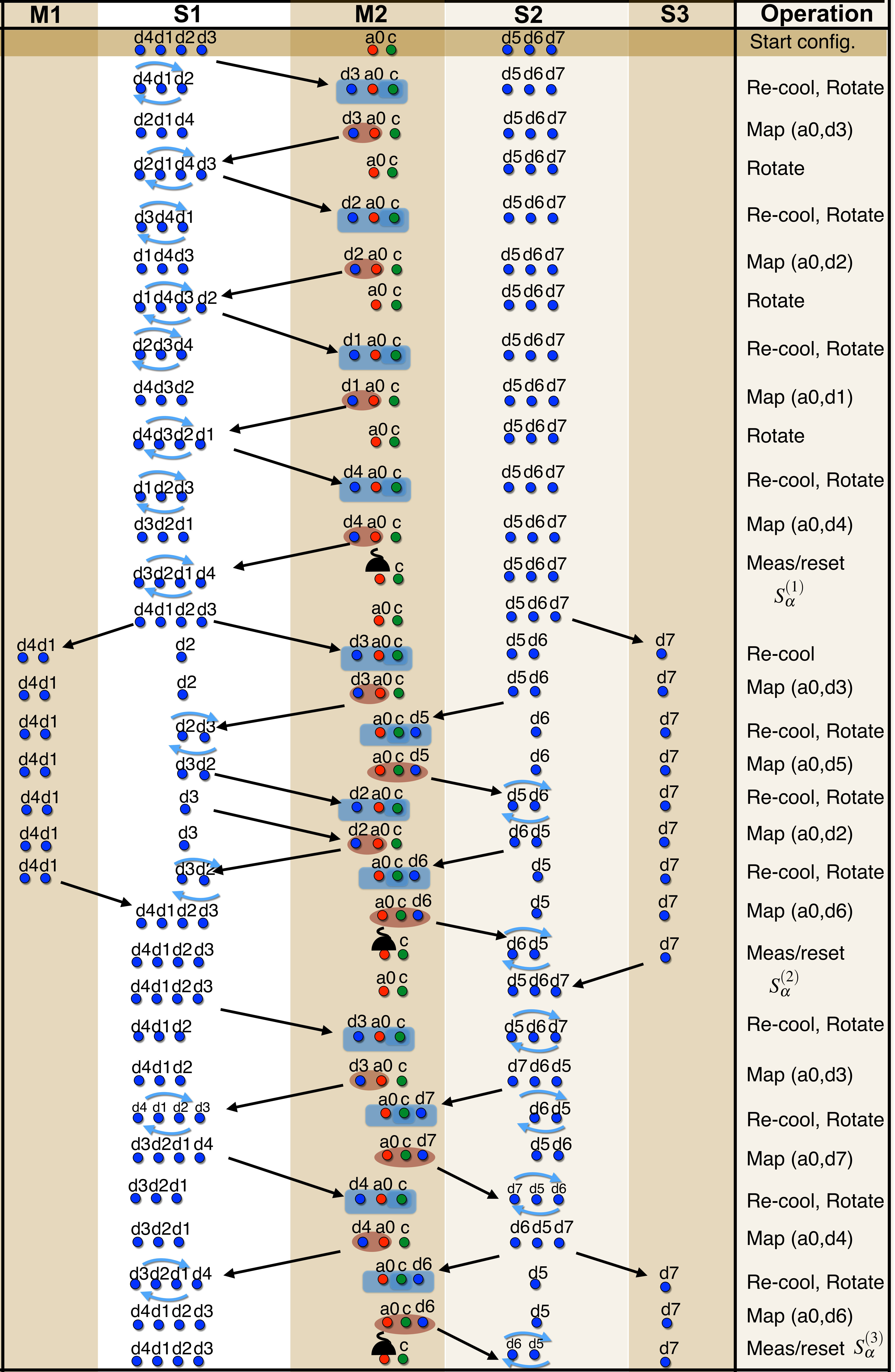}\\
  \caption{\label{Fig:Shuttling_2species_2qb_gates_overview} {\bf Real-space representation of  shuttling-based two-species QEC cycle with 2-qubit MS gates:}  We represent half of a cycle  of QEC for the sequential measurement of $\left\{S_\alpha^{(1)},S_\alpha^{(2)},S_\alpha^{(3)}\right\}$, either for $\alpha=x$ or for $z$, and use the same conventions as in Figs.~\ref{Fig:Shuttling_1species_overview} and \ref{Fig:Shuttling_2species_overview} placing the extra cooling ion (green circle) in the manipulation zone $M_2$. In the rightmost column, we specify the time steps when crystal reconfiguration operations, sympathetic re-cooling, and ancilla measurement/reset, take place. The new mapping functions represented by  red ellipses, consist of the data-ancilla conditional gate $  Map(a0,dj)=U_x^{(a_0,d_j)}$ for a $X$-type stabiliser, or its combination  with local rotations $  Map(a0,dj)=Y_{d_j}(+ \pi/2)U_x^{(a_0,d_j)} Y_{d_j}(- \pi/2)$ for a $Z$-type stabiliser. We recall that these local-rotations can be obtained from the available set of gates in Eqs.~\eqref{eq:rot_xy}-\eqref{eq:rot_z}  by simple refocusing sequences $Y_{d_j}(\theta)=U_{\rm R,\pi/2}(\theta/2)U_{{\rm R}_{d_j},z}(\pi)U_{\rm R,\pi/2}(-\theta/2)U_{{\rm R}_{d_j},z}(-\pi)$.  }
\end{centering}
\end{figure}

Let us now reconsider the same  QEC cycle  with a  two-species ion crystal, where the $7$ data qubits and the additional ancillary qubit are of a different atomic species. This has two important implications: {\it (i)} it reduces the number of crystal reconfigurations that must be implemented by Igor, since the measurement of the ancillary qubit does not need to be performed on an isolated qubit. Even if the physical ions are in the same processing region as the ancilla ion,  the  photons scattered while the ancilla ion is being measured will not be absorbed by them, and thus the encoded state shall not be affected. On the other hand, the photon recoil onto the ion crystal can induce motional excitations that would compromise the fidelities of subsequent MS entangling gates. However,  {\it (ii)} the use of two species allows Igor to apply sympathetic re-cooling of the ion crystal, which minimizes  the number of motional excitations due to recoil or  crystal reconfigurations prior to any MS gate in the stabilizer mappings, and thus reduces the motional error~\eqref{eq:thermal_error} during the entangling gates.

The real-space representation of this 2-species protocol is depicted in Fig.~\ref{Fig:Shuttling_2species_overview}. As advanced previously, the number of crystal reconfiguration operations is highly reduced with respect to the one-species protocol in Fig.~\ref{Fig:Shuttling_1species_overview}. Additionally, intermediate re-cooling steps can also be introduced previous to any stabiliser mapping, such that the fidelity of the corresponding MS entangling gates is not compromised as the QEC cycle proceeds. The circuit and quantum channels representation are somewhat similar to those of the one-species protocol  in Figs.~\ref{Fig:Shuttling_scheme_1species_noRecooling} and~\ref{Fig:Shuttling_circuit_1species_noRecooling}, and will not be presented in detail. However, we remind that they are also important, as they provide a scheduling of the physical operations that would have to be applied in an experiment, and also contain the relevant information for the numerical modeling of Sec.~\ref{sec:numerical_studies}.

\subsubsection{ Shuttling-based, two-species, two-qubit gate protocol}

We now consider a similar shuttling-based  two-species  QEC protocol,  but relying on sequential 2-qubit MS gates for the stabilizer mapping (see Sec.~\ref{sec:non-FT-2-ion-scheme}). We recall that using the combination of a MS gate between a pair of ancilla $a_0$ and data $d_j$ qubits,  followed by a   rotation on the data qubit, one can implement the conditional gates $U_x^{(a_0,d_j)} $ in Eq.~\eqref{eq:cnot_like}. The sequential combination of these gates, together with possible rotations on data qubits $Y_{d_j}(\pm \pi/2)$, leads to a mapping of the stabilizer information into the ancilla qubit (see Fig.~\ref{fig:4-1-readout-with-2-ion-MS-gates}), which can be obtained from its fluorescence $M_0^z$.

\begin{figure*}
 \begin{centering}
  \includegraphics[width=1.95\columnwidth]{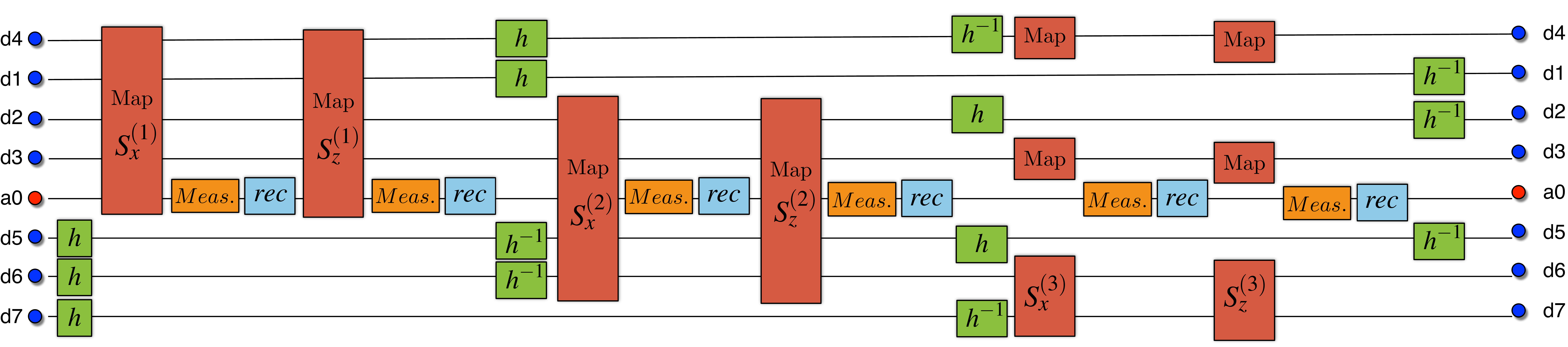}\\
  \caption{\label{Fig:Hiding_scheme} {\bf Circuit representation of  the hiding-based two-species  QEC cycle with multi-qubit MS gates:}   We use the same conventions as in Fig.~\ref{Fig:Shuttling_scheme_1species_noRecooling}. The additional hiding  $h$ and un-hiding operations $h^{-1}$, represented by green boxes, are realized by composite pulse sequences  to coherently map the state of physical qubits from  $4S_{1/2}(m_f=-1/2)$ and  $3D_{5/2}(m_f=-1/2)$ to a set of storage $D$ states (see subsec.~\ref{Sec:DES}).}
\end{centering}
\end{figure*}
 In order to reduce   the  complexity of all the  crystal reconfigurations required to perform the sequential gates of the QEC cycle, Igor shall make use of all of the five regions of the segmented trap in  Fig.~\ref{Fig:ExpHoa} (i.e. two manipulations zones $M_1$ and $M_2$ interspersed between three storage regions $S_1,S_2,S_3$). Moreover, since the ancilla qubit is not measured after each of the conditional  gates $U_x^{(a_0,d_j)} $, Igor needs to re-cool the crystal several times during each stabiliser readout without affecting the ancilla and data qubits. Therefore, we consider that the ancilla and data ions  are of the same species, but equip Igor with an extra ion of a different species $c$ for sympathetic re-cooling of the crystal. Hence, Igor has  $7+1+1$ ions    for his imperfect round of QEC, which he distributes within the segmented trap as  depicted in the starting configuration of Fig.~\ref{Fig:Shuttling_2species_2qb_gates_overview}. The data qubits of a given stabilizer are shuttled one-by-one  from the storage regions onto the manipulation zone $M_2$, where sympathetic re-cooling represented by blue rectangles is applied prior to the data-ancilla mapping $U_x^{(a_0,d_j)} $ or $Y_{d_j}(+ \pi/2)U_x^{(a_0,d_j)} Y_{d_j}(- \pi/2)$ for an $X$- or $Z$-type stabiliser readout, which are labelled as $Map(a_0,d_j)$ and represented by red ellipses. After all the four data qubits of a particular stabiliser have been coupled to the ancilla ion, Igor can proceed to isolate the ancilla-cooling pair of ions in the manipulation zone, and collect the state-dependent fluorescence of the ancilla ion, inferring in this way   the stabilizer information.

As can be seen in this figure, the complexity of the crystal-reconfiguration operations increases considerably, such that idle qubits will suffer more environmental dephasing during the QEC cycle. Therefore, although the multi-qubit errors~\eqref{depolarising_channel_any} of the collective stabiliser mapping of the previous subsections are avoided in this scheme, and the fidelity for 2-qubit MS gates is higher than that of 5-multi-qubit ones (see Table~\ref{tab:summary_gates}), we do not expect any big improvement of the non-fault-tolerant protocol. A possible advantage can only take place if a fully fault-tolerant scheme is implemented.

\subsubsection{  Hiding-based, two-species, multi-qubit gate protocol}

An alternative approach to the shutting-based protocols presented above is to work with a  static ion crystal, but equip Igor with spectroscopic decoupling capabilities. Accordingly, Igor  can  take out a particular  subset of ions   from a given stabilizer mapping by spectroscopically decoupling them, and coherently shelving the physical qubit's population in electronic states that do not couple to the lasers driving gate operations on the qubit transition (see also Secs.~\ref{sec:gate_operations} and \ref{subsec:hiding_error}).

We consider the scenario of a two-species encoding. This allows one to measure the ancilla qubit after the stabiliser mapping, without affecting the state of non-hidden data qubits. In addition, the ancilla qubit, encoded in an ion species different from the data qubits, enables recooling of the entire ion string after each fluorescence ancilla measurement,  minimizing thus motional excitations due to photon recoil. 

The scheme for a complete QEC cycle involves an overall number of 12 hiding and un-hiding operations, each of  which can be  realized by a composite pulse sequence (9 single-ion pulses detailed in Sec.~\ref{Sec:DES}) that  maps the state of data  qubits to a set of storage $D$ states coherently~\cite{nigg-science-345-302}. Rather than displaying the real-space configuration, it is more instructive in this case to give the circuit representation, since no crystal reconfiguration is used, and all ions reside in the same manipulation zone. In Fig.~\ref{Fig:Hiding_scheme}, green boxes depict the hiding/un-hiding pulses, which determine 
which qubits  are involved in the respective stabilizer represented by red boxes. Idle qubits undergo dephasing with a strength according to the time durations of the respective mapping, measurement, recooling and (un-)hiding operations acting on the other qubit(s).

\subsection{\label{switchToFTand2qubitMS} Fault-tolerant trapped-ion  QEC protocols }

In this section, we discuss the trapped-ion protocols for fault-tolerant QEC  with the 7-qubit color code. As emphasized in previous sections, we need to go beyond a single ancillary qubit  by either focusing on the DVS or DVA schemes described in Secs.~\ref{sec:Shor-QEC} and~\ref{sec:Aliferis-diVincenzo-QEC}, respectively. We will consider only  shuttling-based approaches, since the number of spectroscopic decoupling pulses  for a hiding-based scheme would increase dramatically for these fault-tolerant protocols. Therefore, a fully hiding-based fault-tolerant protocol with a static ion crystal is likely to perform worse than the shuttling-based alternatives.

\subsubsection{ Shuttling-based, two-species, DiVincenzo-Shor scheme }

\begin{figure*}
 \begin{centering}
  \includegraphics[width=2.05\columnwidth]{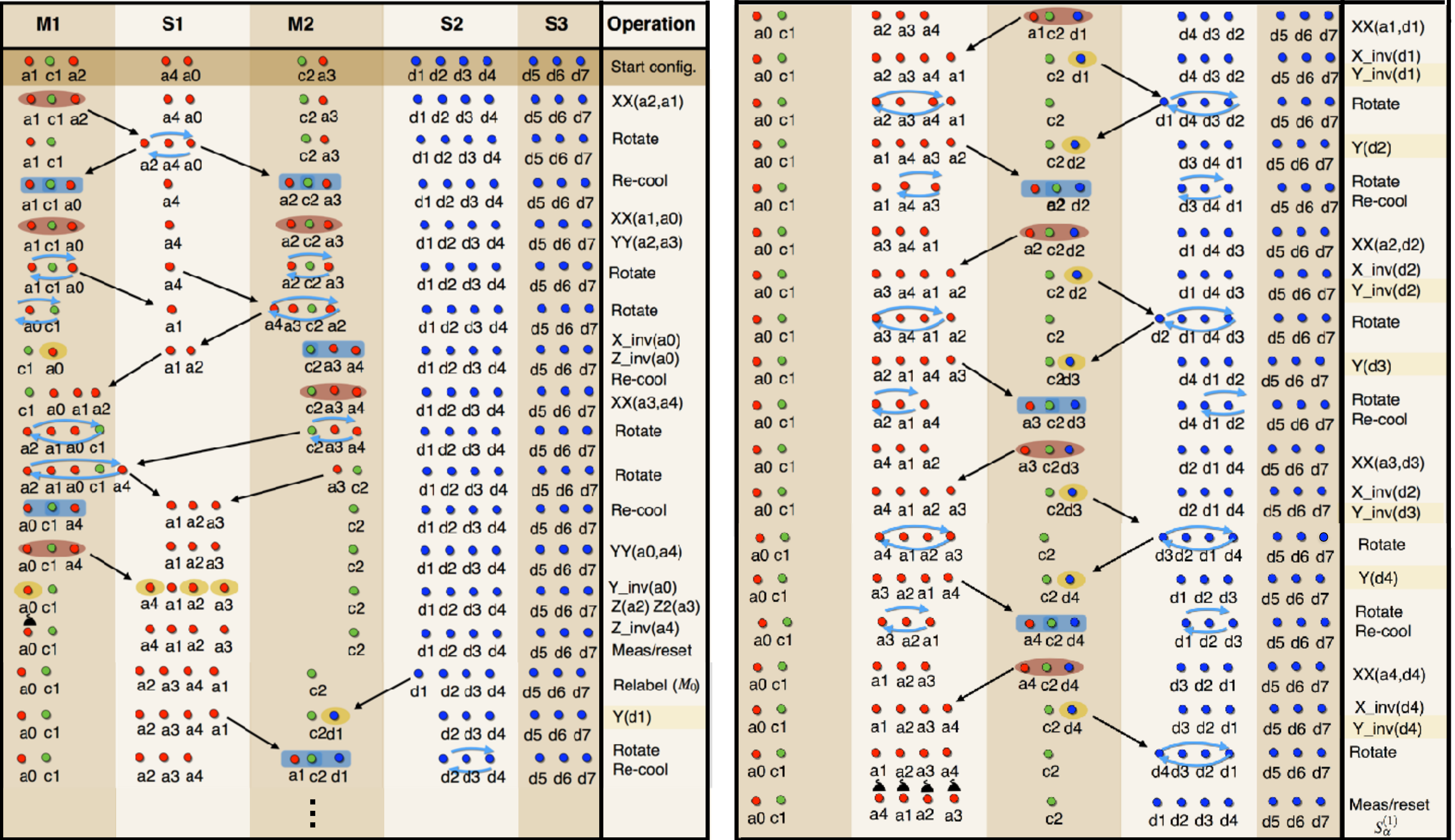}\\
  \caption{\label{Fig:DVS_real_space_scheme}  {\bf Real-space representation of  the shuttling-based two-species QEC cycle with a DiVicenzo-Shor scheme:} We represent the sequence of operations for the fault-tolerant DiVincenzo-Shor readout of a single stabiliser operator  $S_\alpha^{(1)}$ for $\alpha=x$ or $z$ in two panels, and use the same conventions as in Figs.~\ref{Fig:Shuttling_1species_overview}, \ref{Fig:Shuttling_2species_overview}, and~\ref{Fig:Shuttling_2species_2qb_gates_overview}. The extra cooling ions {\fontfamily{phv}\selectfont  c1,c2}  are placed in the manipulation zones $M_1,M_2$, and are again depicted by a green circle. The rightmost columns of both panels contain the sequence of operations that take place. In addition to the ones described in previous figures, we include $X$- and $Y$-type MS gates $XX(i,j)$ and $YY(i,j)$ depicted by red ellipses, as well as single-qubit rotations $ X(j), X\_inv(j)$ corresponding to $X_j(\pm \pi/2)$~\eqref{eq:single_qubit_rot}, and analogously for  $ Y(j), Y\_inv(j)$, and $ Z(j), Z\_inv(j)$, all of which are  depicted by yellow ellipses. We also introduce $Z2(j)$, which corresponds to  $Z_j(\pi)$. Finally, some  $ Y(j), Y\_inv(j)$ rotations on the rightmost columns of each panel are inside a yellow rectangle, which implies that they are only applied for a $Z$-type stabiliser readout ($\alpha=z$). We have also included  a classical relabelling operation, $  Relabel$,  conditional on the measurement outcome $M_0^z$.   }
\end{centering}
\end{figure*}

\begin{figure*}
 \begin{centering}
  \includegraphics[width=2.1\columnwidth]{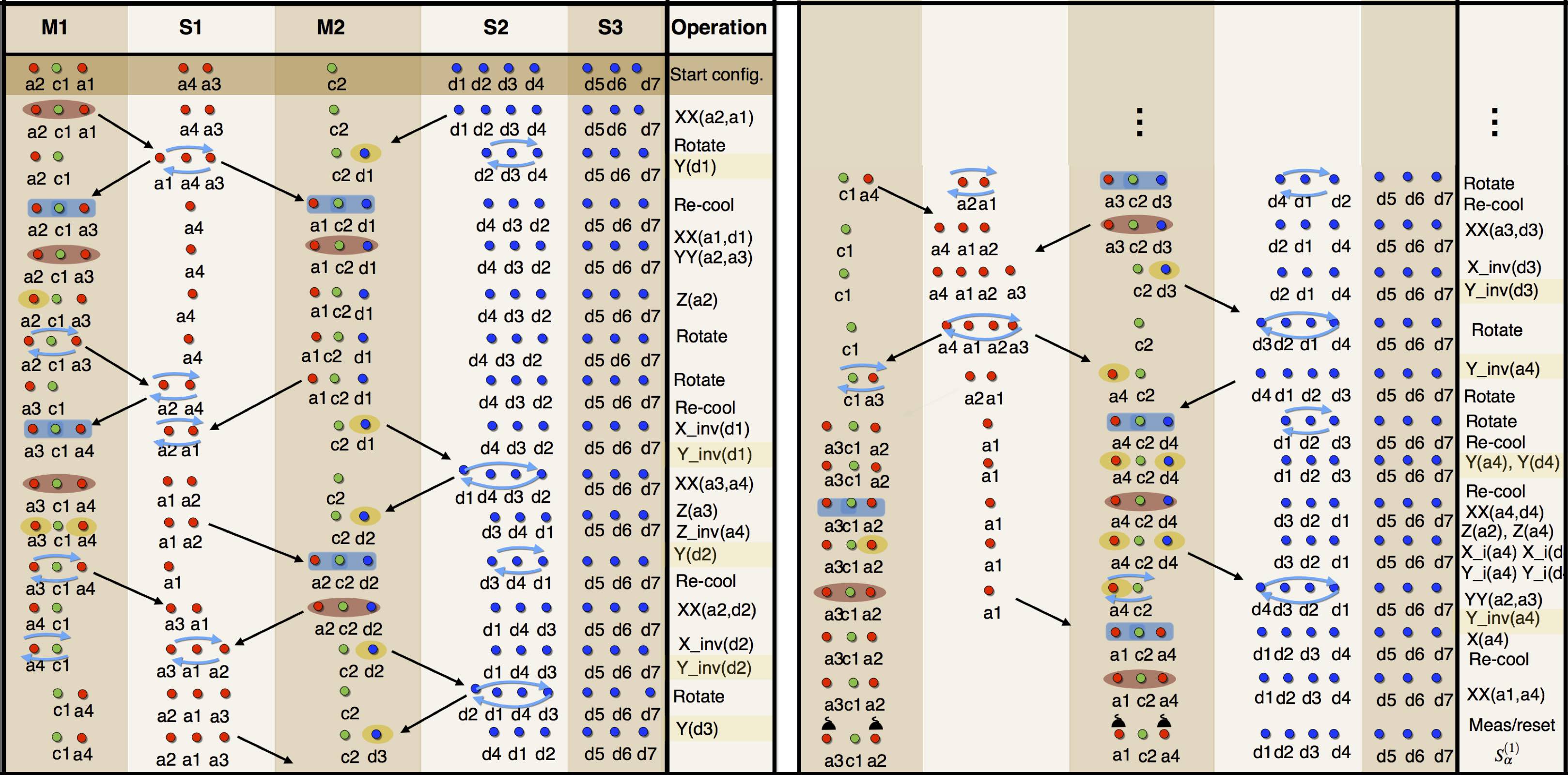}\\
  \caption{\label{Fig:DVA_real_space_scheme} {\bf Real-space representation of  the shuttling-based two-species QEC cycle with a DiVicenzo-Aliferis scheme:}  We represent the   operations for the fault-tolerant DiVincenzo-Aliferis readout of a single  stabiliser   operator $S_\alpha^{(1)}$, either  for $\alpha=x$ or $z$, in two panels, and use the same conventions as in Figs.~\ref{Fig:Shuttling_1species_overview}, \ref{Fig:Shuttling_2species_overview},~\ref{Fig:Shuttling_2species_2qb_gates_overview}, and~\ref{Fig:DVS_real_space_scheme}. }
\end{centering}
\end{figure*}

We first consider the trapped-ion implementation of the DVS scheme, which combines the ancilla encoding and verification of Fig.~\ref{fig:Shor-QEC-state-preparation} with the subsequent transversal coupling to data qubits of Fig.~\ref{fig:Transversal-stab-type-readout} for each of the code stabilizers. To implement a cycle of QEC using these scheme, Igor must be equipped with 5 additional ancillary ions $a_0,\cdots, a_4$. Furthermore, to simplify all the required crystal reconfigurations, we equip Igor with a couple of cooling ions $c_1,c_2$ of a different atomic species such that sympathetic re-cooling can be implemented prior to any entangling MS gate. Igor distributes the 7 data qubits, together with the ancillary and cooling ions, according to the stating configuration of Fig.~\ref{Fig:DVS_real_space_scheme}. In the left panel of this figure, prior to the ancilla  measurement $M_0^z$, we depict the different operations that Igor must apply for the ancilla encoding and verification. If this part of the protocol is successful $M_0^z=+1$, one relabels the ancillary qubits, and proceeds with the rest of the scheme.  The  set of instructions that follow this relabelling correspond to the transversal coupling to the data qubits of Fig.~\ref{fig:Transversal-stab-type-readout}, and ends in the measurement of the remaining ancillas $(M_1^z,M_2^z,M_3^z,M_4^z)$, the parity of which allows Igor to infer the  $S_{\alpha}^{(1)}$ stabiliser information. After re-ordering of the data qubits to move the ions of another stabiliser to the storage region $S_2$, one can repeat the same procedure in Fig.~\ref{fig:Transversal-stab-type-readout} for the next stabilizer, and proceed to complete a full round of QEC. Note that most of these re-ordering operations can be implemented during the measurement period, such that no extra dephasing occurs. To take advantage of the fault-tolerant nature of the scheme, Igor should run two such full rounds of stabiliser readout. If the results coincide, he should then apply a  decoder to determine which error has occurred, and apply a particular  single-qubit $X$- or $Z$-type rotation to the corresponding data  qubit to correct for it. If the measurement results do not match, Igor should apply the full readout scheme once more, and use the decoder on the third set of stabiliser values.

\subsubsection{ Shuttling-based, two-species, DiVincenzo-Aliferis scheme }

Let us now  describe the trapped-ion implementation of the DVA scheme, in which the ancilla encoding, coupling, and verification must occur along the prescription of  Fig.~\ref{fig:DiVincenzo-Aliferis-2ion-MS-gates}. In this case, it suffices to equip Igor with 4 additional ancillary ions $a_1,\cdots, a_4$, and a couple of cooling ions $c_1,c_2$ of a different atomic species. The distribution of these ions within the different zones of the segmented trap is specified in the real-space representation of Fig.~\ref{Fig:DVA_real_space_scheme}, where we also list the operations that Igor must perform for the readout of the $S_{\alpha}^{(1)}$ stabilizer. In contrast to the DVS scheme, the verification step takes place in the final measurement step, and depends on the outcome $(M_3^z,M_4^z)$ of the check ancilla qubits. If Igor obtains $(+1,-1)$, this signals that two errors have propagated into the code space, and Igor must apply $X$- or $Z$-type rotations to the $(j_3,j_4)$ data qubits. Simultaneously, Igor uses the parity of $(M_1^z,M_z^z)$ to infer the eigenvalue information of the stabiliser. Once again, he proceeds with the readout of the remaining stabilizers in a modular fashion, which require an intermediate re-ordering of the data qubits (i.e. bringing the ions belonging to the stabiliser to be measured into the storage zone $S_2$). Once this is achieved, the sequence of operations is again described by the real-space representation of  Fig.~\ref{Fig:DVA_real_space_scheme}.

Once again, to take full advantage of the fault-tolerant nature of the scheme, Igor must perform two or three rounds of stabiliser readout and then apply a minimum-weight decoder to determine which error has occurred, and how to correct it.



\section{\bf Numerical  studies of the performance of trapped-ion QEC protocols}
\label{sec:numerical_studies}

\subsection{Computing resources and numerical  approach}

Having established our criterion for beneficial QEC in Eqs.~\eqref{eq:be_pont} and~\eqref{eq:be_pont_bare}, and described the different trapped-ion protocols together with their quantum channel description in Figs.~\ref{Fig:Shuttling_1species_overview}-\ref{Fig:DVA_real_space_scheme}, let us now describe our numerical approach to assess the performance of these QEC schemes.

Our strategy for numerical analysis is to perform exact simulation of the physical system using pure states in a Monte Carlo method. The results of the simulation are achieved by averaging over the output of at least ten thousands of individual runs; in each run, at each opportunity for an error event the question of whether it occurs (and when relevant, the error's severity) is resolved by drawing a random number. Once sufficiently many results are aggregated, one obtains the same data as would result from a single numerical run using a density matrix. The advantages are twofold: firstly, the memory requirements of the pure state simulation are more modest, allowing for simulation of on the order of 30-40 qubits. A direct simulation of 30 qubits with the density matrix approach would require a matrix of $2^{60}\approx10^{18}$ elements, which is infeasible. This will become crucial when considering fault-tolerant schemes, and also when upgrading these protocols to include larger-distance codes or instances with more than one logical qubit. The second advantage of the Monte Carlo approach is that it is trivial to parallelize, thus one can fully make use of cluster computing resources. 

The hardware used for this work is a cluster of approximately 400 nodes, each of which  is based on a motherboard with two Intel E5-2640v3 CPUs, and between 64 and 256GB of memory. The nodes are connected by Intel TruScale QDR Infiniband, and in principle they can be efficiently used to collectively model a quantum system. However, in practice, the efficient use is to operate the nodes in parallel and independently of one another, aggregating results afterwards according to the Monte Carlo paradigm as outlined above. 

\subsection{Simulation results}

In this subsection, we present the simulation results for all the different QEC protocols discussed above. One type of QEC cycles  is based on the multi-qubit MS gate (see Figs.~\ref{Fig:Shuttling_1species_overview} and~\ref{Fig:Shuttling_2species_overview}), whereas the other is based on the sequential  2-qubit MS gates, including both the non fault-tolerant (see Fig.~\ref{Fig:Shuttling_2species_2qb_gates_overview}) and the fault-tolerant approaches (see Fig.~\ref{Fig:DVS_real_space_scheme} and~\ref{Fig:DVA_real_space_scheme}. Generally, we will use the same Alice-Igor-Bob framework for assessment of the beneficial character of QEC, which has been discussed earlier: Alice prepares a perfect instance of the encoded qubit, and this logical qubit is then subjected to a period of environmental exposure during which Igor may perform one or more cycles of error correction, before Bob assesses the integrity of the qubit by attempting to determine the encoded state (from one of the two choices). For the simulations with multi-qubit MS gates, the initial encoded state was chosen randomly, while for those simulations with 2-qubit MS gates, since they proved to show more encouraging results, each time we let Alice encode only the $|+\rangle$ state, which is the most vulnerable state under dephasing environmental noise. Therefore, Bob's success probability $\mathcal{P}_{\rm B}$ shown in the following two sections has two ranges: the first is from 1 to 0.75, while the second varies from 1 to 0.5, but both cases correspond to the complete decay of the qubit from full coherence to total decoherence. 

For each of these cases, two sets of figures are presented  depending on  the choice of trapped-ion parameters: First, we present a set of figures drawn from the best fidelities that have been reported to date in relevant experiments. An exception is  the single-species shuttling-based scheme and the two-species shuttling-based DVS scheme, where the performance of the MS gates after a few stabilizer mappings is expected to be so bad that one can directly discard the approach (see Tables~\ref{tab:summary_gates}-\ref{table_errors:MS}). 
 Second, we present a set of figures for the future parameters that we anticipate  will be possible to reach in the near future. We refer to these two cases as the ``Current'' and the ``Future'' (or ``anticipated'') performance, respectively.
 
\subsubsection{Shuttling- and hiding-based QEC with multi-qubit MS gates}

\begin{figure}
	\begin{centering}
		\includegraphics[width=0.9\columnwidth]{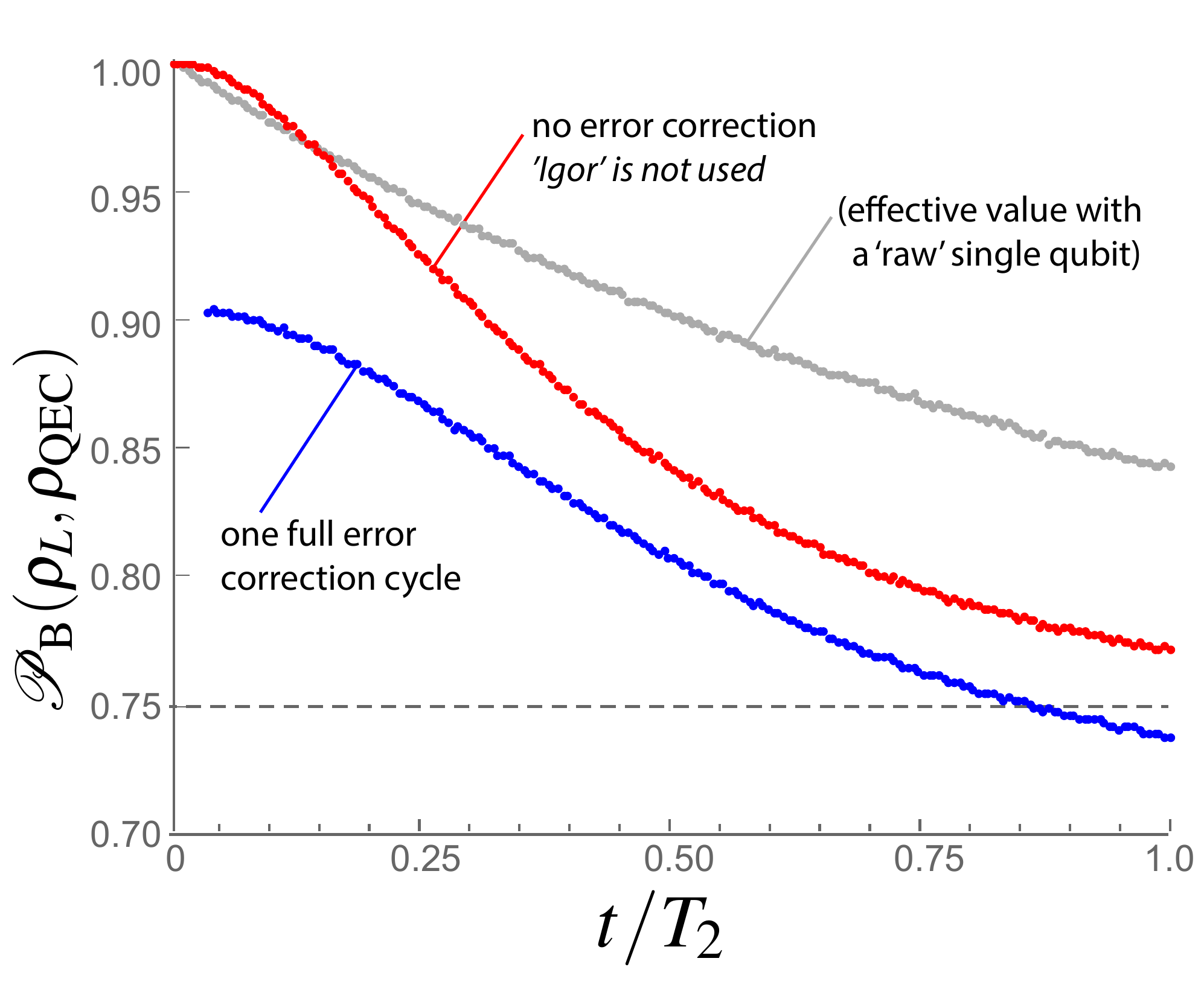}\\
		\caption{\label{Fig:OXF_simulation_hidingNowOPTIMISTIC} {\bf { Success probability $\mathcal{P}_{\rm B}$} under the hiding-based two-species protocol QEC cycle with multi-qubit MS gates} (cf.~Fig.~\ref{Fig:Hiding_scheme}): The parameters underlying the simulation correspond to the \textit{current  values} from Tables~\ref{tab:summary_gates}-\ref{table_errors:MS}. 
			{ Here and in the following figures, the $x$-axis is time in units of the environmental dephasing time $T_2$. }
			The noise model for the imperfect 5-ion MS gate operations corresponds to the quantum channel {\it (i)} of independent depolarizing noise~\eqref{depolarising_channel_iid}. Results show that even for this optimistic noise model,   there is no time window in which the application of an imperfect QEC cycle is beneficial as compared to not applying it (i.e. the regime~\eqref{eq:be_pont} is never attained). The underlying reason is that the current 5-ion MS gate fidelity is insufficient to reach the crossover point. Although not shown in the figure, we note that similar results are found for the Shuttling-based protocol 2, for which reaching the crossover point with current parameters is not possible.  Here and elsewhere,  each random qubit selected for Alice to encode is $U\ket{0}$ where unitary $U$ is formed by selecting three angles $\phi_1$, $\phi_2$ and $\phi_3$ uniformly from $0$ to $2\pi$ and setting $U=\cos\phi_1(\cos\phi_2I+i\sin\phi_2Z)+i\sin\phi_1(\cos\phi_3Y+\sin\phi_3X)$ where $I$ is the identity and $X,Y,Z$ are the Pauli operators.	} 
	\end{centering}
\end{figure}
\begin{figure}
	\begin{centering}
		\includegraphics[width=0.9\columnwidth]{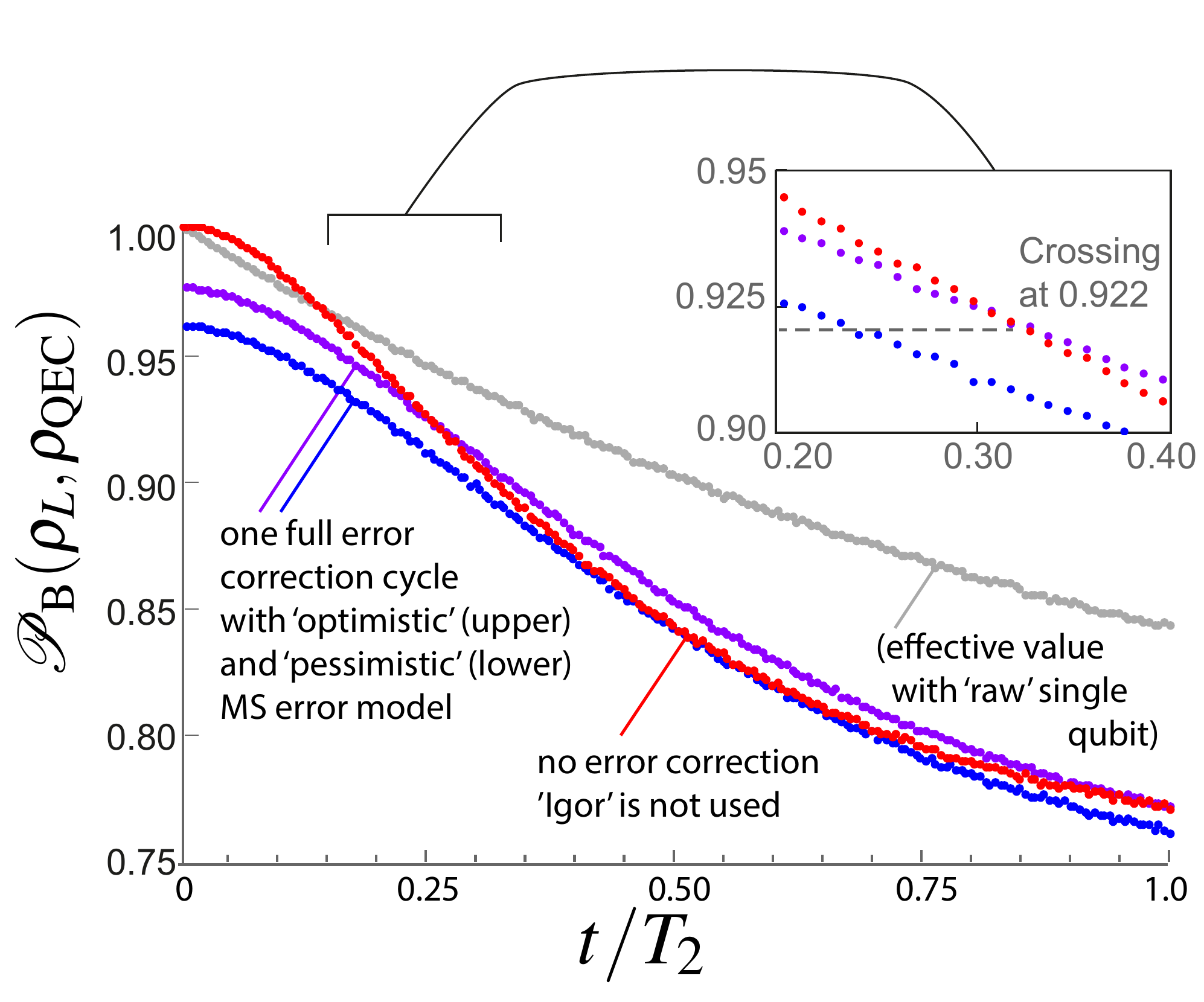}\\
		\caption{\label{Fig:OXF_simulation_1speciesFutureOPTIMISTIC} {\bf { Success probability $\mathcal{P}_{\rm B}$}  under the   single-species shuttling-based QEC cycle with multi-qubit MS gates} (cf.~Fig.~\ref{Fig:Shuttling_circuit_1species_noRecooling}):  The parameters underlying the simulation correspond to the { anticipated improved values} from Tables~\ref{tab:summary_gates}-\ref{table_errors:MS}.  We use two noise models for the imperfect 5-ion MS gate operations involved in the stabilizer mappings, the optimistic model {\it (i)} of independent depolarizing channels~\eqref{depolarising_channel_iid}, and the pessimistic model {\it (iii)} with a multi-qubit depolarizing channel~\eqref{depolarising_channel_any}.  { When we adopt the optimistic noise model and we employ Igor (purple curve) then for $t/T_2>0.3$ there exists a small advantage as compared to not using Igor to correct the logical qubit (red curve). Thus Eq.~\eqref{eq:be_pont} is fulfilled}.  However, when multiple qubit errors are { fully enabled} by the noise model, and we use the multi-qubit depolarizing channel (`Igor', blue data points), the advantage disappears. This highlights the importance of modeling correlated errors appropriately, going thus beyond simplified error models that use the same single-qubit channel after each elementary operation in the quantum circuit.
			For reference, the behavior  of an unencoded, bare physical qubit under the same environmental (dephasing) noise is also shown (grey data points). The inset shows a zoom of the parameter interval in which QEC becomes advantageous: for a total waiting time $\tau$ larger than about 300ms it becomes advantageous to apply an imperfect Igor QEC cycle { but only for  independent depolarizing noise. In this figure and those following the inset shares the same axes labels as the main plot.} }
	\end{centering}
\end{figure}

\begin{figure}
	\begin{centering}
		\includegraphics[width=0.9\columnwidth]{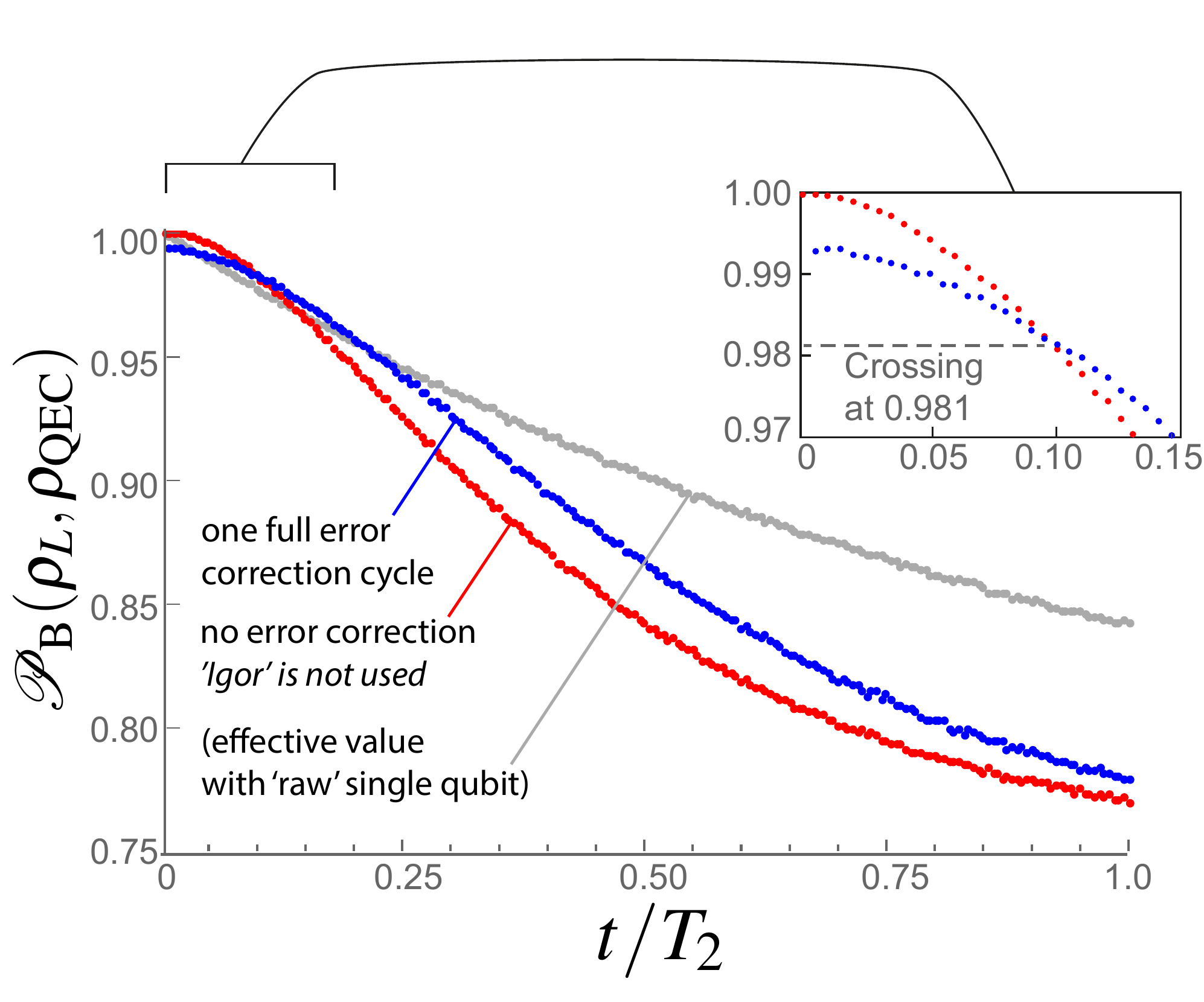}\\
		\caption{\label{Fig:OXF_simulation_2speciesShuttleFuture} {\bf { Success probability $\mathcal{P}_{\rm B}$} under the  two-species shuttling-based QEC cycle with multi-qubit MS gates} (cf. Fig.~\ref{Fig:Shuttling_2species_overview}): The parameters underlying the simulation correspond to the \textit{anticipated improved values} from Tables~\ref{tab:summary_gates}-\ref{table_errors:MS}. The noise model for the imperfect 5-ion MS gate operations corresponds to the worst-case noise model {\it (iii)} of multi-qubit depolarizing noise~\eqref{depolarising_channel_any}. Results show that there exists an ample parameter region (at times larger than about 100ms) in which the application of an imperfect `Igor' QEC cycle becomes advantageous~\eqref{eq:be_pont}  as compared to not applying it. Note that this takes place at $\mathcal{P}_{\rm B}$ values of 0.981, much higher than in the shuttling-based scenario 1, with a marginal gain at $\mathcal{P}_{\rm B}$ values of about 0.92. Note that in the present scheme, for not  too long total times $\tau$, below about 200ms, applying an imperfect QEC cycle is advantageous even as compared to a single, non-encoded physical qubit undergoing dephasing noise of the same strength, such that the more-stringent regime~\eqref{eq:be_pont_bare} for beneficial QEC can also be achieved.} 
	\end{centering}
\end{figure}

\begin{figure}
	\begin{centering}
		\includegraphics[width=0.9\columnwidth]{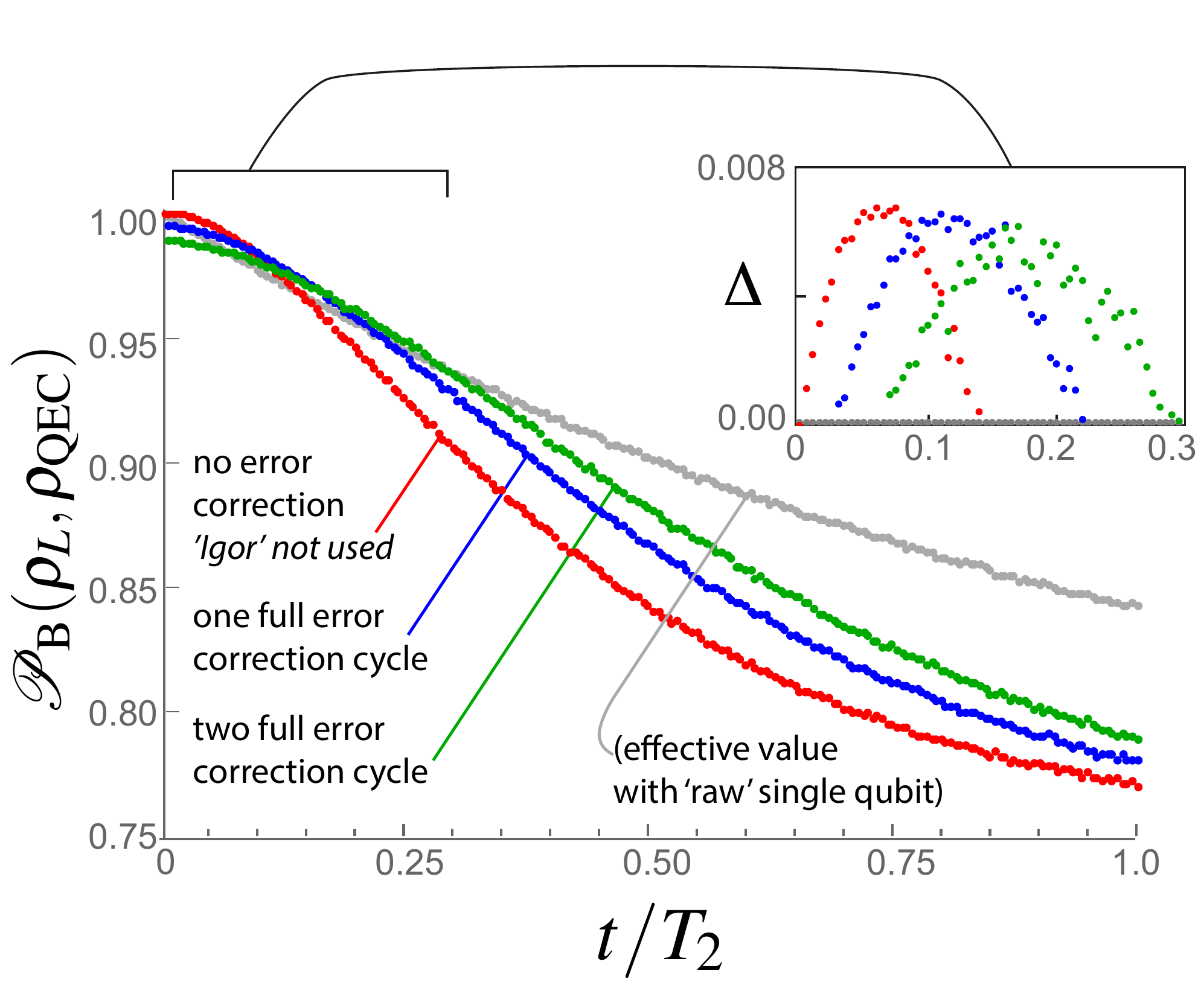}\\
		\caption{\label{Fig:OXF_simulation_2speciesShuttleFuture_multiple_rounds} {\bf { Success probability $\mathcal{P}_{\rm B}$} under  repetitive  two-species shuttling-based QEC cycles   with multi-qubit MS gates} (cf.~Fig.~\ref{Fig:Shuttling_2species_overview}): We consider the same scenario as in Fig.~\ref{Fig:OXF_simulation_2speciesShuttleFuture} with the \textit{ anticipated improved values} of Tables~\ref{tab:summary_gates}-\ref{table_errors:MS}, but the model for the imperfect 5-ion MS gate operations corresponds to the physically-motivated noise model {\it (ii)} of one- and two-qubit depolarizing quantum channel~\eqref{depolarising_channel_1_2}. For a single QEC cycle, direct comparison to the results of Fig.~\ref{Fig:OXF_simulation_2speciesShuttleFuture} show no appreciable difference. Hence, we can conclude that using the more pessimistic noise model with a multi-qubit depolarizing channel~\eqref{depolarising_channel_any}, or using the one with equally-likely one- and two-qubit errors~\eqref{depolarising_channel_1_2} does not make any difference. We also depict the results for two rounds of QEC (green dots), where one sees an increase of the  region~\eqref{eq:be_pont_bare} where QEC is advantageous  compared to a single non-encoded physical qubit with respect to the case with a single QEC round. { The inset plots $\Delta$ which is $\mathcal{P}_{\rm B}$ relative to that for the single-qubit memory. We see that} multiple rounds of QEC allow to sustain the logical qubit for a longer period of time. In the inset, we display the relative { performance} $\Delta$ of the encoded (red), single-cycle QEC (blue), two-cycle QEC (green) with respect to the un-protected physical qubit, which is obtained by subtracting the bare-qubit { $\mathcal{P}_{\rm B}$} of the main panel (grey line), from { $\mathcal{P}_{\rm B}$} for   the different schemes, also in the main panel. As emphasized before, we observe a wider region of advantage for the two cycles of QEC. } 
	\end{centering}
\end{figure}

\begin{figure}
	\begin{centering}
		\includegraphics[width=0.9\columnwidth]{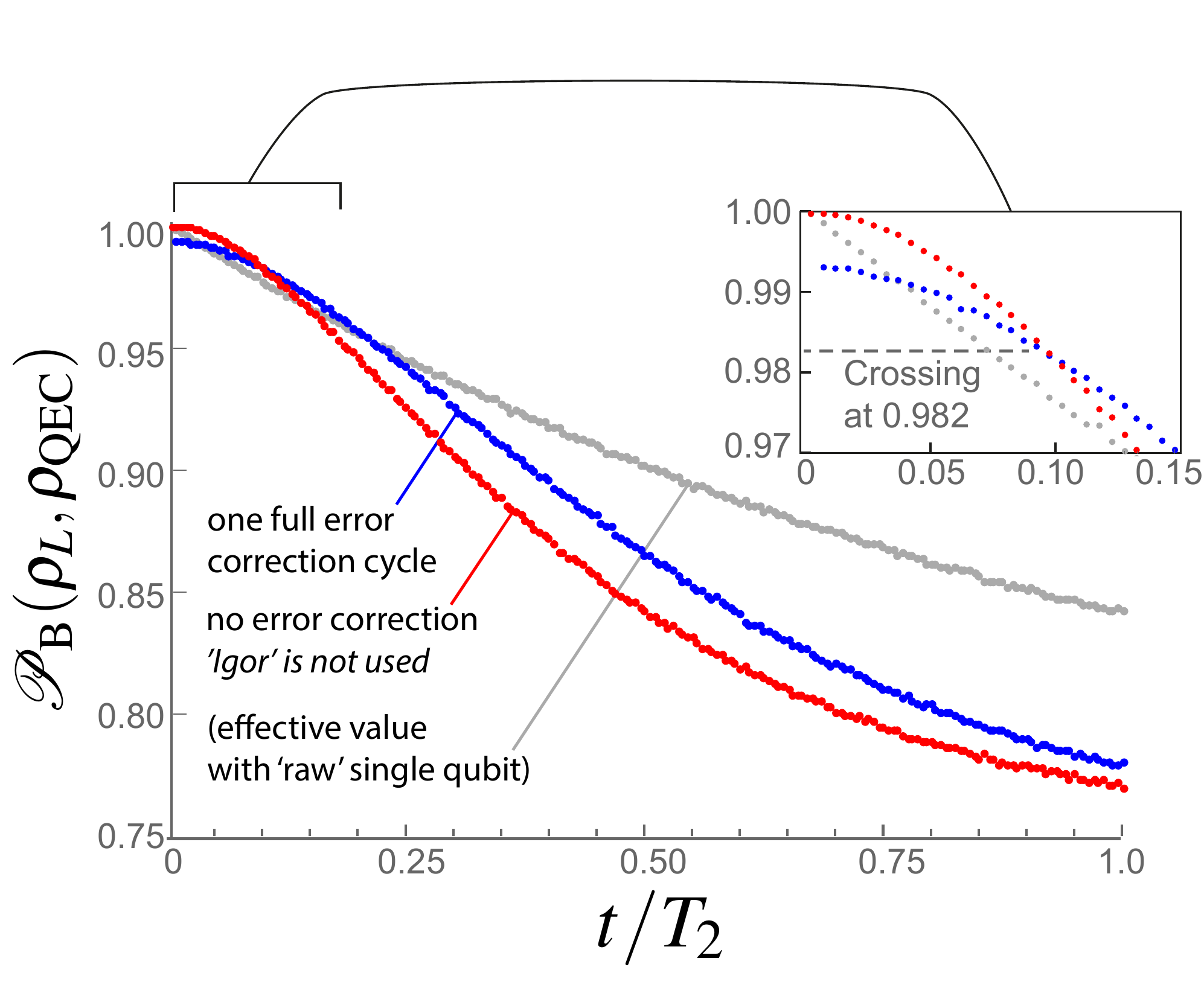}\\
		\caption{\label{Fig:OXF_simulation_hidingFuture} {\bf { Success probability $\mathcal{P}_{\rm B}$} under the hiding-based two-species QEC cycle with multi-qubit MS gates} (cf.~Fig.~\ref{Fig:Hiding_scheme}): The parameters underlying the simulation correspond to the \textit{future improved  values} from Tables~\ref{tab:summary_gates}-\ref{table_errors:MS}.  Experimental capabilities (two species, cooling, etc.) are the same as in Fig.~\ref{Fig:OXF_simulation_hidingNowOPTIMISTIC}, and we use the more challenging (worst-case) noise model {\it (iii)} of multi-qubit depolarizing noise~\eqref{depolarising_channel_any}.
			Results show that there exists a clear parameter window for which the application of an imperfect `Igor' QEC cycle becomes advantageous both as compared to not applying it~\eqref{eq:be_pont}, as well as compared to an unprotected single physical qubit~\eqref{eq:be_pont_bare}. Note that the { $\mathcal{P}_{\rm B}$} value where the QEC cycle crossover towards a beneficial `Igor' takes place is around 0.982, which is very similar to the behavior found for the shuttling-based protocol 2 with future parameters (cf. Fig.~\ref{Fig:OXF_simulation_2speciesShuttleFuture}).	} 
	\end{centering}
\end{figure}

The figures~\ref{Fig:OXF_simulation_hidingNowOPTIMISTIC}-\ref{Fig:OXF_simulation_hidingFuture} in this section constitute the results of our first set of simulations, i.e. those involving MS gates with multiple qubits (5 qubits, specifically). Note that for these simulations, the qubit reset fidelity is assumed to be $5\cdot 10^{-3}$, see Table~\ref{tab:summary_gates}. The results shown in each figure correspond to one of the trapped-ion QEC protocols described in detail in the previous section, and we refer to them here as ``single-species shuttling without cooling'' (shuttling-based protocol 1, in Fig.~\ref{Fig:OXF_simulation_1speciesFutureOPTIMISTIC}), ``dual-species shuttling with cooling'' (shuttling-based protocol 2, in Figs.~\ref{Fig:OXF_simulation_2speciesShuttleFuture} and~\ref{Fig:OXF_simulation_2speciesShuttleFuture_multiple_rounds}) and ``hiding'' (hiding-based protocol, in Figs.~\ref{Fig:OXF_simulation_hidingNowOPTIMISTIC} and~\ref{Fig:OXF_simulation_hidingFuture}). 

Each data point in our figures is the averaged result of at least $40,000$ runs. Each curve is formed from $200$ data points and therefore involves eight million runs in total. For reference, the grey curves show how a single physical qubit would perform if used in the same setting described in Sec.~\ref{sec:QEC_criterion}, i.e. Alice prepares it in a given state and Bob measures it to guess the state (versus the orthogonal state) after the state has been subjected to environmental noise. Moreover, the red curve in each figure shows the performance of the  color code of $7$ physical qubits, but without the error correction provided by `Igor' mid-way through the time $\tau$ during which the encoded state is subjected to environmental noise. The blue curves show how this changes when indeed Igor's cycle is performed. Regrading the criteria for beneficial QEC discussed in Sec.~\ref{sec:QEC_criterion}, we wish to see the blue curve above the red one such that criterion~\eqref{eq:be_pont} is achieved, and ideally even above the grey one, implying that (at least for some choices of duration of the experiment)  equation~\eqref{eq:be_pont_bare} is also fulfilled. If our simulations display such crossing,  we can conclude that advantageous QEC could be achieved in the experiments, given that the particular performance of the different building blocks is realized. We note that in Fig.~\ref{Fig:OXF_simulation_2speciesShuttleFuture_multiple_rounds} there is an additional curve, in green, which shows the effect of applying Igor's correction twice, at 33\% and 67\% of the Alice-Bob time interval; again, one hopes to see the curves associated with Igor surpass the red, or even the grey curves.

From our numerical simulations (see Fig.~\ref{Fig:OXF_simulation_hidingNowOPTIMISTIC} for the hiding approach), we can conclude  that the ``Current'' performance figures for the gate times, fidelities, and so on, would be in general insufficient to prove a beneficial QEC cycle: the curves representing the integrity of the logical qubit at the end of the period $\tau$ are strictly lower when error correction is applied mid-way, versus simply omitting to perform any such correction. This is principally caused by  the complexity of the required circuits, and by the higher error rate of the entangling MS gates as compared to other building blocks of the protocol. Essentially, in these low-distance codes, the MS gates  introduce more noise than can be removed by the QEC cycle. For the shuttling-based approaches with current parameters (not shown in figures), a similar poor performance is found (e.g. for the one-species scheme without re-cooling, the MS gates become so noisy after a few rounds (see Table~\ref{table_errors:MS}), that we can directly discard it from reaching the break-even point~\eqref{eq:be_pont}).

Remarkably,  our results  for the ``Future'' performance are far more encouraging. The least successful hardware variant is QEC protocol based on  a single-species shuttling without re-cooling (cf. Fig~\ref{Fig:OXF_simulation_1speciesFutureOPTIMISTIC}). For this protocol,  the crossing into beneficial effects of the QEC cycle~\eqref{eq:be_pont} occurs only when the total time $\tau$ is such that the logical qubit receives considerable dephasing from the environment.  Moreover, the beneficial effect can vanish entirely when one moves to a more pessimistic model for the MS gate noise including correlations (noise models {\it (ii)} and {\it (iii)} discussed in Sec.~\ref{subsec:Depol_models_MS_gate}). Additionally, the desirable property of outperforming the unprotected physical qubit in our particular task of state discrimination through the QEC cycle~\eqref{eq:be_pont_bare} cannot be achieved.  Fortunately, the results for the ``two-species shuttling'' (cf. Figs.~\ref{Fig:OXF_simulation_2speciesShuttleFuture} and~\ref{Fig:OXF_simulation_2speciesShuttleFuture_multiple_rounds}) and the ``hiding'' protocols (cf. Fig.~\ref{Fig:OXF_simulation_hidingFuture}), using the ``future'' performance numbers, are considerably more encouraging. One sees that the crossing to a beneficial error correction cycle happens early and with a high value of the integrity. Importantly, this implies that multiple rounds of error correction can be beneficially applied, such that the logical qubit can be sustained for a longer time (see Fig.~\ref{Fig:OXF_simulation_2speciesShuttleFuture_multiple_rounds}).

Let us now address if one could obtain still better results  by implementing the QEC using sequential 2-qubit MS gates rather than multi-qubit MS gates. While this would necessitate more gates in total, each gate has a  higher fidelity and, moreover, would propagate errors from the ancillary qubits onto the data qubits in a more restricted fashion. In addition to addressing this question numerically, the sequential 2-qubit MS gates will be an essential ingredient for the realization of fault-tolerant QEC, which we also explore in this section. The necessary analysis and scheduling for this second approach was described above in Section~\ref{switchToFTand2qubitMS}. We now describe the results of the corresponding set of simulations, which employ the 2-qubit MS gate as the entangling operation. For this second set of simulations, we also assumed the better value of $1\cdot 10^{-3}$ for the qubit reset fidelity from Table~\ref{tab:summary_gates}.

\subsubsection{Shuttling-based two-species QEC with 2-qubit MS gates: exploring fault tolerance}

\begin{figure}
	\begin{centering}
		\includegraphics[width=0.95\columnwidth]{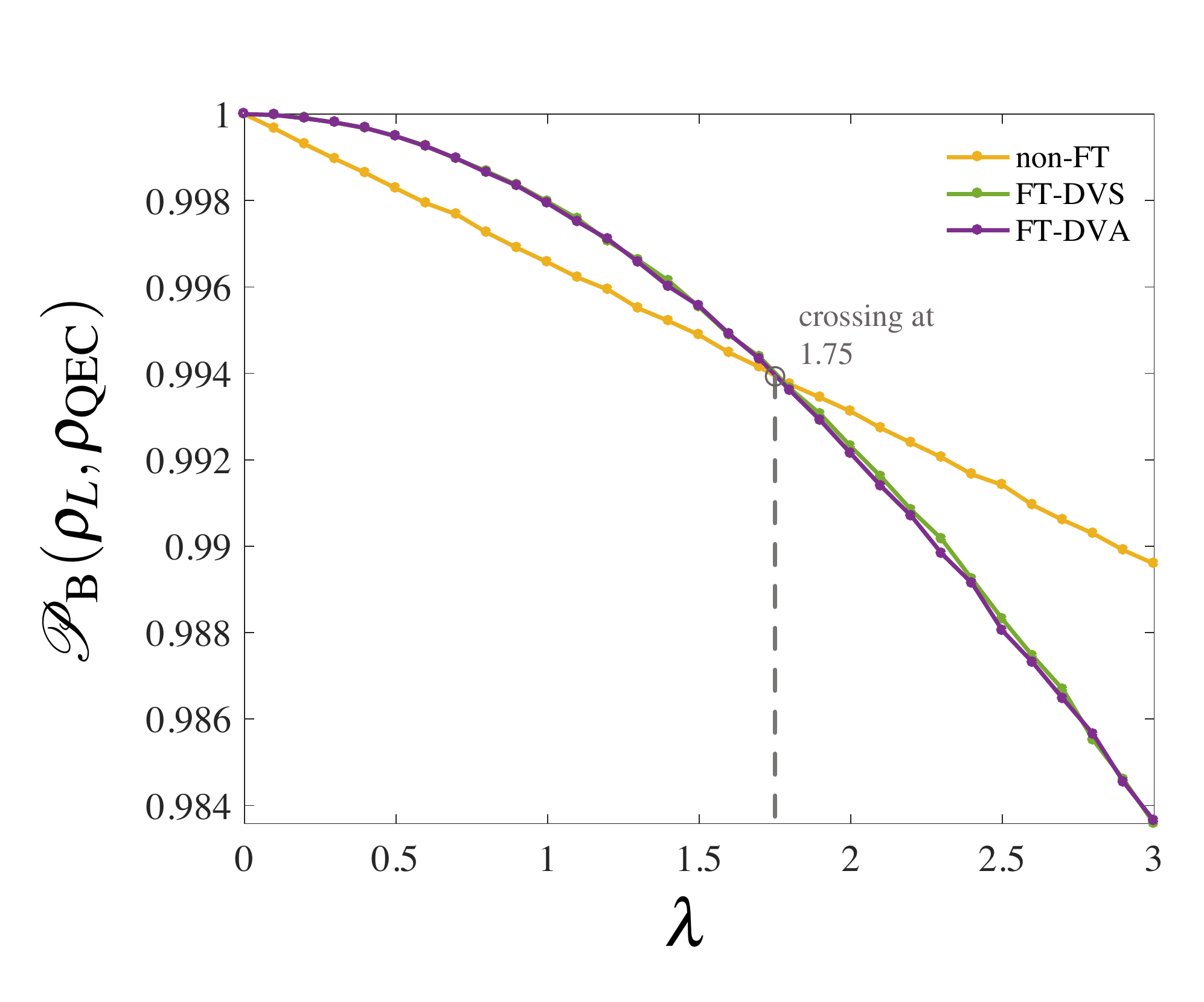}\\
		\caption{\label{Fig:OXF_noenvironment} {{ {\bf Success probability $\mathcal{P}_{\rm B}$ under two-species shuttling-based QEC cycles with only QEC errors:} This graph shows how $\mathcal{P}_{\rm B}(\rho_L,{\rho}_{\rm QEC})$ changes with parameter $\lambda$, which is defined in the main text and adjusts error rates within Igor's cycle. The three plots show: a non-fault-tolerant two-species shuttling-based scheme based on 2-qubit MS gates (cf.~\ref{Fig:Shuttling_2species_2qb_gates_overview}), and the fault-tolerant DV-S scheme (cf.~\ref{Fig:DVS_real_space_scheme}), and the fault-tolerant DV-A scheme (cf.~\ref{Fig:DVA_real_space_scheme}}}. As described in the text, the inverted quadratic curves, as compared to the linear behaviour, are the signature of a correct fault-tolerant circuit. The simulation parameters correspond to the \textit{future improved values} from Tables~\ref{tab:summary_gates}-\ref{table_errors:MS}.  The noise applied for the imperfect 2-ion MS gate operations follows the standard depolarizing model.} 
	\end{centering}
\end{figure}

Before presenting the simulation results to explore the break-even point for the full fault-tolerant QEC protocols, let us first present a  simpler analysis to verify that the DiVicenzo-Shor and DiVicenzo-Aliferis schemes with MS-gates of Sec.~\ref{sec:cnot_alternative} are indeed fault tolerant in the formal sense. To achieve this, the periods  $\tau$ of environmental exposure in the Alice-Igor-Bob scenario are removed. Instead, Alice presents the flawless logical qubit directly to Igor, who performs a redundant  round of error correction, and then directly passes the logical qubit to Bob for the logical state discrimination. Hence, there is no effect of the environmental noise, except during the time when Igor is applying  his imperfect error correction cycle. We introduce a  parameter $\lambda$  to control the severity of the imperfections in the operations applied by Igor, which multiplies the error rates that have been identified as the expected  hardware targets (see Tables~\ref{tab:summary_gates}-\ref{table_errors:MS}). Setting $\lambda=1$ corresponds to assuming that all these targets are exactly met. By plotting { Bob's success probability $\mathcal{P}_{\rm B}$} as a function of $\lambda$, we should see a linearly descending curve for a non-FT protocol (because any single gate failure within Igor's circuit can  { have the consequence of reducing $\mathcal{P}_{\rm B}$}), whereas   a characteristically inverted quadratic curve should arise for a true FT protocol (because it requires two or more gate failures to damage the integrity, the probability of which goes as $\lambda^2$).  The results of our numerical simulations are shown in Figure~\ref{Fig:OXF_noenvironment}. The predicted  shape of the curve is indeed observed through our numerics, thus verifying that the analytically derived protocols based on MS gates, and their translation into the numerical simulation, are correctly fault tolerant. 

{ It is relevant to note that our approach of requiring better-than-breakeven performance, when Igor enters the picture, is closely related to the concept of a pseudo-threshold where a higher level of concatenation outperforms the preceeding level (see e.g. Ref.~\cite{Aho2006}). Moreover, as we noted in the introduction to Section~\ref{sec_qec_protocols}, thresholds in the range of $10^{-4}$ are expected in the context of fault tolerance~\cite{Terhal2009,Aho2006}. Therefore the crossing seen in Fig.~\ref{Fig:OXF_noenvironment} and the corresponding gate fidelities (i.e. $1.75$ times those in Tables~\ref{tab:summary_gates}-\ref{table_errors:MS}) are reassuringly close to expectations. The very recent suggestions of Chao and Reichardt~\cite{ChaoReichardt2017} for smaller ancilla structures in fault tolerant circuits may provide some further boost to the transition.}

\begin{figure}
	\begin{centering}
		\includegraphics[width=1\columnwidth]{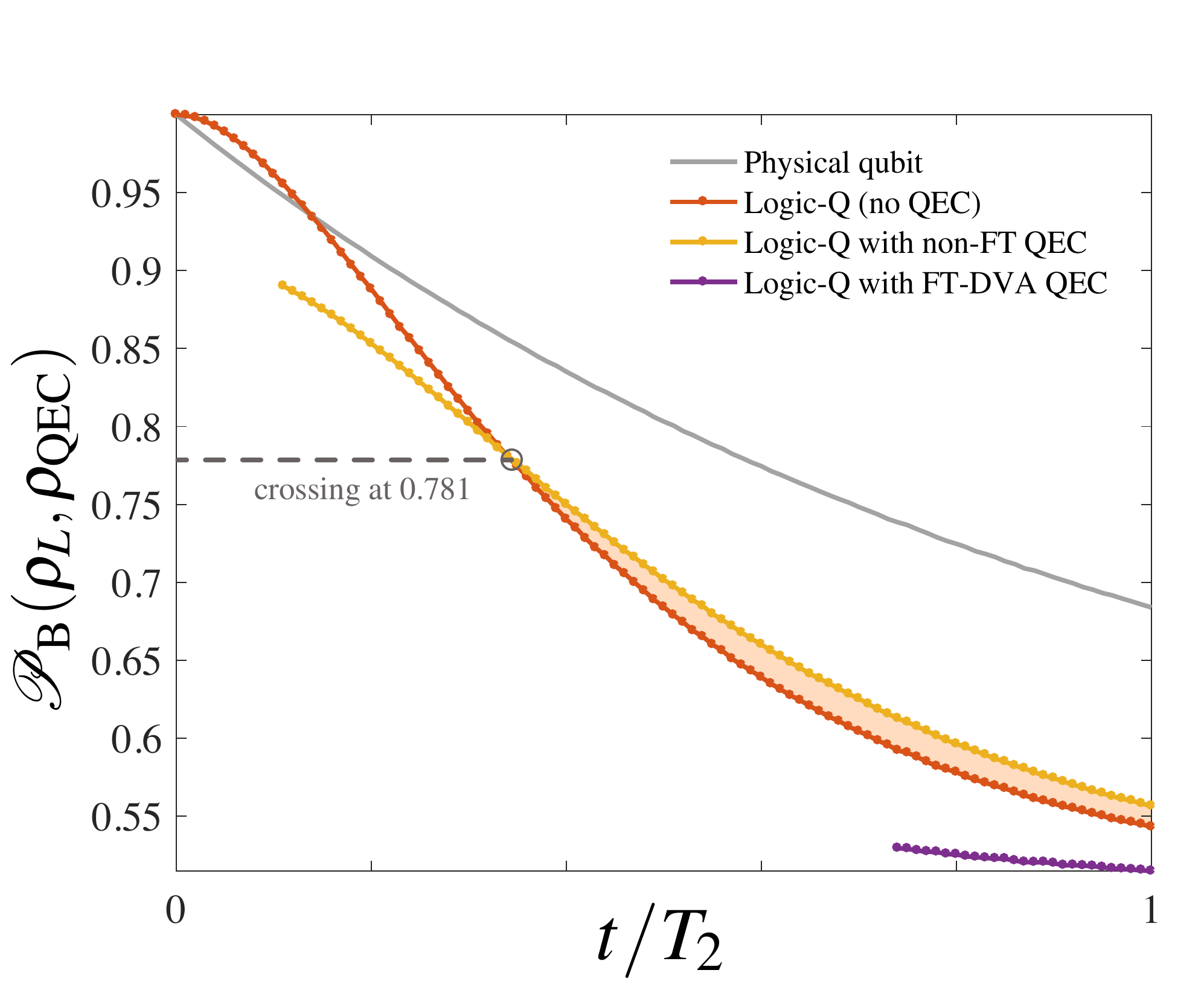}\\
		\caption{\label{Fig:OXF_simulation_2speciesShuttleCurrent} {\bf Today's hardware: { Success probability $\mathcal{P}_{\rm B}$ with}  QEC according to the shuttling-based two species protocol}: The standard Alice-Igor-Bob protocol with the parameters underlying the simulation corresponding to the \textit{current values} from Tables~\ref{tab:summary_gates}-\ref{table_errors:MS}. The red curve shown here { shows Bob's performance in the absence of Igor, so that with errors occur only due to the environment} (see Table~\ref{table:first_criterion}). For reference, the { equivalent plot} for a single physical qubit with the same environmental noise is also shown (grey curve). We see that { using a logical qubit and the} non-FT QEC cycle (yellow) can produce a small positive { effect; the shaded region indicates this beneficial region.}. However, the fully FT protocol (purple line), when it is  possible (see main text), is always inferior.   } 
	\end{centering}
\end{figure}

Having thus verified the nature of the circuits, we can proceed to assess their performance when there are finite periods of exposure to the environment. First, we simulate using the \textit{current values} of operational infidelities from Tables~\ref{tab:summary_gates}-\ref{table_errors:MS}. The results are shown in Fig.~\ref{Fig:OXF_simulation_2speciesShuttleCurrent}. In this Figure, and the remaining figures in this section, the underlying protocol is that of 2-species shuttling with re-cooling (as described in previous sections). Typically, for the figures in this section, each data point is aggregated from one million numerical experiments, and each full curve involves about 100 data points; thus a curve represents approximately $10^8$ numerical experiments.

Figure~\ref{Fig:OXF_simulation_2speciesShuttleCurrent} shows two reference lines for the  criteria of beneficial QEC: the grey line { indicates the performance with} a single physical qubit~\eqref{eq:be_pont_bare}, while the red line shows the { performance with} an encoded logical qubit when no error correction is performed by Igor~\eqref{eq:be_pont}. As one would expect, for very short periods of environmental exposure, the red line  lies above the grey one, since the probability for two (or more) errors affecting the encoded data qubits is much smaller than the  single-qubit  error probability affecting the bare qubit.  The yellow line shows the performance of the  non-FT QEC  protocol that employs sequential 2-qubit MS gates (cf. Fig.~\ref{Fig:Shuttling_2species_2qb_gates_overview}). As can be observed,  while this line  never beats the  { $\mathcal{P}_{\rm B}$ for} the bare  physical qubit~\eqref{eq:be_pont_bare}, it does indeed exceed the { performance of} the encoded qubit~\eqref{eq:be_pont} once the environmental exposure is severe. Note that this result  indicates the superior performance of the QEC schemes with two-ion MS gates  versus the previous ones based on multi-ion MS gates. We recall that for the corresponding multi-ion MS circuits, no clear advantage could be seen using current hardware performance i.e. there was no { equivalent} to the yellow-red crossing  of Fig.~\ref{Fig:OXF_simulation_2speciesShuttleCurrent}. Let us remark that this result is  non-trivial, since the complexity of the circuits using sequential 2-ion MS gates increases considerably with respect to the schemes that exploit multi-qubit MS gates (compare  Fig.~\ref{Fig:Shuttling_2species_2qb_gates_overview} to Fig.~\ref{Fig:Shuttling_2species_overview}). In any case, the break-even point~\eqref{eq:be_pont} is achieved when the integrity of the qubit has already decayed considerably. In order to take full advantage of QEC, improving on this feature, we will now consider the DV-S and DV-A schemes for fault-tolerant QEC.

The purple line of Fig.~\ref{Fig:OXF_simulation_2speciesShuttleCurrent}, just visible in the lower right corner of the plot, is the performance of the DiVicenzo-Aliferis (DV-A) protocol. The line does not exist over the majority of the graph simply because there is insufficient time to perform a full error correction cycle due to the long circuit depth associated to Fig.~\ref{Fig:DVA_real_space_scheme}. { The time required for Igor's actions is summarised in Table~\ref{tab:timeCosts}. Note that for the FT approaches, it is necessary to evaluate each stabilizer more than once in order to control measurement errors; Igor performs the checks twice, and then a third time if the first two outcomes do not agree. For the Shor FT scheme, it is necessary to prepare and verify an ancilla state prior to stabilizer measurement. If the verification fails we must restart the ancilla preparation. In our simulations we allow for up to four such restarts; the probability that more are needed is negligible even for current technologies. 
	
	Notice that in Fig.~\ref{Fig:OXF_simulation_2speciesShuttleCurrent}, even when} the total protocol time  is large enough to permit Igor to act, the performance  of the DV-A scheme is very poor, and { none} of the break-even points for useful QEC can be reached. The conclusion from this set of simulations is that a device built with the `current' performance numbers could suffice for a basic demonstration of QEC, but could not possibly make a successful demonstration of a fault { tolerant} QEC code. Fortunately, this picture changes as we move to the future  performance figures.

\begin{figure}
	\begin{centering}
		\includegraphics[width=1\columnwidth]{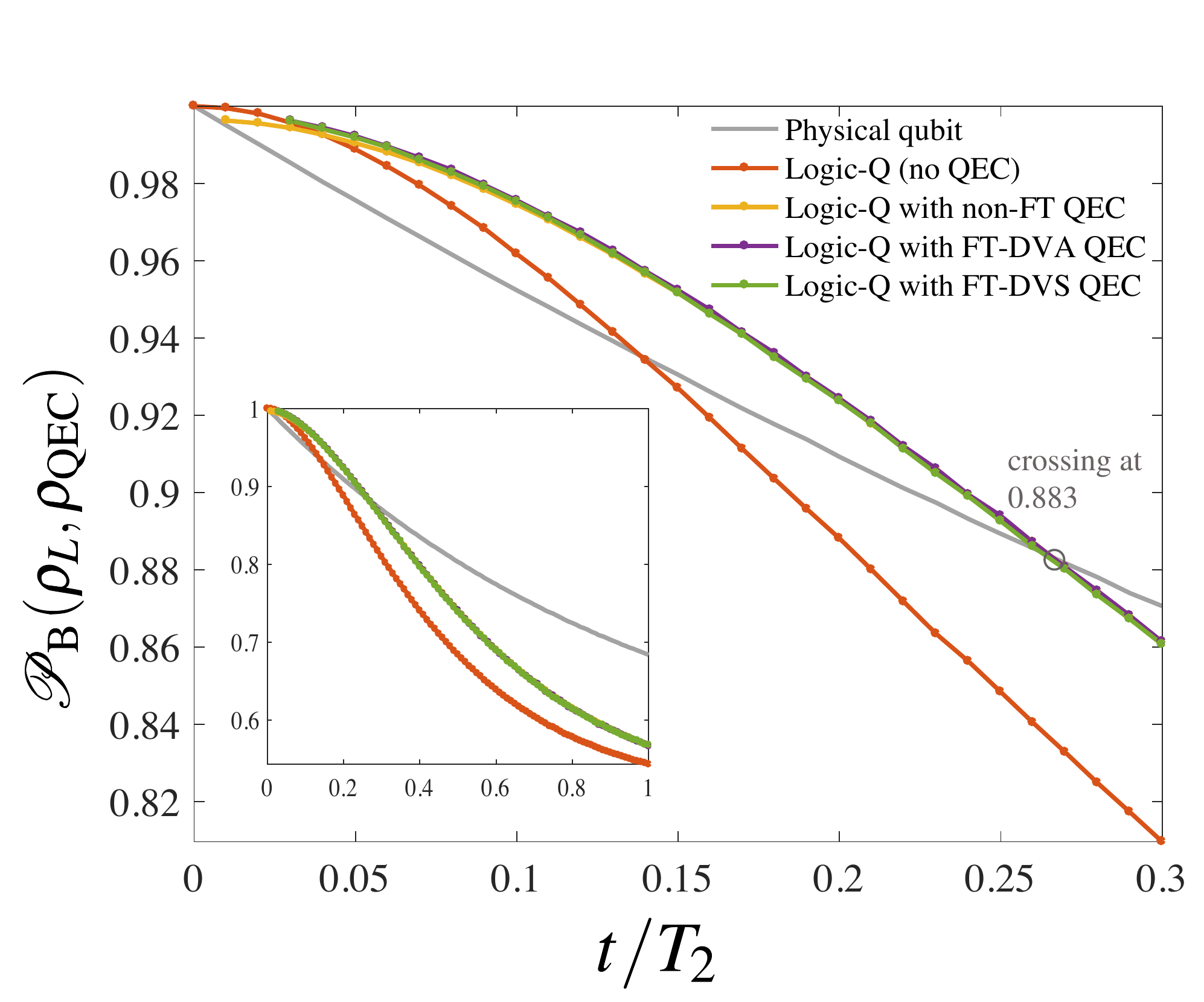}\\
		\caption{\label{Fig:OXF_simulation_2speciesShuttleFuture-m} {\bf Future hardware: { Success probability $\mathcal{P}_{\rm B}$ with}  QEC according to the shuttling-based two species protocol}: The scenario simulated in this case is the same as in Fig.~\ref{Fig:OXF_simulation_2speciesShuttleCurrent}, except that the parameters applied correspond to the \textit{future improved  values} from Tables~\ref{tab:summary_gates}-\ref{table_errors:MS}.  Performance is obviously profoundly improved, as discussed in the main text.} 
	\end{centering}
\end{figure}

\begin{figure}
	\begin{centering}
		\includegraphics[width=1\columnwidth]{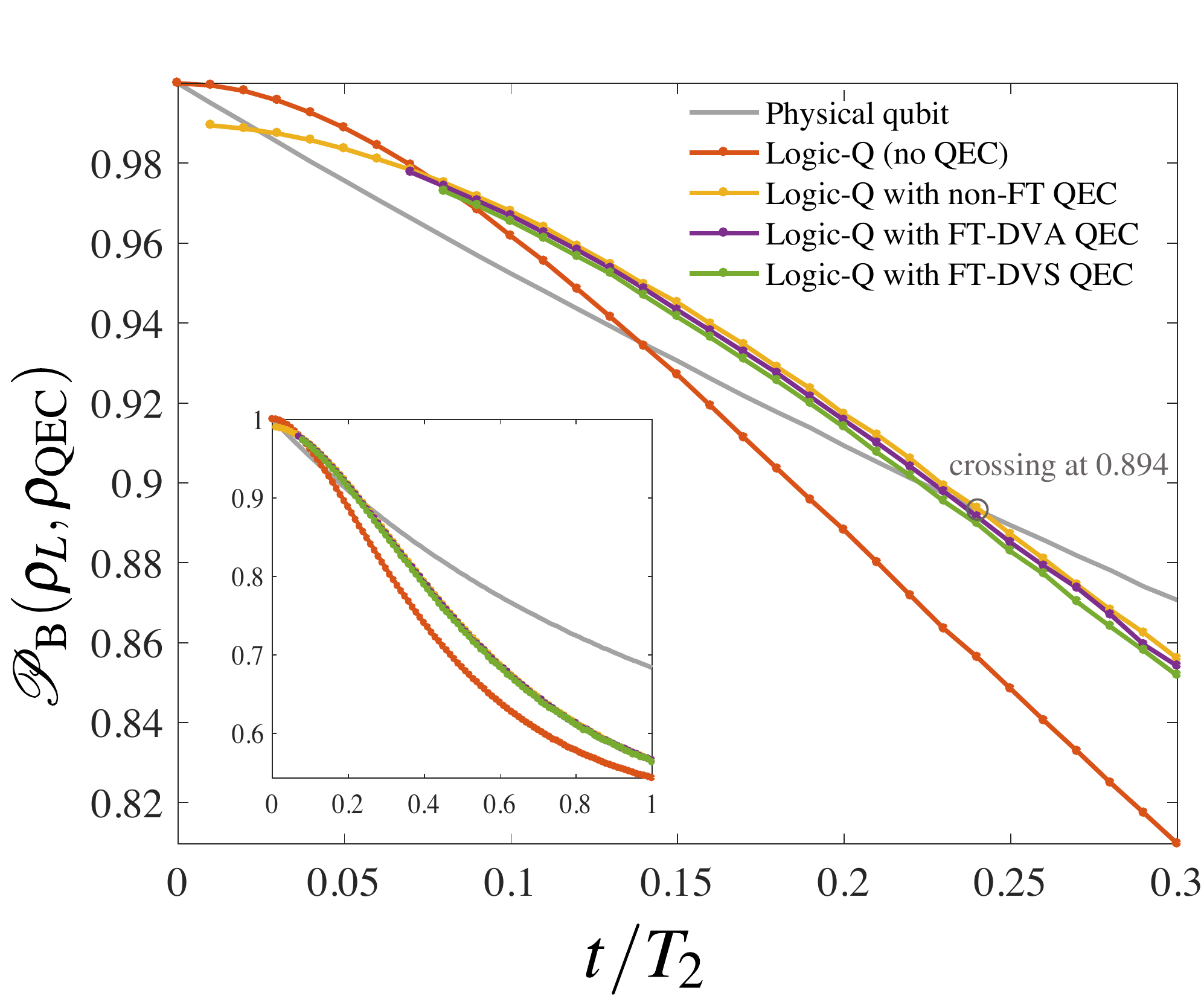}\\
		\caption{\label{Fig:OXF_simulation_2speciesShuttleFuture3timesWorse} {\bf Inferior future hardware: { Success probability $\mathcal{P}_{\rm B}$ with}  QEC according to the shuttling-based two species protocol}: The error model and scenarios simulated here are the same with Fig.~\ref{Fig:OXF_simulation_2speciesShuttleFuture-m}. The operational numbers are scaled to yield a three times worse performance with respect to those of the \textit{future improved  values}: each operation takes three times longer, and the gate fildelities  are three times worse { (the environmental dephasing rate remains unchanged)}. As discussed in the main text, while the performance here is degraded versus the preceding figure, it is nevertheless sufficient to largely demonstrate the goal of beneficial QEC.}
	\end{centering}
\end{figure}

Figure~\ref{Fig:OXF_simulation_2speciesShuttleFuture-m} shows the results of repeating the simulations in Fig.~\ref{Fig:OXF_simulation_2speciesShuttleCurrent} but now with the \textit{future anticipated  values} from Tables~\ref{tab:summary_gates}-\ref{table_errors:MS}. One can see that the performance is profoundly improved. Now, the lines for the non-FT Igor, and both the DV-A and DV-S fault-tolerant protocols, lie almost on top of one another. Moreover, they both beat  the ``no-Igor'' red line, fulfilling Eq.~\eqref{eq:be_pont} for the whole period of time studied numerically. Additionally, they also lie above  the ``single physical qubit'' grey line, fulfilling Eq.~\eqref{eq:be_pont_bare}, over a wide range of values of the environmental exposure time. Finally, we note that  for longer times, one could use multiple rounds of error correction. Although the non-FT curve and the two FT curves seem nearly identical, the latter do outperform the non-FT protocol for small levels of environmental error (see top left of the figure). This is consistent with our expectations from Fig.~\ref{Fig:OXF_noenvironment}, where we learned that when Igor's hardware performs exactly at the future anticipated level (i.e. $\lambda=1$) then the FT protocols are superior to the non-FT one.

\begin{table}
	\centering
	{
		\begin{tabular}{|c|c|c|} \hline
			Protocol & Total time (current) (ms) &  Total time (anticipated) (ms) \\\hline\hline
			non-FT &   21.2 & 5.9 \\ \hline
			FT DVA &  49.3$\times n$  & 14.3$\times n$ \\ \hline
			FT DVS & 46.0$\times n$ + 3.7$\times m$ & 13.1$\times n$ + 1.0$\times m$ \\ \hline
		\end{tabular}
		\caption{{\bf Time required for one cycle of error correction with both current and future hardware}, assuming the shuttling-based two species protocol. Parameter $n$ is the number  rounds of error correction applied (usually $n=2$, but $n=3$ when the initial two rounds disagree). Parameter $m$ is the number of {\it additional} attempts at preparing  the GHZ ancilla state beyond the minimum, due to detection of  error(s) when verifying the ancilla. On average, $m=0.66$ for current hardware and $m=0.02$ for future hardware. }
		\label{tab:timeCosts}
	}
\end{table}

The curves in Figure~\ref{Fig:OXF_simulation_2speciesShuttleFuture-m} are so encouraging, that it may turn out that the trapped-ion hardware development does not need to reach the expected values  of Tables~\ref{tab:summary_gates}-\ref{table_errors:MS} in order to achieve the goal of beneficial  QEC. In order to test this feature, we have  tripled the error rates in all operations, and analyzed the performance for the QEC procedures. The results are shown in Fig.~\ref{Fig:OXF_simulation_2speciesShuttleFuture3timesWorse}. We see that now there is a slight variation in performance with the non-FT Igor marginally superior to FT-DV-A, which in turn is marginally superior to FT-DV-S. (Note that it is to be expected that the FT circuits are now inferior to the non-FT circuit, since by tripling the error we are now at the far right of the range in Fig.~\ref{Fig:OXF_noenvironment}.) In Fig.~\ref{Fig:OXF_simulation_2speciesShuttleFuture3timesWorse} we also see that the non-FT Igor is quicker to perform than DV-A which in turn is quicker than DV-S, since the curves are not plotted when there is insufficient Alice-Bob time interval for a complete Igor cycle. Additionally the figure shows that the crossing of the curves with error correction and the physical qubit occurs only slightly  earlier in Fig.~\ref{Fig:OXF_simulation_2speciesShuttleFuture3timesWorse} (0.894) than that in Fig.~\ref{Fig:OXF_simulation_2speciesShuttleFuture-m}. {\em  The conclusion from this figure is that, even using a system with three times greater operational infidelities with respect to the expected future estimates of our work,  one  could nevertheless support a strong demonstration of beneficial QEC.} This is  a very encouraging for the near-term development of trapped-ion QEC.

\begin{figure}
	\begin{centering}
		\includegraphics[width=1\columnwidth]{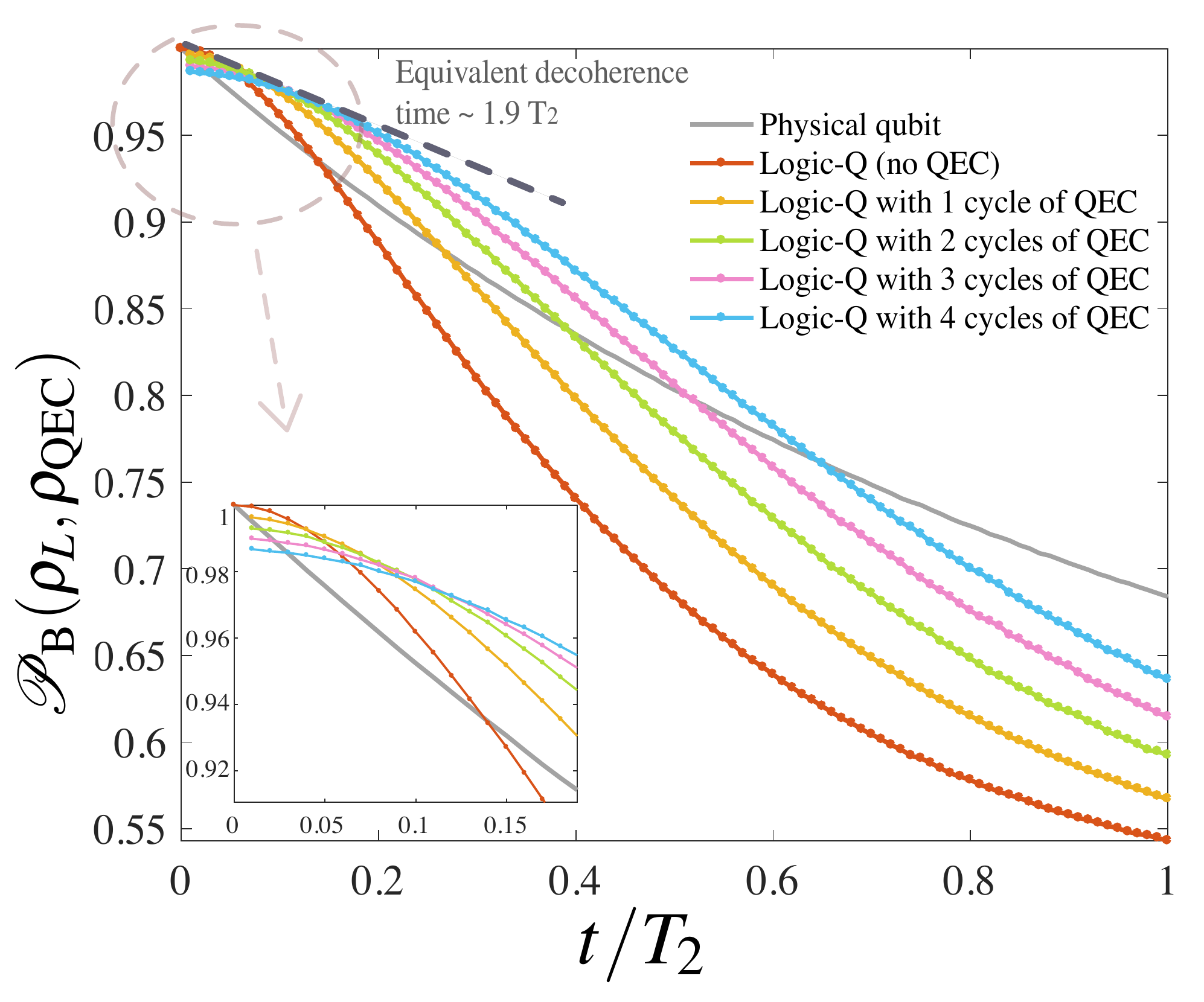}\\
		\caption{\label{Fig:OXF_simulation_2speciesFutureShuttleMultiIgor} {\bf Multiple cycles of future hardware: { Success probability $\mathcal{P}_{\rm B}$ with}  QEC according to the shuttling-based two species protocol}: The parameters underlying the simulation correspond to the \textit{future improved values} from Tables~\ref{tab:summary_gates}-\ref{table_errors:MS}. The QEC method used here is the non fault-tolerant one (cf.~\ref{Fig:Shuttling_2species_2qb_gates_overview}), as it takes the shortest possible time, and thus has more potential if more error correction cycles are to be applied within one round. Results show that the application of more QEC cycles sustain the logical qubit for a longer time, as depicted by the grey dashed line, which is drawn based on the outline of the curves representing that with error correction. As discussed in the main text, the dashed line { allows us to infer the {\it effective}} rate of decay of integrity of the logical qubit.
		} 
	\end{centering}
\end{figure}

{ In our final set of simulations, shown in Figure~\ref{Fig:OXF_simulation_2speciesFutureShuttleMultiIgor}, we consider} multiple rounds of error correction considering the future expected levels of infidelity. These results show that, by several rounds of QEC, one can  protect the logical qubit at a level that is superior to a single physical qubit over a sustained period of time. By the simple principle of selecting the number of Igor cycles according to the total Alice-Bob interval, we find that the coherence time of the logical qubit can reach values nearly twice as big as  the raw physical qubit -- a very significant alteration that should be easily observed and which demonstrates the `encoded qubit alive' goal clearly. { Of course, a factor of two is far from sufficient to achieve large scale quantum computing, but this is to be expected since we are employing only one layer of a small code. For further error suppression, one would either concatenate the Steane code (recursively replacing each data qubit with a full logical qubit, through at least three or four levels) or else one would scale using topological techniques. As the Steane code is smallest instance of the 2D Color Code, the latter would be an attractive option -- but such considerations are beyond the scope of the present paper.  }

{ This concludes our review of our numerical simulations. It is worth reiterating that in every case we presume that Alice and Bob are perfect since we are interested in the integrity of the memory itself, separate from the creation or analysis of the logical qubit. One may wonder whether this  presents difficulties for experimental tests, since in reality Alice, Igor and Bob are merely phases of a single experiment and presumably suffer the same error rates. An analysis of this point is beyond the scope of the present paper, but it has been considered in a subsequent work~\cite{Xiaosi2017}. The encouraging conclusion is that, for a broad class of error models (including the typical ones), experimental evaluation of integrity is possible even with noisy Alice and Bob. The key observation is that typically it suffices for Alice to randomly choose between a fixed subset of possible states to send; then, if one can find a fault tolerant circuit for Alice to use to prepare each such state, and a corresponding fault tolerant analysis method for Bob, one finds that the noisy nature of their actions has relatively little impact on the measured integrity.
	
	Finally, we remark that while the error models considered here have been stochastic, the framework we have introduced applies to any form of noise and therefore a full exploration of coherent and even non-Markovian noise is an interesting prospect. A recent paper has highlighted the potential for coherent errors to impact QEC performance in a qualitatively different way~\cite{Barnes2017}.
}

\section{\bf Conclusions and outlook}
\label{sec:conclusions}

In this work, we have presented a detailed description of  current/future experimental  capabilities for the implementation of  topological QEC  with trapped-ion crystals.  We have also described the characteristics of the main sources of noise and imperfections in the experiments. Based on this discussion, we have introduced a complete trapped-ion toolbox for QEC, including a discussion of fault-tolerant designs based on the characteristics of the trapped-ion  set of available gate operations. Using this toolbox, we have presented  different protocols to implement a QEC cycle based on the 7-qubit color code, which  exploit either crystal reconfiguration, or spectroscopic decoupling/re-coupling techniques. We have  derived  effective error models for the different building blocks of these QEC cycles, which are in close connection with the experimental sources of noise, and go well-beyond the simplified standard approaches  that use  quantum channels  affecting all QEC operations with the same probability. Using these effective models for the current and expected performance of the QEC building blocks, we have performed extensive numerical simulations to determine the experimental conditions that are required for these QEC protocols to become beneficial, a fundamental and necessary condition for any future  implementation of QEC. Moreover, we have also assessed the conditions for the encoded logical qubit to outperform the physical unprotected qubit for a particular quantum-information task.

From this  study, and in light of our numerical results, we can draw the following conclusions. The  performance of the 7-qubit trapped-ion color code for the single- or two-species, shuttling- or hiding-based, QEC protocols with multi-qubit MS gates, and assuming the "current'' performance of the experimental building blocks, is inadequate to achieve the break-even point of beneficial QEC. Therefore, the trapped-ion hardware must be improved. Using the operational criteria~\eqref{eq:be_pont}-\eqref{eq:be_pont_bare} introduced in this work, we have been able to assess and quantify the required experimental  improvements towards the QEC goal, and present realistic values  of the different building blocks   that must be achieved (see Tables~\ref{tab:summary_gates}-\ref{tbl:shuttlingops}).  Our numerical  results for  the ``future'' expected improvements show that the crossing onto a beneficial QEC cycle can indeed be achieved with either the shuttling- or the hiding-based protocols with multi-qubit MS gates, and occurs  at much earlier times and with a much higher value of the integrity of the encoded qubit. Therefore, we conclude that it will be of primary importance  to incorporate and optimize the QEC building blocks towards the values introduced in Tables~\ref{tab:summary_gates}-\ref{tbl:shuttlingops} for the success of trapped-ion implementations of the  QEC color code  with  two-species ion crystals. Moreover, we have observed a clear advantage of the QEC schemes based on sequential 2-qubit MS gates, especially in the context of the fault-tolerant designs. Therefore, our studies show that the natural next step in the progress towards trapped-ion fault-tolerant  QEC will be to upgrade the syndrome extraction routines according to the schemes hereby introduced. We finally note that for the presented protocols, not only is the necessary condition for a beneficial QEC cycle~\eqref{eq:be_pont} fulfilled, but that we have also  shown that the encoded qubit can perform better than the unprotected qubit~\eqref{eq:be_pont_bare}. Moreover, we have shown that repetitive QEC cycles can sustain the integrity of the logical qubit for increasing periods of time, provided that the above break-even point is achieved.


\acknowledgements
 The research is based upon work supported by the Office of the Director of National Intelligence (ODNI), Intelligence Advanced Research Projects Activity (IARPA), via the U.S. Army Research Office Grant No. W911NF-16-1-0070. The views and conclusions contained herein are those of the authors and should not be interpreted as necessarily representing the official policies or endorsements, either expressed or implied, of the ODNI, IARPA, or the U.S. Government. The U.S. Government is authorized to reproduce and distribute reprints for Governmental purposes notwithstanding any copyright annotation thereon. Any opinions, findings, and conclusions or recommendations expressed in this material are those of the author(s) and do not necessarily reflect the view of the U.S. Army Research Office. 

We also acknowledge support by U.S. A.R.O. through Grant No. W911NF-14-1-010. A.B.  acknowledges support from Spanish MINECO Projects FIS2015-70856-P,  and CAM regional research consortium QUITEMAD+.   P. S., T. M. and R. B. acknowledge support from the Austrian Science Fund (FWF), through the SFB FoQus (FWF Project No. F4002-N16) and the Institut f\"{u}r Quanteninformation GmbH. 

 

\end{document}